\documentclass[tradiabstract]{aa} 

\usepackage{graphicx}
\usepackage{txfonts}

\usepackage{epsf}
\usepackage{rotating}
\usepackage{graphics}

\def \kms{km/s}
\def \ms{m\,s$^{-1}$}
\def \1s{$1\,\sigma$}

\def \t0{T$_0$}

\def\kp{p}
\def\lp{k}

\begin{document}

   \title{The SOPHIE search for northern extrasolar~planets\thanks{Based 
            on observations collected with the SOPHIE  spectrograph on the 1.93-m telescope at
            Observatoire de Haute-Provence (CNRS), France, by the SOPHIE Consortium 
            (programs 07A.PNP.CONS to 15A.PNP.CONS). 
The full version of the SOPHIE measurements (Table~\ref{table_rv})
is only available in electronic form at the CDS via anonymous ftp to 
cdsarc.u-strasbg.fr (130.79.128.5)
or via http://cdsweb.u-strasbg.fr/cgi-bin/qcat?J/A+A/TBC.}}
            
   \subtitle{X. Detection and characterization of giant planets by~the~dozen}

   \author{G.~H\'ebrard\inst{1,2}, 
	L.~Arnold\inst{2},
	T.~Forveille\inst{3}, 
	A.\,C.\,M.~Correia\inst{4,5},
	J.~Laskar\inst{5},
	X.~Bonfils\inst{3},
	I.~Boisse\inst{6},
	R.\,F.~D\'{\i}az\inst{7},
	J.~Hagelberg\inst{8},
	J.~Sahlmann\inst{9},
	N.\,C.~Santos\inst{10,11},
	N.~Astudillo-Defru\inst{3,7},
	S.~Borgniet\inst{3},
	F.~Bouchy\inst{6}, 
	V.~Bourrier\inst{7,1},
	B.~Courcol\inst{6}, 
	X.~Delfosse\inst{3},
	M.~Deleuil\inst{6}, 
	O.~Demangeon\inst{6}, 
	D.~Ehrenreich\inst{7},
	J.~Gregorio\inst{12}, 
	N.~Jovanovic\inst{13,14}, 
	O.~Labrevoir\inst{15}, 
	A.-M.~Lagrange\inst{3},
	C.~Lovis\inst{7},
	J.~Lozi\inst{13}, 
	C.~Moutou\inst{6,16}, 
	G.~Montagnier\inst{1,2}, 
	F.~Pepe\inst{7},
	J.~Rey\inst{7},
	A.~Santerne\inst{10},
	D.~S\'egransan\inst{7},
	S.~Udry\inst{7}, 
	M.~Vanhuysse\inst{17}, 
	A.~Vigan\inst{6}, 
	P.\,A.~Wilson\inst{1}
}

  \offprints{G. H\'ebrard (hebrard@iap.fr)}

   \institute{
Institut d'Astrophysique de Paris, UMR7095 CNRS, Universit\'e Pierre \& Marie Curie, 
98bis boulevard Arago, 75014 Paris, France 
\email{hebrard@iap.fr}
\and
Observatoire de Haute-Provence, CNRS, Universit\'e d'Aix-Marseille, 04870 Saint-Michel-l'Observatoire, France
\and
Universit\'e J.~Fourier (Grenoble 1)/CNRS, Laboratoire d'Astrophysique de Grenoble (LAOG, UMR5571), France
\and
CIDMA, Departamento de F\'{\i}sica, Universidade de Aveiro, Campus de Santiago, 3810-193 Aveiro, Portugal
\and
ASD/IMCCE, CNRS-UMR8028, Observatoire de Paris,  PSL, UPMC, 77 Avenue Denfert-Rochereau, 75014 Paris, France
\and
Aix Marseille Universit\'e, CNRS, LAM (Laboratoire d'Astrophysique de Marseille) UMR 7326, 13388 Marseille, France
\and
Observatoire de Gen\`eve,  Universit\'e de Gen\`eve, 51 Chemin des Maillettes, 1290 Sauverny, Switzerland
\and
Institute for Astronomy, University of Hawaii, 2680 Woodlawn Drive, Honolulu, HI 96822, USA
\and
European Space Agency, European Space Astronomy Centre, P.O. Box 78, Villanueva de la Canada, 28691 Madrid, Spain
\and
Instituto de Astrof{\'\i}sica e Ci\^encias do Espa\c{c}o, Universidade do Porto, CAUP, Rua das Estrelas, 4150-762 Porto, Portugal
\and
Departamento\,de\,F{\'\i}sica\,e\,Astronomia,\,Faculdade\,de\,Ci\^encias,\,Universidade\,do\,Porto,\,Rua\,Campo\,Alegre,\,4169-007\,Porto,\,Portugal
\and
Atalaia Group, CROW-Observatory Portalegre, Portugal
\and
National Astronomical Observatory of Japan, Subaru Telescope, 650 North A'Ohoku Place, Hilo, HI, 96720, USA
\and
Department of Physics and Astronomy, Macquarie University, NSW 2109, Australia
\and
Centre d'Astronomie, Plateau du Moulin \`a Vent, 04870 Saint-Michel-l'Observatoire, France
\and
Canada France Hawaii Telescope Corporation, Kamuela, HI 96743, USA
\and
Oversky, 47 All\'ee des Palanques, 33127 Saint-Jean-d'Illac, France
}

   \date{Received TBC; accepted TBC}
      
\abstract{We present new radial velocity measurements of eight stars secured with the spectrograph 
SOPHIE at the 193-cm telescope of the Haute-Provence Observatory
allowing the detection and characterization of new giant extrasolar planets. 
The host stars are dwarfs of 
spectral types between F5 and K0
and magnitudes between 6.7 and 9.6; the planets have minimum masses 
$M_\textrm{p} \sin i$ between 0.4 to 3.8~M$_\mathrm{Jup}$ and orbital periods of several days to several months.
The data allow only single planets to be discovered around the first six stars (HD\,143105, HIP\,109600, HD\,35759, 
HIP\,109384, HD\,220842, and HD\,12484), but one of them shows 
the signature of an additional substellar
companion in the system. The seventh star, HIP\,65407, allows the discovery of two giant planets, 
just outside the 12:5 resonance in weak mutual interaction.
The last star, HD\,141399, was already known to host a four-planetary system; our additional data and 
analyses allow new constraints to be put on it. 
We present Keplerian orbits of all systems, together with dynamical analyses of the two multi-planetary~systems.
HD\,143105 is one of the brightest stars known to host a hot Jupiter, 
which could allow numerous follow-up studies to be conducted despite this is not a transiting system.
The giant planets HIP\,109600b, HIP\,109384b, and HD\,141399c are located in the habitable zone of their host star.
}

\keywords{planetary systems -- techniques: radial velocities -- stars: individual: HD\,143105, HIP\,109600, HD\,35759, HIP\,109384, HD\,220842, HD\,12484, HIP\,65407, HD\,141399}

\authorrunning{H\'ebrard et al.}
\titlerunning{Detection and characterization of giant planets by the dozen}

\maketitle

\section{Introduction}
\label{sect_intro}

Since its installation in late 2006 at the 193-cm telescope of the Haute-Provence Observatory (OHP), the 
spectrograph SOPHIE is used among other programs to perform a volume-limited survey of giant extrasolar 
planets (Bouchy et al.~\cite{bouchy09}). The goal of that survey is to improve the statistics on the exoplanet 
parameters and their hosting stars by increasing the number of known Jupiter-mass planets, as well 
as identify particular cases allowing interesting follow-up studies, e.g. on dynamics for multi-planetary 
systems, or on structure characterization if they turn out to transit their host star.
Indeed, numerous questions concerning giant planets remain open, 
first of all their formation processes which could operate through 
core-accretion (e.g. Mordasini et al.~\cite{mordasini09}; Guilera et al.~\cite{guilera10})
or disk-instability (e.g. Cai et al.~\cite{cai10}; Boss~\cite{boss11})
models. Their subsequent evolution is subject to debates as well, 
concerning for example their migration processes 
(e.g. Lin et al.~\cite{lin96}; Levison et al.~\cite{levison06};  Carter-Bond et al.~\cite{carter12})
or the existence of a population of inflated close-in 
planets, likely related to interactions with their host stars, e.g. though tides or irradiation 
(e.g. Baraffe et al.~\cite{baraffe14}).
The physics of their internal structures (e.g. Liu et al.~\cite{liu15})
and atmospheres (e.g.~Sing et al.~\cite{sing16})
are also far to be fully understood.
Thus, detecting and characterizing new giant planets provide constraints on the models of their formation 
and evolution, and allows the exploration of the diversity of planetary systems to be~pursued.

Among the  $\sim2000$ exoplanets known so far according the Exoplanet 
Encyclopaedia\footnote{http://exoplanet.eu.}, only about 1100 have a mass measurement, 
most of them measured from radial velocities. Less than 700 have a mass measured with an accuracy 
better than $\pm30$\,\%, and less than 280 have known both mass and radius 
(Han et al.~\cite{han14})\footnote{http://exoplanets.org.}. Accurate radial velocity measurements remain 
an efficient and powerful technique for research and characterization of exoplanetary systems.
By comparison to transiting surveys, they are particularly powerful to detect multi-planetary systems and
to construct the orbital periods distribution on its longer part. They also provide the planetary masses 
distribution, allowing super-Earths, giant planets, and brown dwarfs populations to 
be distinguished and~studied.

\begin{figure}[h!] 
\begin{center}
\includegraphics[width=\columnwidth]{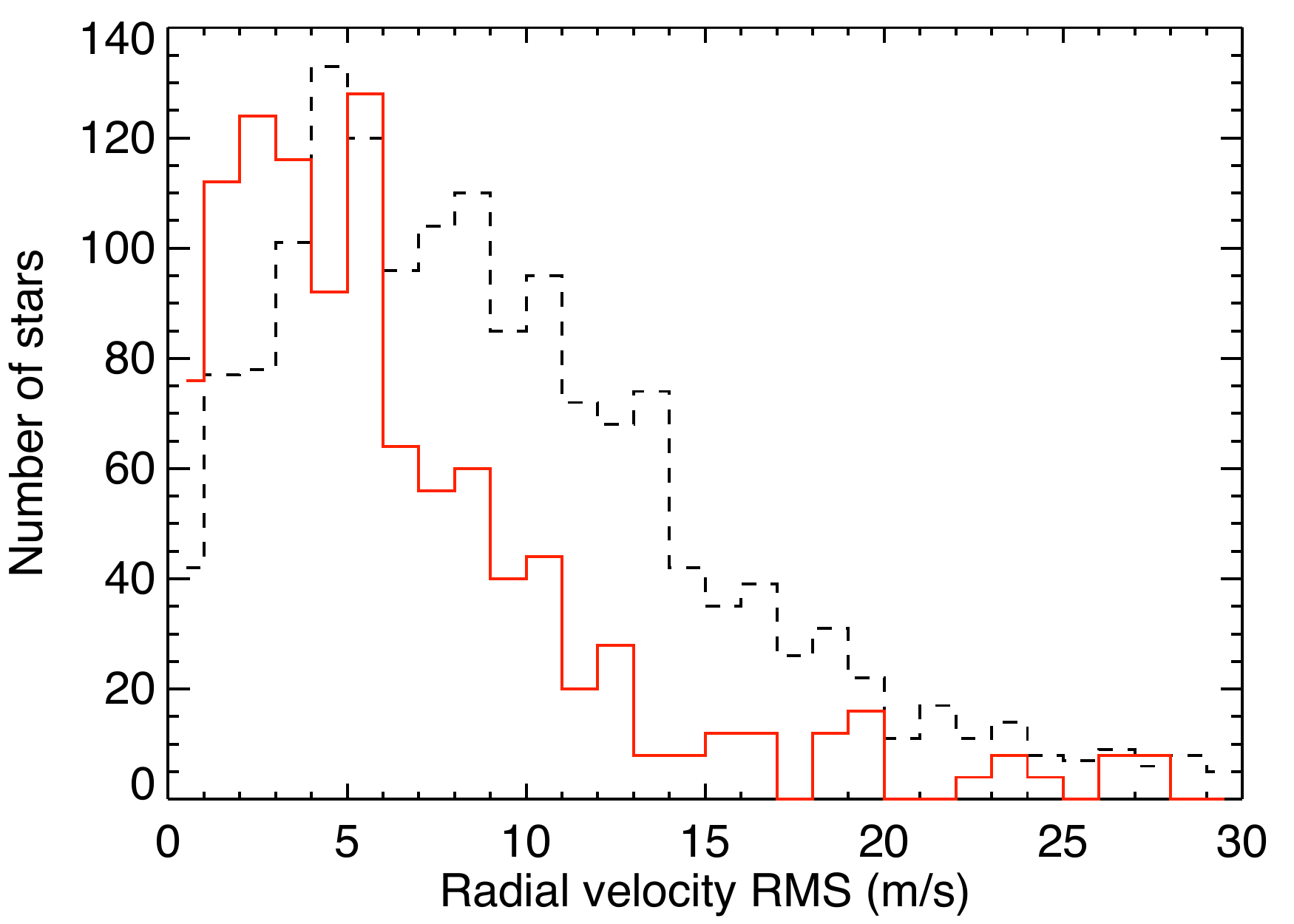}
\caption{Histogram of the radial velocity dispersions for 1900 stars observed with SOPHIE since 2006 
as part of the giant planets survey (dashed, black). The typical accuracy is of the order of $\pm7$\,m/s.
The histogram plotted in solid, red line selects only the measurements secured after the SOPHIE improvement 
done in June 2011; it shows an improvement of the radial velocity accuracy at a level of $\pm3.5$\,m/s.
To allow a better comparison between both histograms, 
the $y$-scale of the red histogram has been multiplied by a factor four.}
\label{fig_histo}
\end{center}
\end{figure}

In its current version, the catalog of the SOPHIE survey for giant planets includes 2300 dwarf stars 
at distance $<60$\,pc and with 
$B-V$ between 0.35 and 1.0. Most of them (about 1900) have already been  observed with SOPHIE. 
Figure~\ref{fig_histo} shows histograms of the radial velocities dispersion of all the stars with at least two 
SOPHIE measurements. Since its upgrade in June 2011, the typical 
accuracy of that program was reduced from $\pm7$\,m/s down to $\pm3.5$\,m/s. 
The main reason for that amelioration is the SOPHIE fiber-link scrambling properties improvement
due to the implementation of pieces of octagonal-section fibers (Bouchy et al.~\cite{bouchy13}).
Such accuracy allows Jupiter-mass planets to be detected on periods up to several years, or Saturn-mass planets 
on periods up to several months.

That SOPHIE survey has already allowed the detection and characterization of several planets 
(Da Silva et al.~\cite{dasilva08}; Santos et al.~\cite{santos08}; Bouchy et al.~\cite{bouchy09};
H\'ebrard et al.~\cite{hebrard10}; Boisse et al.~\cite{boisse10}; Moutou et al.~\cite{moutou14};
D\'{\i}az et al.~\cite{diaz16a})
and of more massive companions in the planet-brown dwarf boundary 
(D\'{\i}az et al.~\cite{diaz12}; Wilson et al.~\cite{wilson16}). 
Here we present the discovery of six new, single planets in orbit around the stars 
HD\,143105, HIP\,109600, HD\,35759, HIP\,109384, HD\,220842, and HD\,12484, and 
of a two-planet system around HIP\,65407. We also present new measurements 
and analyses of the multi-planetary system orbiting HD\,141399, previously announced by 
Vogt et al.~(\cite{vogt14}) while we were independently monitoring it with SOPHIE.

The SOPHIE observations and data reduction are presented in Sect.~\ref{sect_observations}. 
We determine the parameters of the host stars in Sect.~\ref{sect_stellar_properties}, then derive and 
discuss the planetary systems properties in Sects.~\ref{sect_single_planet_systems} 
and~\ref{sec_multi_planet_systems} for single and multi-planetary systems, respectively, before 
concluding in Sect.~\ref{sect_conclusion}.

\begin{figure*}[] 
\begin{center}
\includegraphics[scale=0.34]{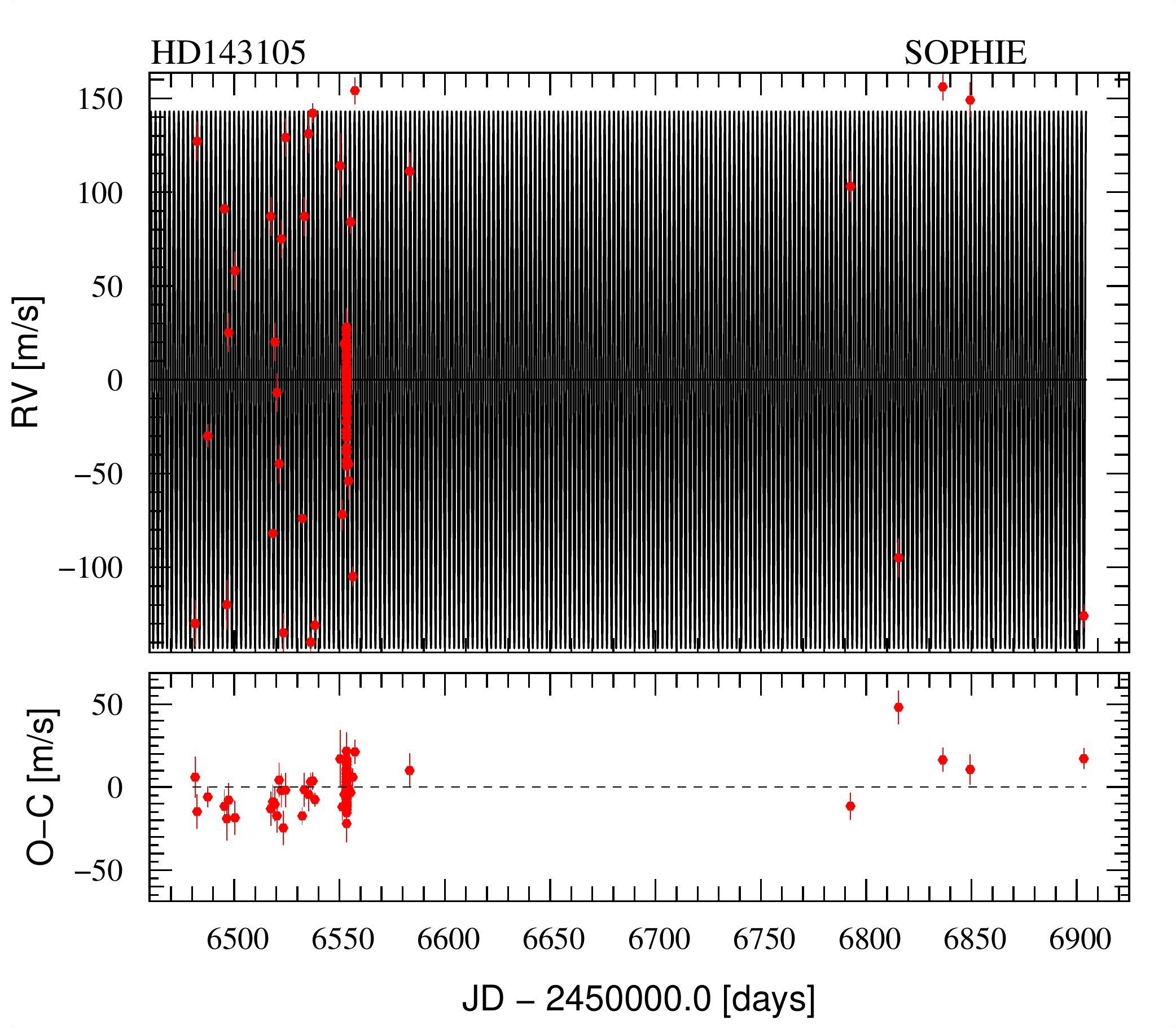}
\includegraphics[scale=0.34]{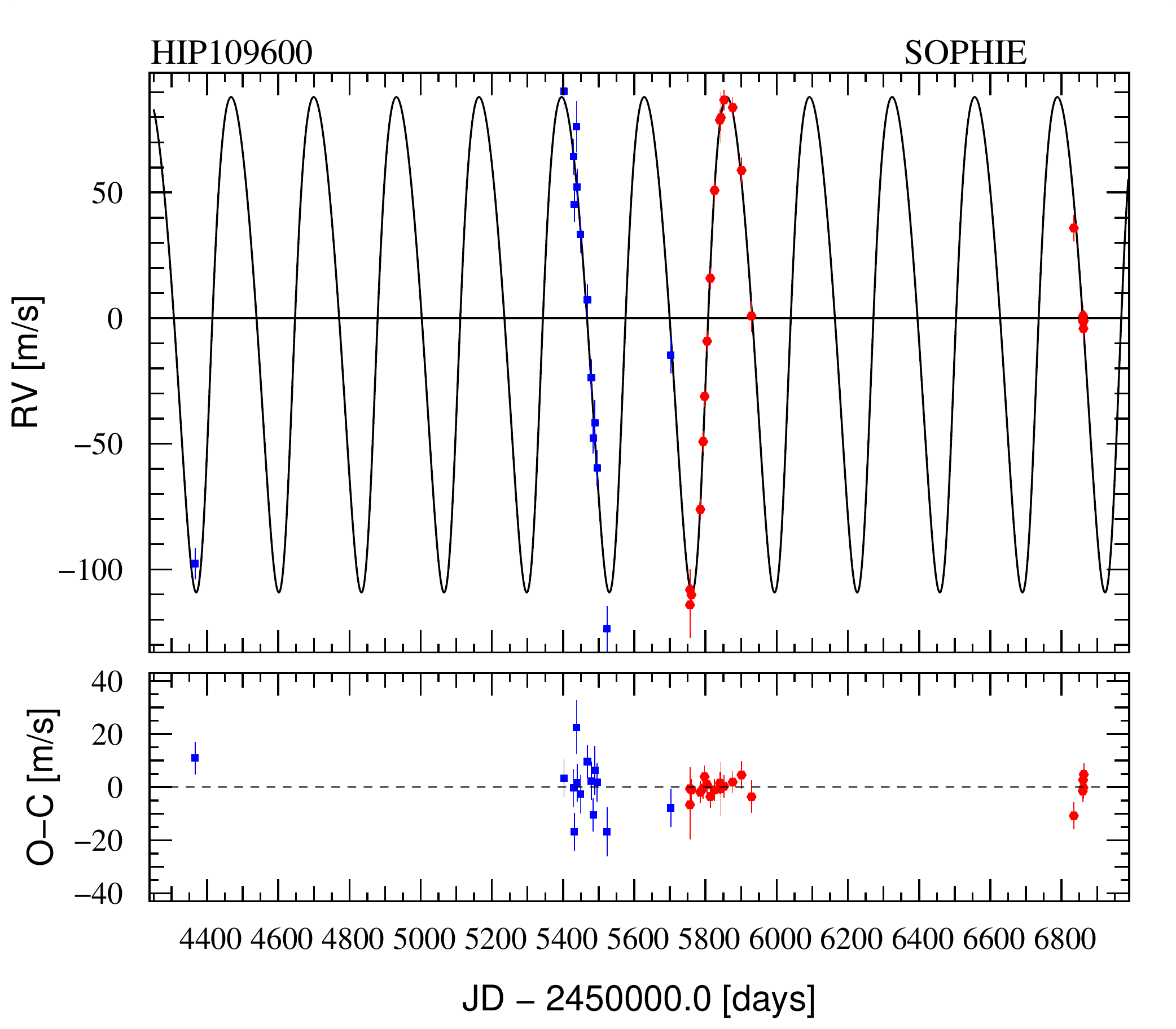}
\includegraphics[scale=0.34]{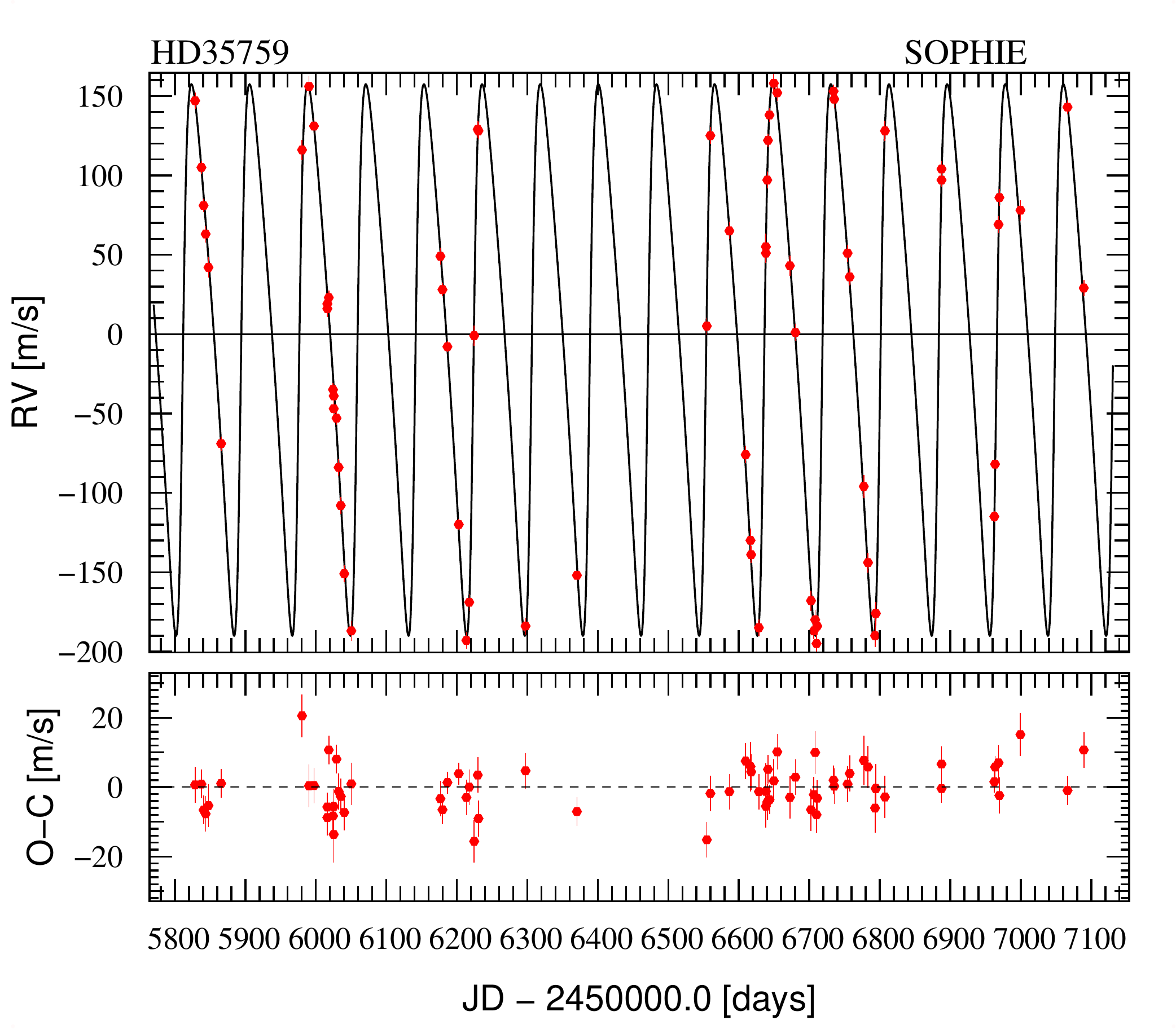}
\includegraphics[scale=0.34]{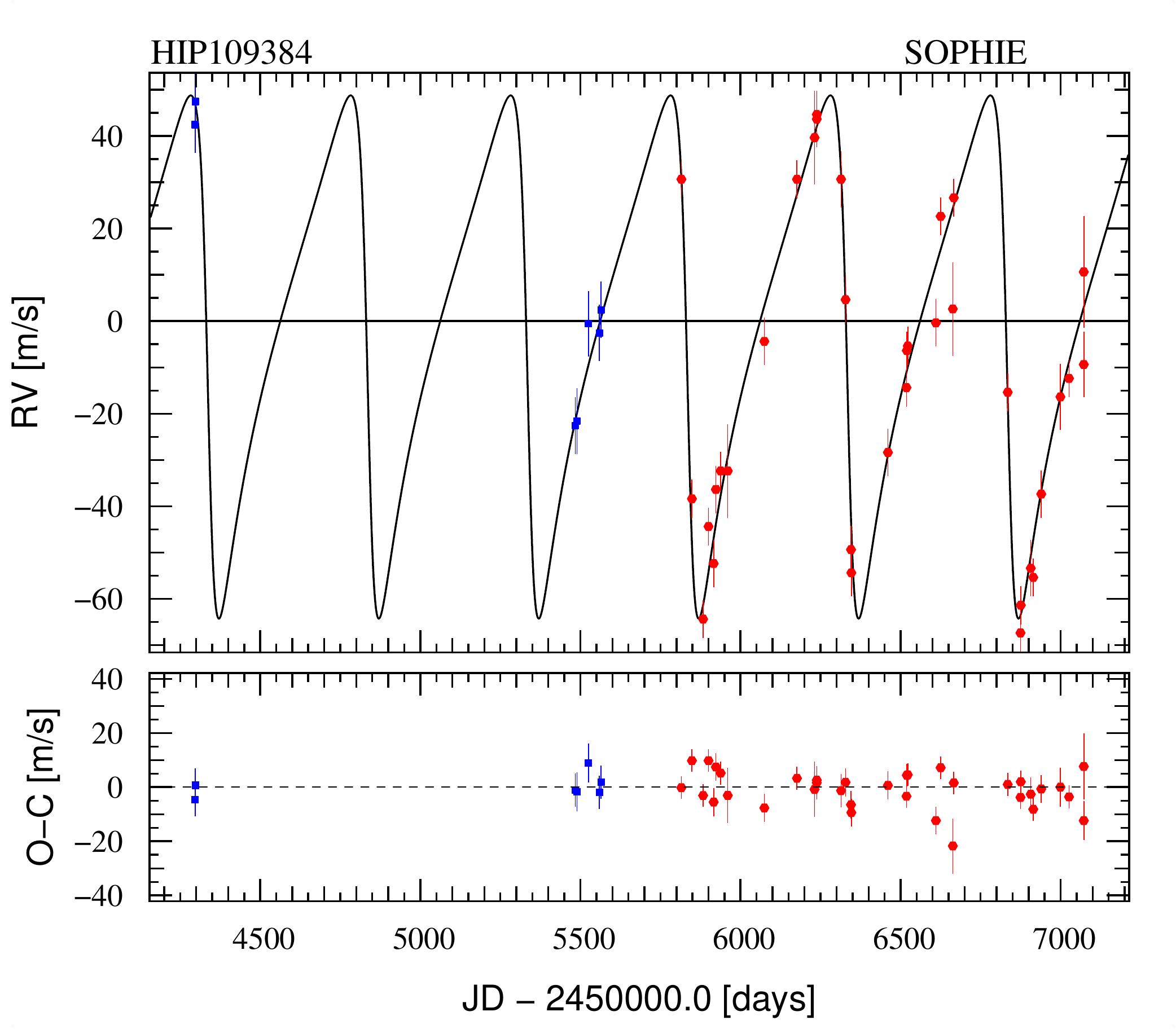}
\includegraphics[scale=0.34]{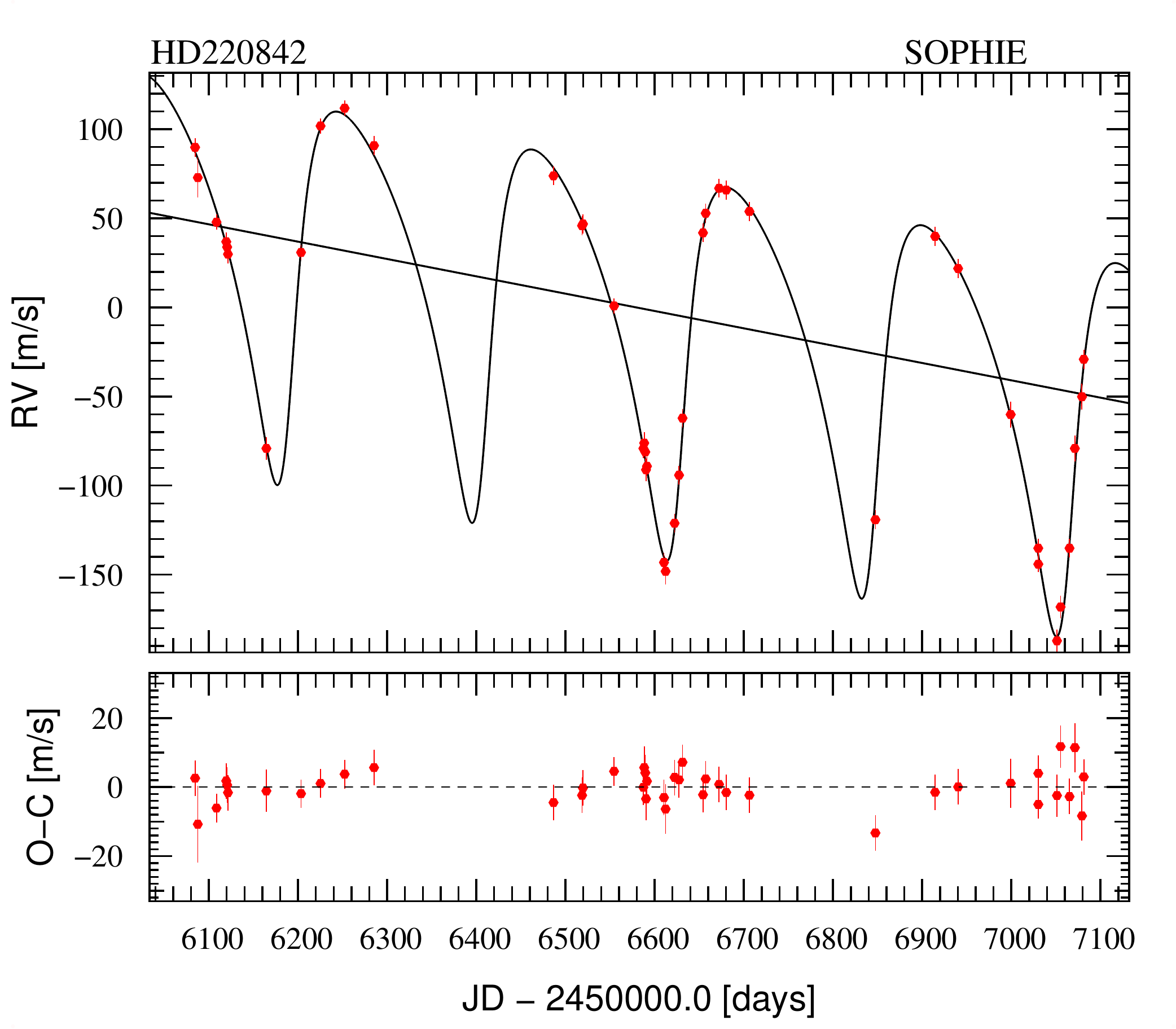}
\includegraphics[scale=0.34]{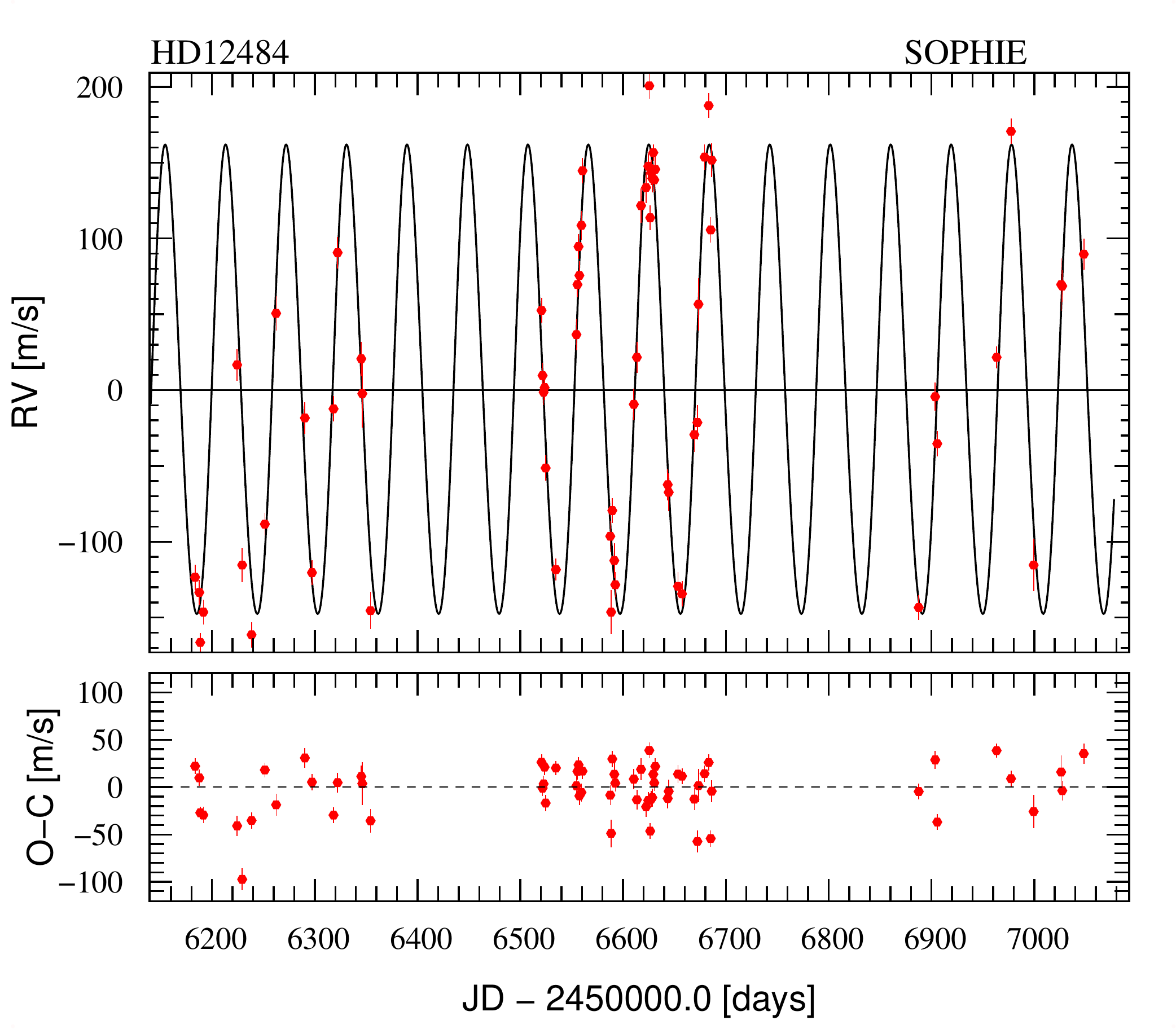}
\includegraphics[scale=0.34]{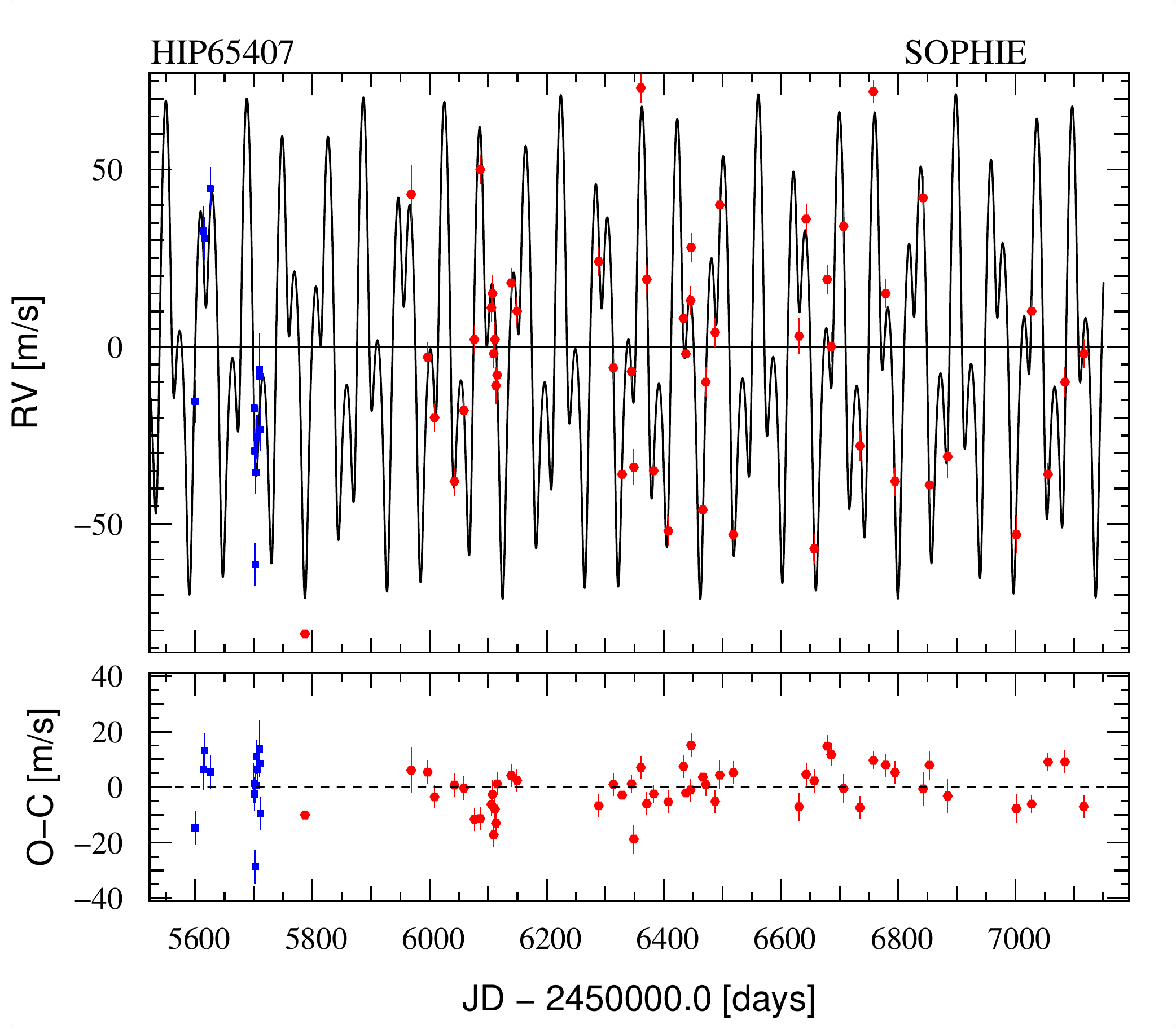}
\includegraphics[scale=0.34]{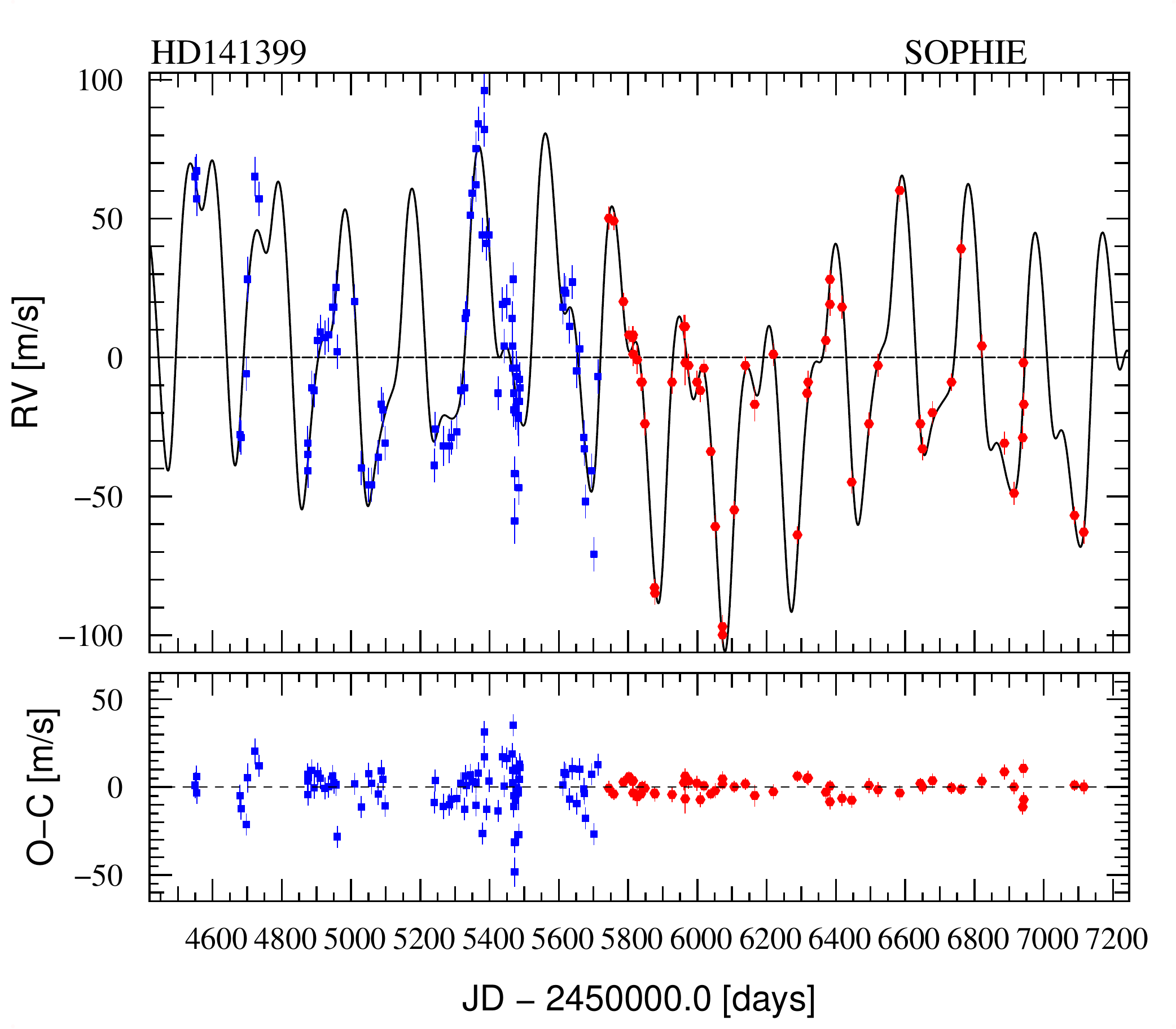}
\caption{Radial velocity SOPHIE measurements of the eight stars
as a function of time
(before/after the June-2011 upgrade are plotted in blue/red), and Keplerian fits.
The orbital parameters corresponding to the Keplerian
fits are reported in Tables~\ref{table_parameters_single} and~\ref{table_parameters_multi}. 
On each panel, the residuals with 1-$\sigma$~error bars are plotted below the fit.
}
\label{fig_omc}
\end{center}
\end{figure*}

\begin{table}[h]
  \centering 
  \caption{Radial velocities (barycentric referential) and bisector span measured with SOPHIE 
  (full table available electronically).}
  \label{table_rv}
\begin{tabular}{cccccl}
\hline
\hline
BJD$_{\rm UTC}$ & RV & $\pm$$1\,\sigma$  & Bis & star  & instrument \\
-2\,400\,000 & (km\,s$^{-1}$)  & (km\,s$^{-1}$) & (km\,s$^{-1}$)  \\
\hline
54366.3748	& -70.117 & 0.006 & -0.001	&	HIP\,109600	&	SOPHIE \\
55402.6067	& -69.929 & 0.007 & 0.004	&	HIP\,109600	&	SOPHIE \\
55429.5443	& -69.955 & 0.007 & 0.004	&	HIP\,109600	&	SOPHIE \\
55431.4808	& -69.974 & 0.007 & 0.011	&	HIP\,109600	&	SOPHIE \\
\ldots & \ldots & \ldots  & \ldots  & \ldots  & \ldots \\
\ldots & \ldots & \ldots  & \ldots  & \ldots  & \ldots \\
56941.2460	& -21.541 & 0.005 & -0.033	&	HD\,141399	&	SOPHIE+ \\
56942.2807	& -21.556 & 0.004 & -0.040	&	HD\,141399	&	SOPHIE+ \\
57088.6956	& -21.596 & 0.003 & -0.025	&	HD\,141399	&	SOPHIE+ \\
57115.5526	& -21.602 & 0.004 & -0.029	&	HD\,141399	&	SOPHIE+ \\
\hline
\end{tabular}
\end{table}

\section{SOPHIE spectroscopy and radial velocities}
\label{sect_observations}

We observed the eight stars with SOPHIE at the OHP 1.93-m telescope.
SOPHIE is a cross-dispersed, environmentally stabilized echelle spectrograph dedicated to 
high-precision radial velocity measurements (Perruchot et al.~\cite{perruchot08}; 
Bouchy et al.~\cite{bouchy09}). Observations were secured in fast reading mode of 
the detector and \textit{high-resolution} mode of the spectrograph, allowing the resolution 
power $\lambda/\Delta\lambda=75000$ to be reached. 
Depending of the star, the spectra were obtained over timespans of two to eight years. Their typical 
exposure times are of a few minutes, in order to reach a signal-to-noise ratios per pixel 
at 550~nm around 50. The total dataset presented here represents nearly 600 exposures 
for a total integration time exceeding 60 hours on the sky.
For all the exposures the first optical fiber aperture was used for starlight whereas the second 
one was put on the sky to evaluate the sky background pollution, especially from Moonlight. 
Wavelength calibrations were made at the beginning and the end of each observing night, and  
with a $\sim2$-hour frequency during the nights. This allows the drift of the instrument to 
be interpolated at the time of each exposure.
Four stars were observed both before and after the SOPHIE upgrade of June 2011
(Sect.~\ref{sect_intro}; Bouchy et al.~\cite{bouchy13}). 
As the upgrade introduces significant differences, we considered for each of them two distinct 
datasets, namely SOPHIE and SOPHIE+, depending whether the data were acquired before 
or after the upgrade, respectively. 
The four remaining stars were only observed after the upgrade (SOPHIE+~only).

We used the SOPHIE pipeline (Bouchy et al.~\cite{bouchy09}) to extract the spectra from the 
detector images, cross-correlate them with numerical masks which produced clear cross-correlation 
functions (CCFs), then fit the CCFs by Gaussians to get the radial velocities 
(Baranne et al.~\cite{baranne96}; Pepe et al.~\cite{pepe02}). The bisector spans of each CCF 
were also computed following Queloz et al.~(\cite{queloz01}). All the 39 SOPHIE spectral 
orders (covering the wavelength range 3872~\AA\ -- 6943~\AA) 
were used for the cross-correlations. The spectra significantly polluted by sky background
(in particular Moonlight)
were corrected following the method described e.g. by Bonomo et al.~(\cite{bonomo10}). 
Each exposure was then corrected from the interpolated drift of the instrument. In the case 
of the SOPHIE+ data, we also applied the ``RV constant master" correction presented 
by Courcol et al.~(\cite{courcol15}) which takes into account for small additional instrumental~drifts.

\begin{table*}[th!]
  \centering 
\caption{Adopted stellar parameters for the eight host stars (see Sect.~\ref{sect_stellar_properties}).}
  \label{table_stellar_parameters}
\begin{tabular}{lcccccccc}
\hline
Parameter  			&	HD\,143105		&	HIP\,109600		&	HD\,35759			&	HIP\,109384		&	HD\,220842		&	HD\,12484			&	HIP\,65407		&	HD\,141399		\\ 
\hline
$m_v$                		&	$6.75 \pm 0.01$	&	$9.16 \pm 0.02$	&	$7.74 \pm 0.01$	&	$9.63 \pm 0.02$	&	$7.99 \pm 0.01$	&	$8.17 \pm 0.01$	&	$9.42 \pm 0.02$	&	$7.20 \pm 0.02$	\\ 
Spectral~type$\dagger$	&	F5				&	G5				&	G0				&	G5				&	F8				&	F8				&	K0				&	K0				\\ 
$B-V$          			&	$0.52 \pm 0.01$ 	&	$0.75 \pm 0.03$ 	&	$0.59 \pm 0.01$ 	&	$0.79 \pm 0.04$ 	&	$0.58 \pm 0.02$ 	&	$0.66 \pm 0.02$ 	&	$0.78 \pm 0.04$ 	&	$0.73 \pm 0.04$ 	\\   
Parallax [mas]			&	$20.55\pm0.34$	&	$17.12\pm0.93$	&	$13.83\pm0.76$	&	$17.84\pm0.85$	&	$16.03\pm0.73$	&	$19.61\pm0.97$	&	$16.74\pm1.28$	&	$27.65\pm0.37$	\\ 
Distance [pc]     		&	$48.7 \pm 0.8$ 		&	$58.6 \pm 3.2$ 		&	$72.5 \pm 4.0$ 		&	$56.2 \pm 2.7$ 		&	$62.5 \pm 2.8$ 		&	$51.1 \pm 2.5$ 		&	$55.5 \pm 4.6$ 		&	$36.2 \pm 0.5$ 		\\ 
$v\sin i_\star$ [\kms]		&	$9.1 \pm 1.0$		&	$2.7 \pm 1.0$		&	$3.5 \pm 1.0$		&	$2.7 \pm 1.0$		&	$3.4 \pm 1.0$		&	$8.2 \pm 1.0$		&	$2.8 \pm 1.0$		&	$2.9 \pm 1.0$		\\ 
${\rm [Fe/H]}$ 			&	$+0.15 \pm 0.04$	&	$-0.12 \pm 0.02$	&	$+0.04 \pm 0.02$	&	$-0.26 \pm 0.03$	&	$-0.17 \pm 0.02$	&	$+0.05 \pm 0.02$	&	$+0.25 \pm 0.04$	&	$+0.35 \pm 0.03$	\\ 
$\log{R'_\mathrm{HK}}$	&	$-5.00 \pm 0.13$	&	$-5.07 \pm 0.21$	&	$-5.36 \pm 0.33$	&	$-5.02 \pm 0.19$	&	$-5.17 \pm 0.24$	&	$-4.43 \pm 0.10$	&	$-4.60 \pm 0.10$	&	$-5.26 \pm 0.24$	\\ 
$P_\mathrm{rot}$ [days]	& 	$11 \pm 3$		&		$42 \pm 11$	&		$28 \pm 9$ 	&		$43 \pm 11$	&		$22 \pm 6$	&		$7 \pm 4$		&		$19 \pm 7$	&		$49 \pm 12$	\\
$T_{\rm eff}$ [K]		&	$6380 \pm 60$		&	$5530 \pm 35$		&	$6060 \pm 30$		&	$5180 \pm 45$		&	$5960 \pm 20$		&	$5920 \pm 30$		&	$5460 \pm 50$		&	$5600 \pm 40$		\\ 
$\log g$ [cgs] 			&	$4.37 \pm 0.04$	&	$4.45 \pm 0.03$	&	$4.24 \pm 0.03$	&	$4.43 \pm 0.10$	&	$4.24 \pm 0.02$	&	$4.65 \pm 0.05$	&	$4.47 \pm 0.08$	&	$4.28 \pm 0.05$	\\ 
Mass~$[\rm{M}_{\odot}]$	&	$1.51 \pm 0.11$	&	$0.87 \pm 0.06$	&	$1.15 \pm 0.08$	&	$0.78 \pm 0.06$	&	$1.13 \pm 0.06$	&	$1.01 \pm 0.03$	&	$0.93 \pm 0.07$	&	$1.07 \pm 0.08$	\\ 
\hline
\multicolumn{7}{l}{$\dagger$: From Simbad.} \\
\end{tabular}
\end{table*}

The derived radial velocities and bisector spans are reported in Table~\ref{table_rv}. Typical uncertainties 
are between 3 and 10~m/s depending on the stars properties and actual signal-to-noise of the 
spectra. This includes photon noise and wavelength calibration uncertainty, the latest being 
estimated to the order of 2~m/s for each exposure. For the exposures obtained before the 
June 2011 upgrade, we also quadratically added a 5-m/s additional uncertainty due to 
poor scrambling properties at that time (Bouchy et al.~\cite{bouchy13}). The final 
uncertainties do not include any ``jitter'' due to stellar activity. 
The uncertainties reported here are significantly larger than those obtained with SOPHIE+ 
by Courcol et al.~(\cite{courcol15}) on stars of similar magnitudes and spectral types 
to detect low-mass planets. This is due to three main reasons. 
Firstly, the exposure times and obtained signal-to-noise ratios are smaller here, so the 
radial velocity uncertainty due photon noise is larger. 
Secondly, as the exposure times are shorter, the short-term stellar intrinsic variations 
are less efficaciously averaged out. 
Finally, the SOPHIE observations of Courcol et al.~(\cite{courcol15}) are done with 
simultaneous wavelength calibration in the second entrance aperture, which is not the case here. 
This illustrates the different strategies adopted with a given instrument to reach distinct 
goals. The present program aiming at detecting giant planets, it could be done at lower 
accuracy. In addition, it does not need perfect weather conditions and could continue even 
in cases of thin clouds in the sky, large seeing, or significant sky background pollution due 
to Moonlight or even twilight.

The SOPHIE radial velocities of the eight stars are plotted in Fig.~\ref{fig_omc} as a function of time. 
They all show variations well over the expected dispersion due to the measurement accuracy and 
the possible stellar jitter. The bisectors of the CCFs (Table~\ref{table_rv}) are stable for each object; they show 
dispersions well below that of the radial velocities variations. 
An anticorrelation between the bisector and the 
radial velocity is usually the signature of radial velocity variations induced by stellar activity
(see, e.g., Queloz et al.~\cite{queloz01}; Boisse et al.~\cite{boisse09}). 
Lomb-Scargle periodograms of the bisectors do not show any significant feature either. 
Similarly, other indicators linked to stellar activity, including CCF widths or 
\ion{Ca}{ii} emission index $\log{R'_\mathrm{HK}}$, do not show any significant peaks in 
their periodograms. All of this indicates  the radial velocity variations are mainly due to Doppler shifts of the 
stellar lines rather than stellar profile variations. This leads to concluding that reflex motions due to 
companion(s) are the likely causes of the stellar radial velocity variations. The amplitudes 
of the variations put the companions in the planet-mass regime.

\section{Properties of the eight planet-host stars}
\label{sect_stellar_properties}

Table~\ref{table_stellar_parameters} summarizes the stellar parameters of the eight planet-host stars.
Magnitudes, $B-V$, and parallaxes are from Hipparcos (Perryman et al.~\cite{perryman97}).
For each star we constructed an averaged spectrum using all the SOPHIE spectra unaffected by Moon pollution, 
and we performed spectral analyses from them. We used the method presented in Santos et 
al.~(\cite{santos04}) and Sousa et al.~(\cite{sousa08})
to derive the effective temperatures $T_{\rm eff}$, the 
surface gravities $\log g$, and the metallicities ${\rm [Fe/H]}$. 
The metallicities computed using the CCF calibration presented by Boisse et al.~(\cite{boisse10}) agree 
with the above ones, but we did not adopt these values as they are less accurate (typically $\pm0.09$\,dex)
than those obtained from stellar spectroscopic analysis. 
Using those derived spectroscopic parameters as input, stellar masses were derived 
from the calibration of Torres et al.~(\cite{torres10}) with a correction following Santos 
et al.~(\cite{santos13}). Errors were computed from 10,000 random values of the stellar 
parameters within their error bars and assuming Gaussian distributions.
We derived the projected rotational velocities
$v\sin i_\star$ and their uncertainties
from the parameters of the CCF using 
the calibration of Boisse et al.~(\cite{boisse10}).

\begin{figure}[b!] 
\begin{center}
\includegraphics[width=6cm]{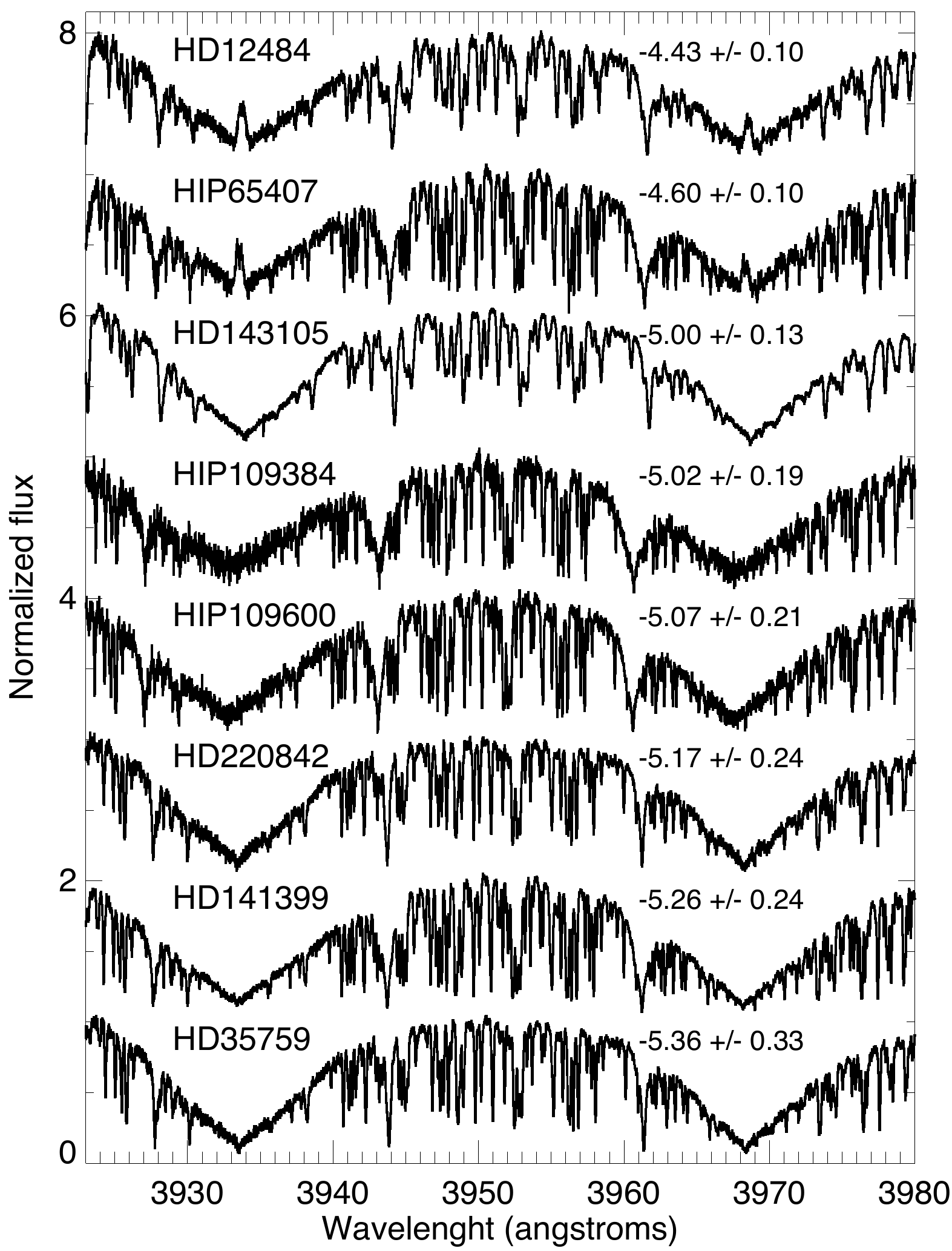}
\caption{H and K \ion{Ca}{ii} lines of the eight planet-host stars on the averaged SOPHIE~spectra. 
The stars are sorted by increasing activity from bottom to top, with the $\log{R'_\mathrm{HK}}$ values
overplotted on the right. The flux of HD\,35759 is the correct one; each star plotted above is 
successively shifted by 1 on the $y$-scale for visibility reasons.
Only HD\,12484 and HIP\,65407 show visually significant chromospheric emissions.}
\label{fig_caII}
\end{center}
\end{figure}

The eight planet hosts are dwarf 
stars of spectral types between F5 and K0. 
Three stars are metal-poor (HIP\,109600, HIP\,109384, and HD\,220842) and three
are metal-rich (HD\,143105, HIP\,65407, and HD\,141399), the two remaining ones 
having Solar metallicities.
They show standard, slow rotations with $v\sin i_\star$ of the order of 2 to 4\,km/s, with 
the two exceptions of HD\,143105 and HD\,12484 showing faster rotations with $v\sin i_\star$ 
of the order of 8 to 9\,km/s.

The activity levels of the stars were evaluated from the \ion{Ca}{ii} emission index $\log{R'_\mathrm{HK}}$. 
The reported values and uncertainties are the averages and dispersions of values measured on each epoch 
for a given star.
We note that the stars with the highest values of $\log{R'_\mathrm{HK}}$ are those 
for which that index shows the less dispersion with time. This could be surprising 
as active stars are expected to have larger variations of $\log{R'_\mathrm{HK}}$  than 
quiet stars. Here the larger uncertainties in $\log{R'_\mathrm{HK}}$ values of 
inactive stars is due to the larger dispersion they show due to the lower signal-to-noise 
ratios in the center of the cores of their H and K \ion{Ca}{ii} lines. Active stars have 
stronger flux in those lines, which implies larger signal-to-noise ratios and smaller 
dispersions.
The three stars that present the lowest $\log{R'_\mathrm{HK}}$ also have lower $\log g$ values. 
This could be a sign that these stars are beginning their evolution to be sub-giants, for which the 
$\log{R'_\mathrm{HK}}$ calibration is not effective.
As shown in Fig.~\ref{fig_caII}, only two stars, namely HD\,12484 and 
HIP\,65407, show detectable chromospheric emissions. They are the 
two only ones among the eight stars to have $\log{R'_\mathrm{HK}}$
above $-4.7$ and considered here as active. According to
Santos et al.~(\cite{santos00b}), their levels of activity should induce stellar jitters of the 
order of 20 and 10\,m/s, respectively. The low activity levels of the six remaining stars 
should imply stellar jitters of a few m/s at most. 

According to Noyes et al.~(\cite{noyes84}) and Mamajek \& Hillenbrand~(\cite{mamajek08}), 
these levels of activity imply stellar rotation periods $P_\mathrm{rot}$ around 11, 7, and 20~days for HD\,143105, 
HD\,12484, and HIP\,65407, respectively, and between 20 and 60~days for the five less active stars.
The uncertainties related to these $P_\mathrm{rot}$ estimations are difficult to quantify. 
Those reported in Table~\ref{table_stellar_parameters} are estimations taking into account 
the uncertainties on our measured $\log{R'_\mathrm{HK}}$
and a $\pm 0.16$ uncertainty on the Rossby numbers (Mamajek \& Hillenbrand~\cite{mamajek08}).
These estimations of stellar rotation periods roughly agree with the upper 
limits derived from the $v\sin i_\star$ (Bouchy et al.~\cite{bouchy05a})
except for HIP\,109600, HIP\,109384, and HD\,141399.   
They are different from the orbital periods measured below for the planets, 
so there are no hints that the periodic signals in radial velocities could be due to  stellar activity modulated by 
star rotation.
One possible exception is the inner planet of HIP\,65407, which is discussed below in Sect.~\ref{Keplerian_fit_HIP65407}.

Stellar activity could also causes low-frequency radial-velocity variations due to magnetic cycles 
(see, e.g. Wright et al.~\cite{wright08}; Fares et al.~\cite{fares09}; 
Moutou et al.~\cite{moutou11}; Carolo et al.~\cite{carolo14}).
For active stars, this can induce radial velocity variations up to 20 to 30~m/s. 
The shortest known magnetic cycles have periods of about two years 
and most of them are longer (22~years for the Sun). The periodicities we report below for 
our two active stars are shorter than 100~days, so they are unlikely to be due to magnetic cycles.
Magnetic cycles have also been identified in inactive stars, with amplitudes up to a few m/s 
(Lovis et al.~\cite{lovis11}). Four of our six inactive stars show periodicities shorter than 240~days, 
so they are unlikely to be caused by magnetic cycles. Our two remaining inactive stars (namely HIP\,109384 
and HD\,141399) show radial velocity variations with longer periods. According Lovis et al.~(\cite{lovis11}), 
the amplitude of their magnetic cycles should be of the order of 1~m/s or below.
Thus, magnetic cycles are unlikely to cause the larger variations we report.

\begin{figure*}[] 
\begin{center}
\includegraphics[scale=0.42]{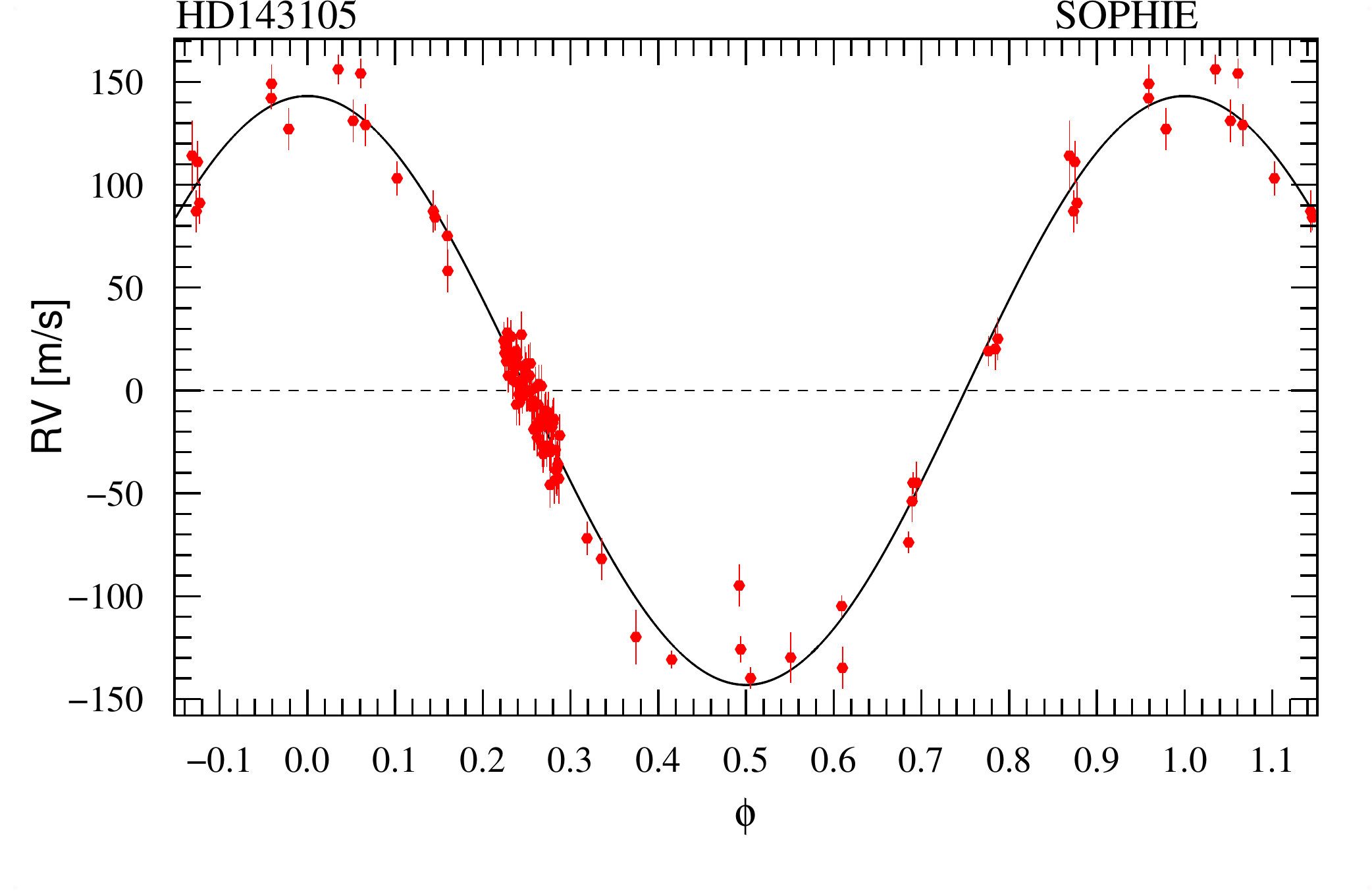}
\includegraphics[scale=0.42]{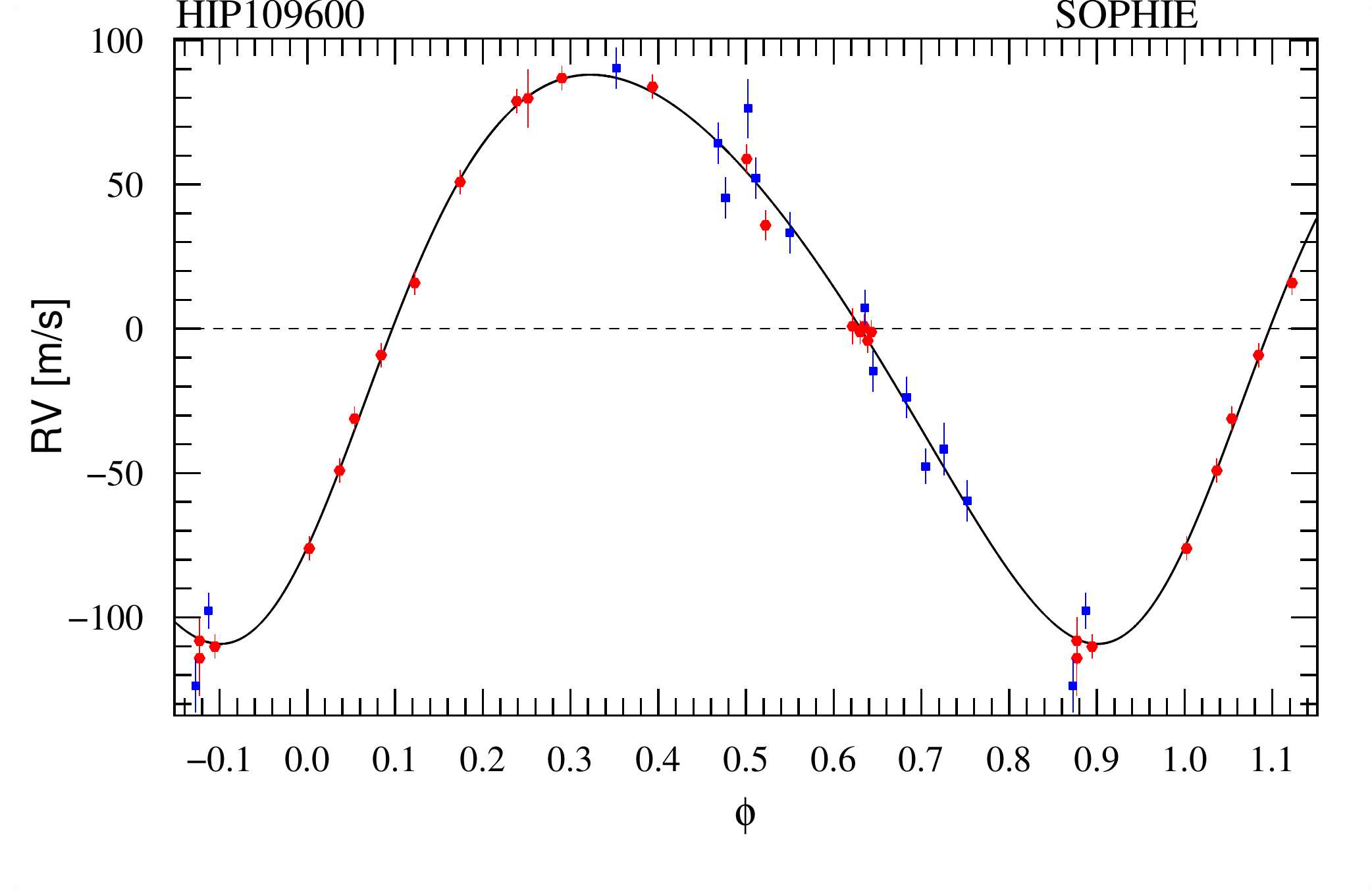}
\includegraphics[scale=0.42]{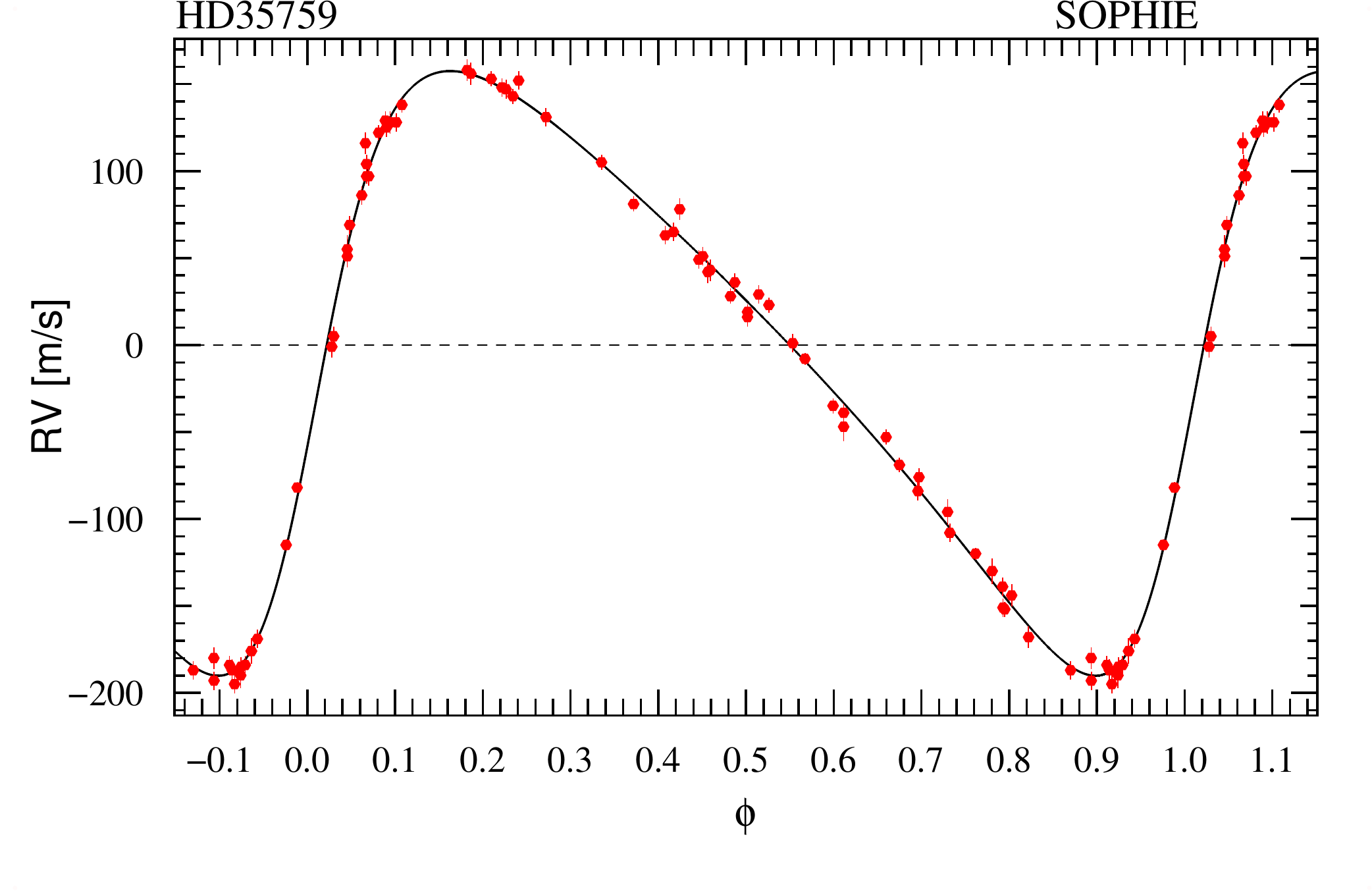}
\includegraphics[scale=0.42]{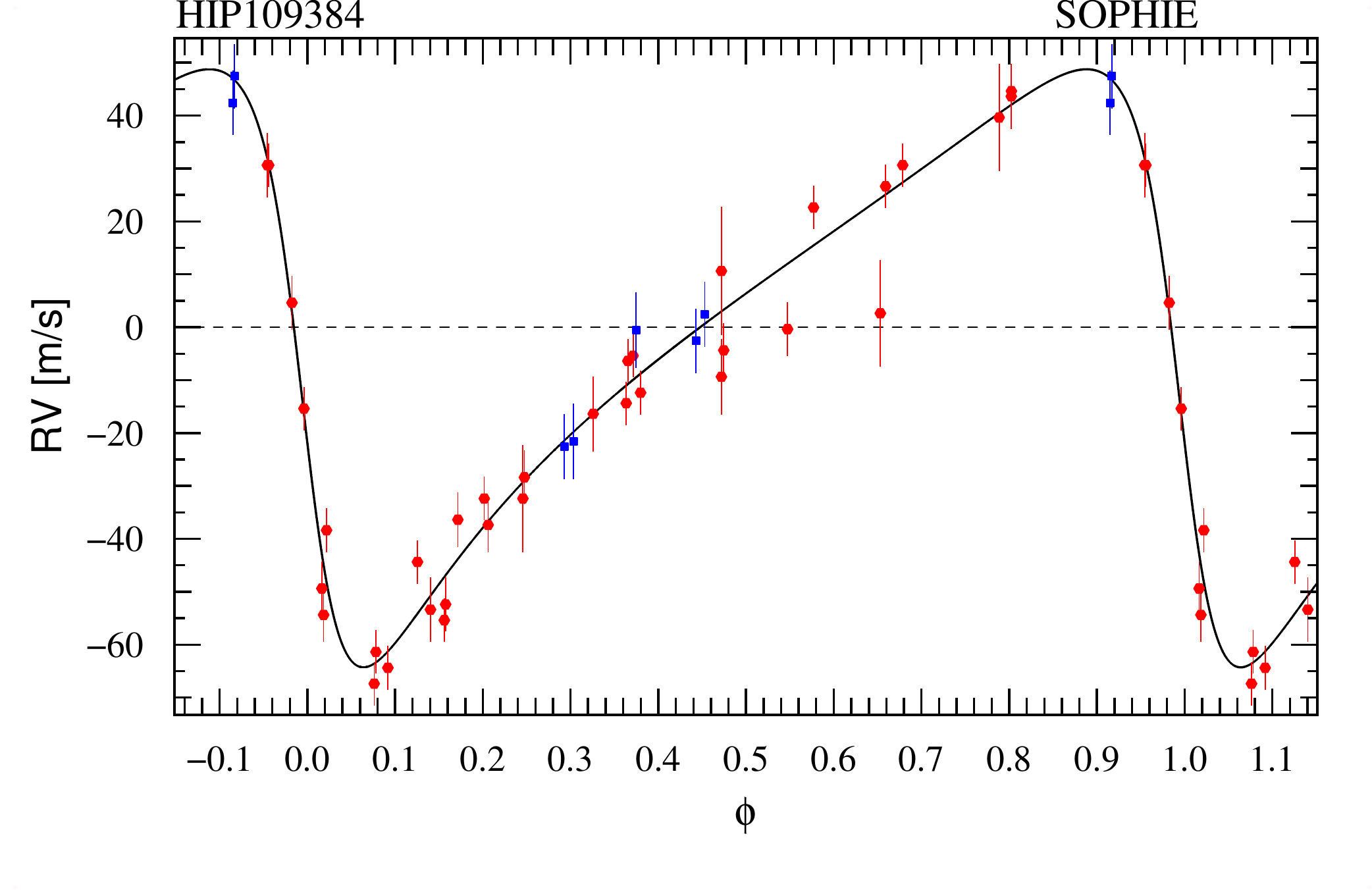}
\includegraphics[scale=0.42]{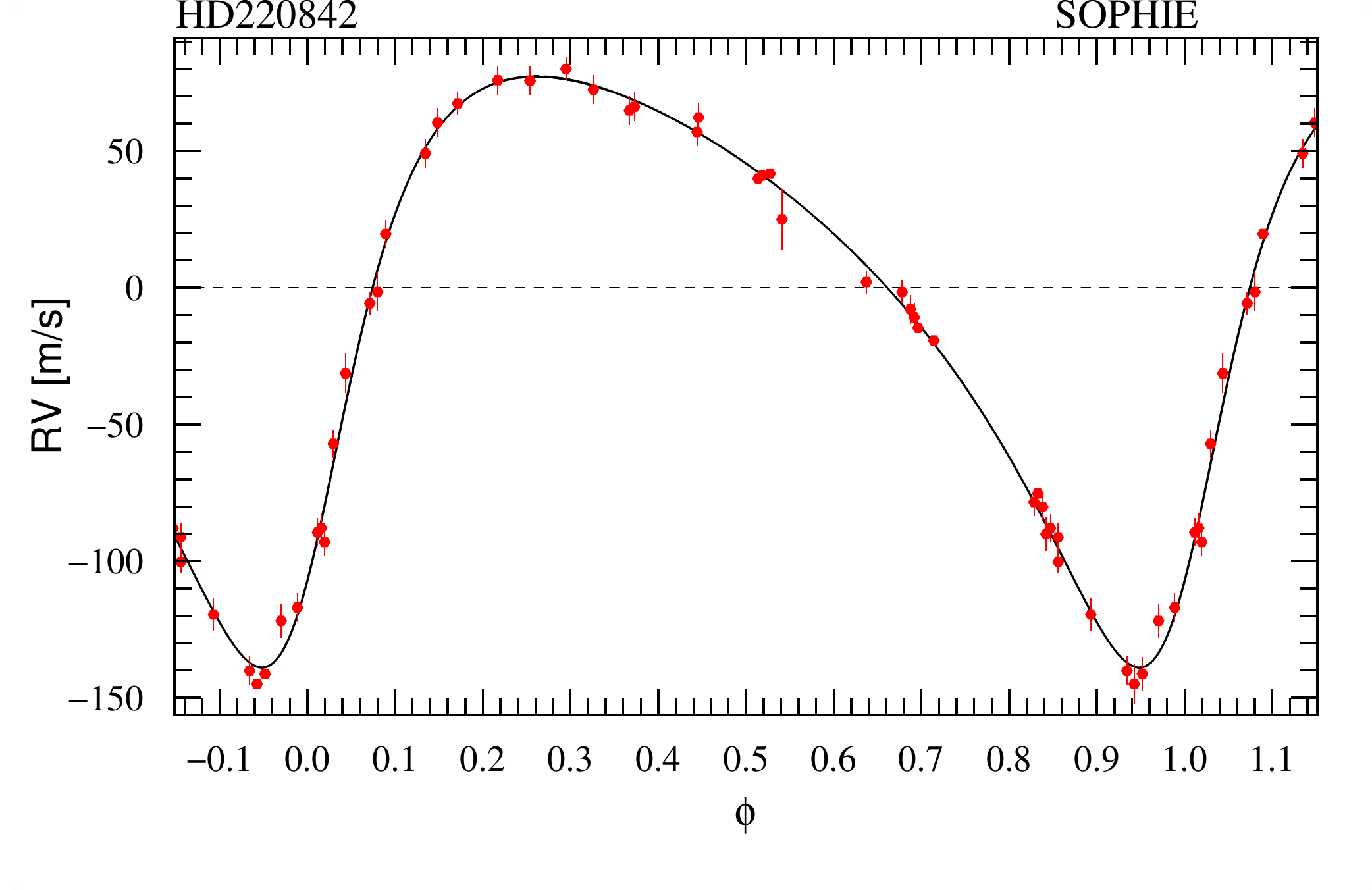}
\includegraphics[scale=0.42]{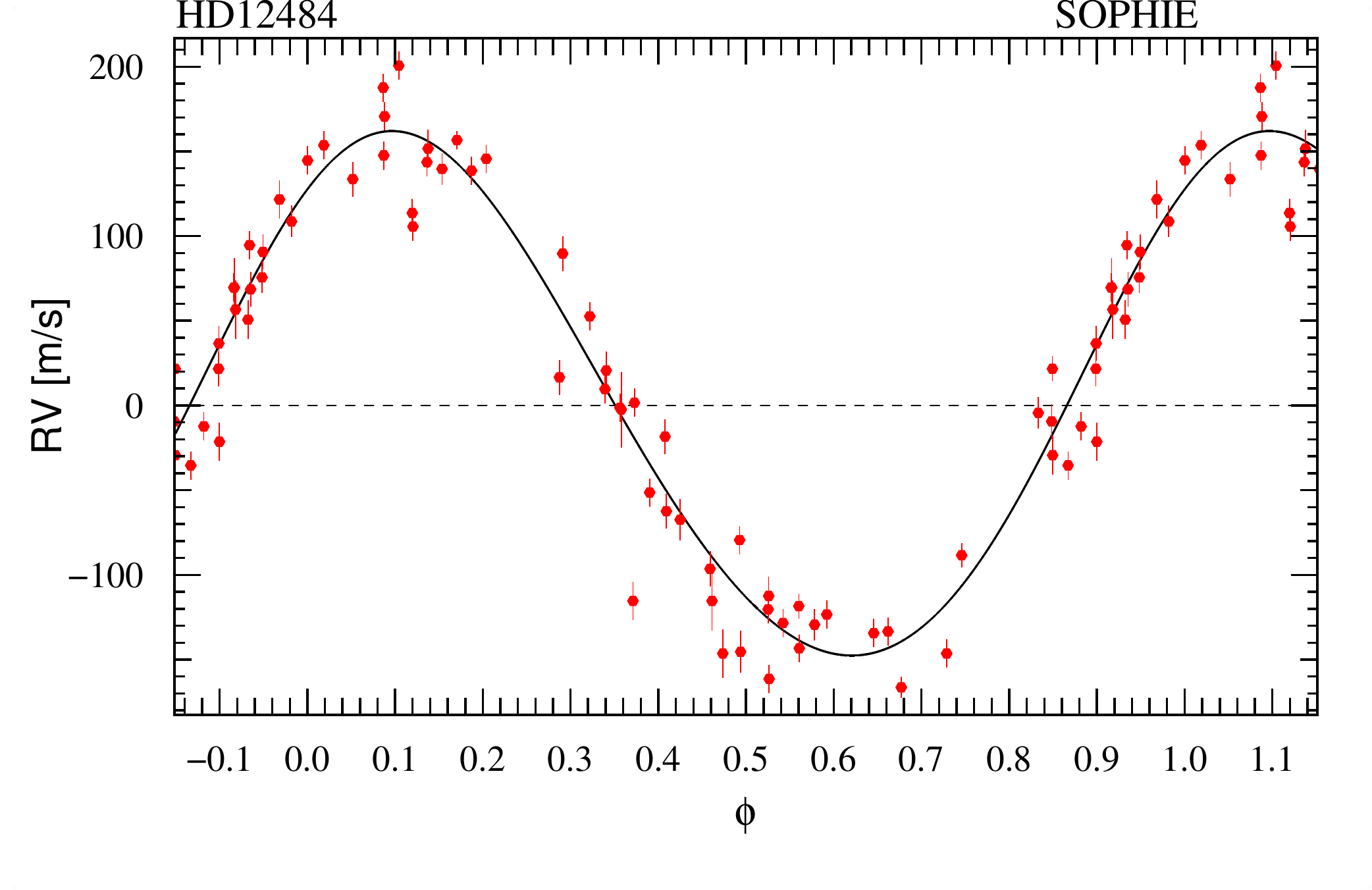}
\caption{Phase-folded radial velocity curves for the six stars with a single-planet detection.
The SOPHIE data obtained before/after the June-2011 upgrade are plotted in blue/red 
with 1-$\sigma$~error bars, and the Keplerian fits are the solid lines. 
The orbital parameters corresponding to the Keplerian fits are reported in Table~\ref{table_parameters_single}. 
In the cases of HD\,220842 the linear drift is subtracted here.
The HD\,143105b transit detection attempt of 2013-09-17 is shown around $\phi=0.25$.}
\label{fig_orb_phas1}
\end{center}
\end{figure*}

\section{Single-planet systems detection and characterization}
\label{sect_single_planet_systems}

\subsection{Method}
\label{sect_method}

We describe here the orbit analyses of the six stars presenting only a single characterized planet:
HD\,143105, HIP\,109600, HD\,35759, HIP\,109384, HD\,220842, and HD\,12484.
For each of them the radial velocities present periodic variations which were already identified 
since the first $\sim10$ to $\sim20$ SOPHIE spectra were obtained. 
The following measurements allowed the periods to be confirmed for each object, and the parameters to 
be~refined. We fitted the six final radial velocity sets with single Keplerian orbits using the procedure 
{\tt yorbit} (S\'egransan et al.~\cite{segransan11}; Bouchy et al.~\cite{bouchy16}). The six parameters of the orbits 
are free to vary during the fits. In cases of objects with both SOPHIE and SOPHIE+ data, we introduced one 
extra free parameter in order to take into account for the possible radial velocity shift between both datasets.
The radial velocity shifts we found (see below) agree with the values found by 
Moutou et al.~(\cite{moutou14}) and Wilson et al.~(\cite{wilson16}) between SOPHIE and SOPHIE+
(see however the particular case of HIP\,109384 in Sect.~\ref{sect_HIP109384}). 
Finally, we allowed radial velocity linear drifts in addition to the Keplerian motions which added one extra free
parameter for each object. Only one significant drift was detected
(at more than $3\sigma$), in the case of HD\,220842.
The free shift between SOPHIE and SOPHIE+ is unlikely to hide any possible actual stellar drift, 
except maybe in the case of HIP\,109384 (Sect.~\ref{sect_HIP109384}).

Lomb-Scargle periodograms of the datasets show clear peaks at the periods of each orbit. Some additional peaks 
are also visible, corresponding to aliases or secondary lobes due to the sampling. The additional peaks all present 
smaller power than the main peaks, and Keplerian fits at these periods are clearly less good than the Keplerian 
fits made at the periods corresponding to the main peaks. So there are no ambiguities in the periods of the six detected 
planets. Periodograms of the radial velocities residuals after Keplerian fits show no significant peaks. The current 
data show no significant detections of extra companions in addition to the six single planets presented here, 
with the exception of the linear drift detected in HD\,220842.

The derived orbital parameters of the six single planets are reported in Table~\ref{table_parameters_single} together 
with error bars which were computed from Monte~Carlo and residuals permutation experiments
both based on error bars on each radial velocity measurement. 
They provide similar error bars with no significant asymmetries or skews. 
We finally adopt the most conservative uncertainties.
Table~\ref{table_parameters_single} includes 
the orbital period $P$,
the eccentricity $e$,
the argument of periastron $\omega$,
the time  $T_0$ at periastron (or time of possible transits for circular or nearly circular orbits of HD\,143105b and HD\,12484b),
the semi-amplitude $K$,
le sky-projected mass of the planet $M_\textrm{p} \sin i$,
the semi-major axis $a$,
the systemic radial velocities $V_{r}$,
the number $N$ of radial velocities and their time span,
the dispersion $\sigma_\mathrm{O-C}$ of the residuals to the Keplerian fit,
and the additional radial velocity drift (or its~limit).
The uncertainties on $M_\textrm{p} \sin i$ and $a$ are mainly due to the 
uncertainties on the host-star masses.  The fits and their 
residuals are plotted as a function of time in the six upper panels of Fig.~\ref{fig_omc}; the phase-folded plots 
are~displayed in Fig.~\ref{fig_orb_phas1}. 

The Keplerian fits of the radial velocities provide measurements of the sky-projected mass $M_\textrm{p} \sin i$ 
of the planetary companions while the inclination $i$ of the orbit remains unknown. Large masses for the companions 
could not formally be excluded from the radial velocities in cases of particularly small inclinations.
In order to identify possible face-on systems we searched for signatures of orbital motion in Hipparcos astrometric data.
Our  stars have each between 100 and 150 Hipparcos measurements covering several periods of each planet, 
with a typical accuracy of a few mas.
They were fitted with 5-parameter models using the orbital parameters obtained from the radial velocities, 
following the method described in details by Sahlmann et al.~(\cite{sahlmann11}).
This method has proven to be reliable in detecting astrometric orbital signatures (e.g. Sahlmann et al.~\cite{sahlmann11}, 
~\cite{sahlmann13}; D\'{\i}az et al.~\cite{diaz12}; Wilson et al.~\cite{wilson16}). Here we detected no orbital motions.
The corresponding upper limits $M_{\textrm{p}{\mathrm{\,max}}}$ on the companion masses are reported in the 
bottom line of Table~\ref{table_parameters_single}. They are not really constraining. Any companion with a true mass 
around one Solar mass or above would have been detected as a secondary peak in the CCFs or as a variation in 
their bisectors,~which is not the case here. We can however conclude here that none of our companions is an obvious 
binary in a face-on~system.

The six single planetary systems are discussed~below.

\subsection{HD\,143105}
\label{sect_HD143105}

\subsubsection{Keplerian fit} 
\label{sect_keplerian_fit_hd143105}

HD\,143105 (BD$\,+69^{\circ}825$, HIP\,77838) is a rotating star ($v\sin i_\star = 9.1 \pm 1.0$\,\kms) but does not show 
high activity level ($\log{R'_\mathrm{HK}} =-5.00 \pm 0.13$).
Simbad classifies it as an F5 star but our temperature measurement ($T_{\rm eff} = 6380 \pm 60$\,K;
Sect.~\ref{sect_stellar_properties}) 
suggests the slightly later spectral type F7.
We started observing it in July 2013, after the SOPHIE+ upgrade. 
The star rapidly showed the typical 
signature of a hot Jupiter, with variation of $K\sim150$\,m/s and a period near 2.2\,days. We followed that 
star for 14 months in order to confirm the detection, constrain the properties of the system, and detect 
any possible additional companion. The final dataset includes $N=100$ radial velocity measurements 
that we fitted with Keplerian models. They did not provide a significant detection of 
any eccentricity: we measured $e=0.021 \pm 0.016$ and $\omega = 40 \pm 70^{\circ}$.
The 3-$\sigma$ upper limit we derived is $e<0.07$. In agreement with the data and the fact that hot Jupiters are expected 
to be circularized, we assumed a circular orbit for the final fit. The derived radial velocity 
semi-amplitude is $K=144.0 \pm 2.6$\,m/s corresponding to a mass 
$M_\textrm{p} \sin i =1.21\pm 0.06$\,M$_\mathrm{Jup}$.
The planet HD\,143105b orbits a slightly metal-rich star, 
in agreement with the tendency found for stars 
harboring Jupiter-mass planets (see, e.g., Santos et al.~\cite{santos05}).

The residuals around the Keplerian fit show a dispersion of 10.6\,m/s.
That quite large value is not due to  the  activity level of the star, which is low, 
nor the assumption for a circular orbit.
That dispersion is mainly explained by the uncertainties of the radial velocities which 
is 9.2\,m/s on average. The relatively low level of radial velocity accuracy obtained for that star by 
comparison to the other ones is due to the broadening of the lines cause by the stellar rotation. Moreover, 
HD\,143105 is the hottest among the eight stars studied here and granulation effects might be higher.
No significant linear drift nor particular structures were found in the residuals as a function of the~time.

\subsubsection{Transit search} 

Following 55\,Cancri, HD\,143105  is the second brightest star known to host 
a planet with a period shorter than three days (55\,Cancri\,e being a super-Earth).
Since HD\,143105 is  bright and the transit probability as high as 14\,\%\ for its hot Jupiter, 
that system was particularly interesting for photometric follow-up studies. So after 
the first SOPHIE observations on that star in July 2013 which rapidly showed it hosts a hot Jupiter, 
we started a transit search the following month using three different telescopes: Oversky, CROW, and~ROTAT.

The Oversky Observatory (35\,cm f/11 Schmidt-Cassegrain) telescope unfortunately had bad weather and the 
photometry could not be used to constrain any transit model. The CROW observatory in Portugal (28\,cm f/10 
Schmidt-Cassegrain, KAF1603ME CCD, 25\,sec exposures, $Ic$ filter, defocused images) produced three photometric 
series on August 15, August 26, and September 17, 2013. The dispersions are 4.0, 3.7, and 4.6~mmag,
respectively, and 2.3, 2.1, and 2.7~mmag when data are binned to have a point approximatively every 120~sec. 
The ROTAT telescope at OHP (60\,cm f/3.2 Newton, KAI11000M CCD) was also used for three nights in 2013: 
August 26 (4-sec exposures, Luminance filter, defocused images), September 6 (20-sec exposures, $R$ filter, 
defocused images), and September 17 (4.5-sec exposures, Luminance filter, defocused images). 
On September 6, the dispersion is 10.7~mmag, and 7.8~mmag when binned to have a point approximatively 
every 120~sec. On September 17, cirrus were present sometimes and the dispersion is 20.1~mmag, 
and 8.1~mmag when~binned.

\begin{figure}[b!] 
\begin{center}
\hspace{-.4cm}
\includegraphics[scale=0.49]{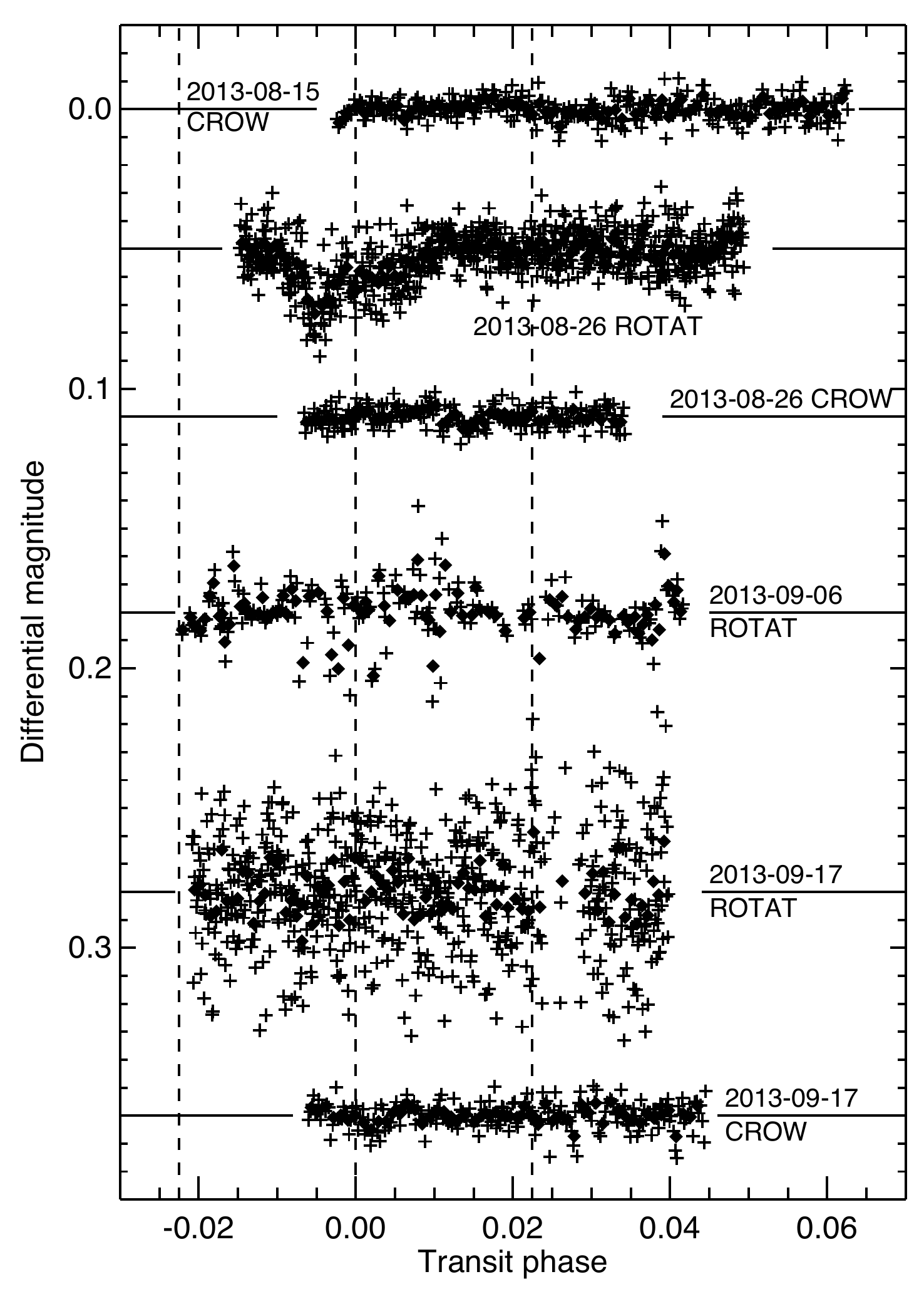}
\includegraphics[scale=0.47]{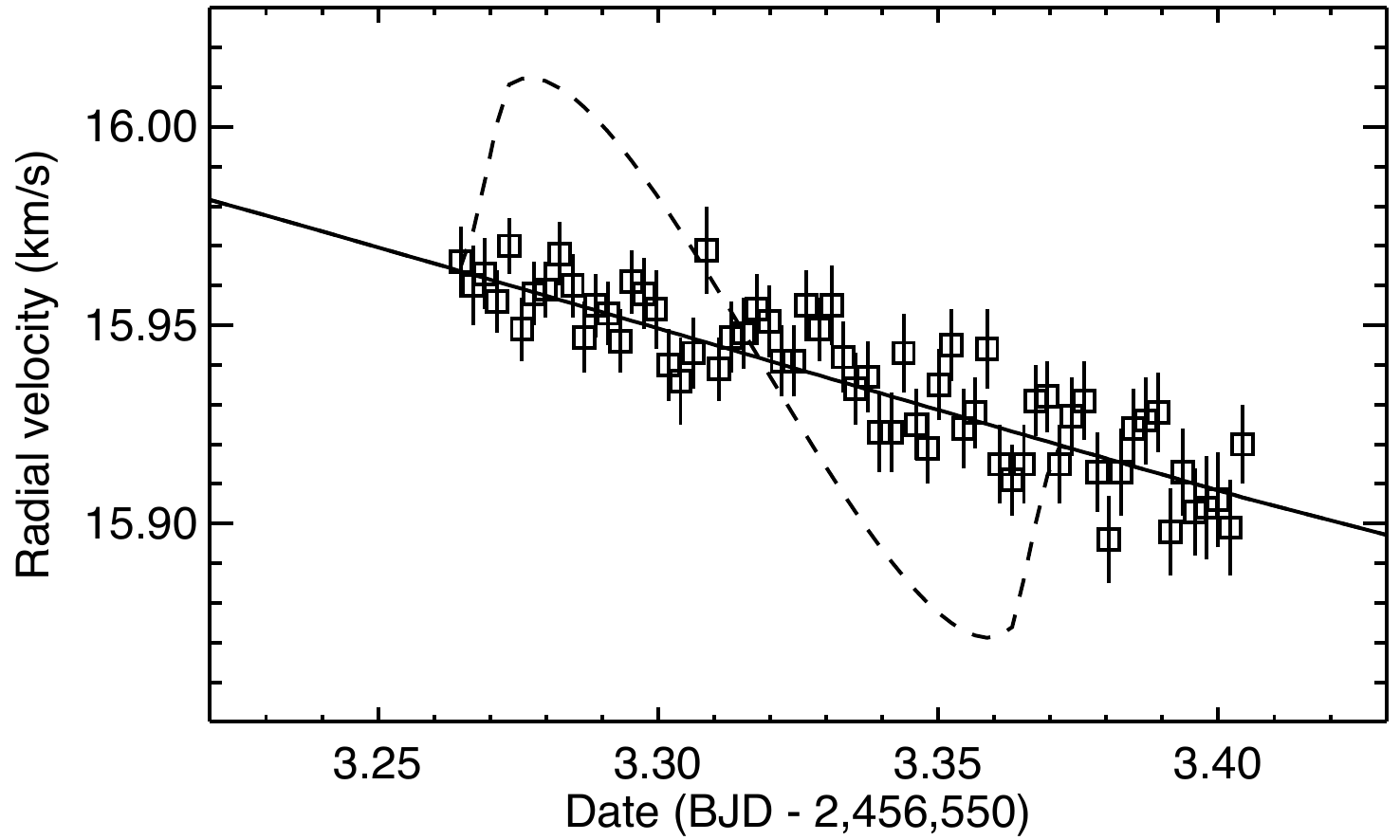}
\caption{Search for transits of HD\,143105b. 
\textit{Top: photometry}.
The differential photometry is plotted as a function of the phase of the searched transits.
The telescopes (CROW or ROTAT) and the dates are indicated for each of the six observations, 
which have been shifted in $y$-axes for clarity.
The horizontal lines represents the mean of the data (or mean outside the event for 2013-08-26 with ROTAT).  
The original data (plus signs) have been binned to have a point approximatively every 120~seconds (diamond signs) 
to help their comparison.
The vertical dashed lines represent the hypothetical ingress, center, and egress phases for central transits. 
The  transit-like event observed on 2013-08-26 with ROTAT was not confirmed by other 
observations.
\textit{Bottom: spectroscopy}.
SOPHIE radial velocities measured during a possible transit (2013-09-17).
The expected Rossiter-McLaughlin anomaly is shown in dotted line (in case of aligned, prograde orbit); it is not detected.
The solid line shows the Keplerian model with no transit 
(Figs.~\ref{fig_omc} and~\ref{fig_orb_phas1},~Table~\ref{table_parameters_single}).}
\label{fig_photom}
\end{center}
\end{figure}

The light curves are shown in Fig.~\ref{fig_photom} (upper panel). 
Differential photometry has been measured and the data have been detrended with 
smooth second order polynomials. 
Assuming a stellar radius $1.15 \,\rm{R}_{\odot}$ typic of a F5V star, values for $a$ and $P$ 
from Table~\ref{table_parameters_single},  and a central transit (impact parameter $b=0$), we evaluate 
the transit maximum duration to $\sim2.37$~hours, i.e. between transit phases $-0.022$ and $+0.022$.
The uncertainty on the epoch of each transit mid-time is of the order of $\pm 10$~minutes 
(see Table~\ref{table_parameters_single}).
A light drop of about 15~mmag has been observed with ROTAT for about one~hour on August 26 
at the expected time of transit (these data were not detrended as they show that feature). 
Outside that event, the dispersion is 7.3~mmag, or 3.1~mmag when binned. 
Comparison stars (significantly fainter) 
do not show such event, so the hint for a transit detection was high. However, that event has 
not been observed at the other 
dates. Even more, the CROW observation secured on August 26 did not 
show the feature seen with ROTAT the same night.
We conclude thus that no transit deeper than 10~mmag was detected. 
The event observed with ROTAT remains unexplained. A possibility could be a sporadic 
event due to stellar activity accidentally occurring at the time of of possible~transit.

On September 17 the reality of the transit was still unclear, so we 
attempted a spectroscopic detection of the transit with 
SOPHIE through the Rossiter-McLaughlin effect (e.g. H\'ebrard et 
al.~\cite{hebrard11}). As the star is rotating 
($v\sin i_\star = 9.1 \pm 1.0$\,\kms), the amplitude of the 
Rossiter-McLaughlin anomaly is expected to be as high 
as $\sim100$\,m/s and thus easily detectable. No deviation from 
the Keplerian curve was detected during that night however
(see upper-left panel of Fig.~\ref{fig_orb_phas1} and lower panel of Fig.~\ref{fig_photom}).
So, even if a transiting planet on a polar orbit and a null 
impact parameter would produce no Rossiter-McLaughlin signal, 
this scenario remains unlikely and we keep the conclusion that the planet is not transiting. 
The radial velocities obtained during the September-17 night are used in the Keplerian 
fit presented above in Sect.~\ref{sect_keplerian_fit_hd143105}.

\begin{figure}[b!]
\begin{center}
\hspace{-.4cm}
\includegraphics[width=\columnwidth]{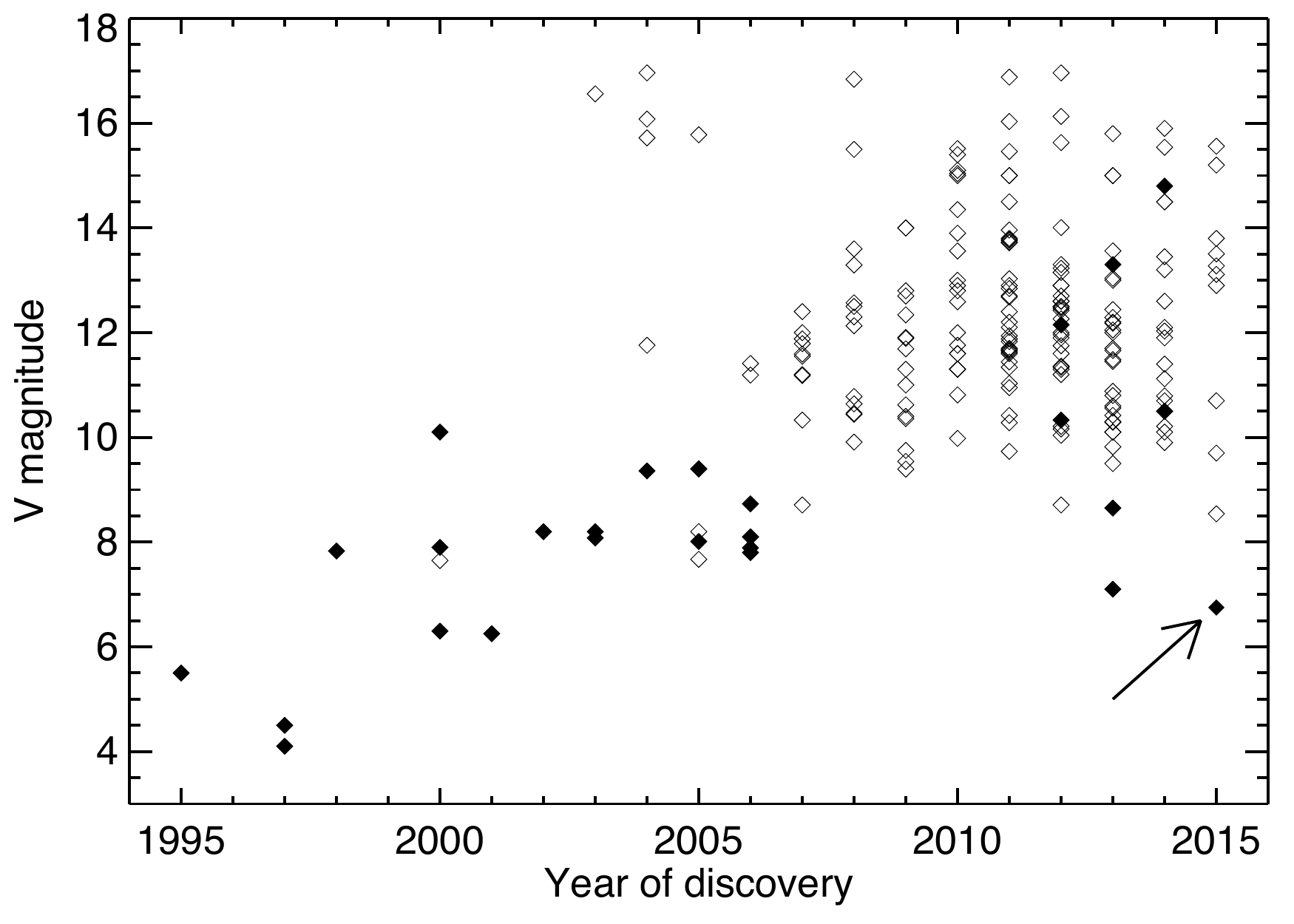}
\caption{Magnitudes of known hot-Jupiters host stars as a function of the planet discovery year.
Transiting systems are empty symbols whereas non-transiting systems detected only in radial velocities are filled.
Hot Jupiters are selected from exoplanets.org as planets with masses $>0.2$\,M$_\mathrm{Jup}$
and periods $<6$\,days (220 planets). The host star of the new hot Jupiter HD\,143105b 
is indicated with an arrow.
}
\label{fig_HJ_magn}
\end{center}
\end{figure}

\subsubsection{Hot Jupiters around bright stars} 

With a magnitude $m_v = 6.75 $, 
HD\,143105 is one of the brightest stars known to host a hot Jupiter. 
Figure~\ref{fig_HJ_magn} shows 
the $V$ magnitudes of hot-Jupiter host stars as a function of the year of their discovery, 
distinguishing transiting planets from those only characterized from radial velocities.
Here hot Jupiters are defined as planets with $M_\textrm{p} \sin i > 0.2$\,M$_\mathrm{Jup}$ and 
orbital periods shorter than 6~days, as suggested by mass and period distributions of known planets.
In the continuation of the three first discovered hot Jupiters, 51\,Peg\,b, Tau\,Boo\,b, and Ups\,And\,b
(Mayor \&\ Queloz~\cite{mayor95}; Butler et al.~\cite{butler97}), the first decade of radial velocity 
surveys allowed about a dozen hot Jupiters to be discovered around stars brighter than $V=8$. 
Three of them afterward turned out to transit, namely HD\,209458b, HD\,189733b, and HD\,149026b
(Charbonneau et al.~\cite{charbonneau00}; Bouchy et al.~\cite{bouchy05b}; Sato et al.~\cite{sato05}).
The transits in front of these three bright stars allowed numerous and important follow-up studies.
The second decade allowed a plenty of hot Jupiters to be detected directly in transit from ground-
and space-based photometric surveys in front of fainter stars. 
Still, Fig.~\ref{fig_HJ_magn} shows that radial velocity surveys continue to detect hot Jupiters in 
front of bright stars, and thus they are not all known. The transit probability being around 10\,\%\ 
for a hot Jupiter, it might remain transiting hot Jupiters to discover in front of bright stars, which 
would be of high interest for follow-up studies. They could be detected from radial velocity surveys, 
or later from the future TESS and PLATO~satellites. It is possible that HD\,143105b has been missed 
by some previous radial velocity surveys in case they avoided observing fast rotators. As shown here however, 
a stellar rotation of the order of $v\sin i_\star \sim 9$\,km/s does not prevent at all hot Jupiters detection
from radial velocities.

\begin{table*}[th!]
  \centering 
\hspace{-3cm}
  \caption{Fitted Keplerian orbits and planetary parameters for the six single-planet system, with 1-$\sigma$ error bars.}
  \label{table_parameters_single}
\begin{tabular}{l|cccccc}
\hline
\hline
Parameters  								&	HD\,143105b			& HIP\,109600b			&	HD\,35759b			&	HIP\,109384b			&	HD\,220842b			&	HD\,12484b		 \\
\hline		
$P$ 				[days]					& $2.1974 \pm 0.0003$		& $232.08 \pm 0.15$			& 	$82.467 \pm 0.019$		& $499.48 \pm 0.32$			& $218.47 \pm 0.19$			& $58.83 \pm 0.08$		\\
$e$										& $<0.07$					& $0.163 \pm 0.006$			& 	$0.389 \pm 0.006$		& $0.549 \pm 0.003$			& $0.404 \pm 0.009$			& $0.07 \pm 0.03$		\\
$\omega$ 		[$^{\circ}$]					& $-$					& $-131.3 \pm 1.6$			& 	$-104.0 \pm 1.3$		& $104.49 \pm 0.37$			& $-134.9 \pm 1.7$			& $-35 \pm 40$			\\
$T_0$		 	[BJD]$^\ddagger$			& $56531.344 \pm 0.007	$	& $56017.1 \pm 1.1$			& 	$56469.72 \pm 0.24$	& $56337.02 \pm 0.34$		& $56624.8 \pm 0.8$			& $56699 \pm 5$		\\
$K$				[\ms]						& $144.0 \pm 2.6$			& $98.6 \pm 0.5$			& 	$173.9 \pm 1.3$		& $56.53 \pm 0.22$			& $108.1 \pm 1.2$			& $155 \pm 5$			\\
$M_\textrm{p} \sin i$	[M$_\mathrm{Jup}$]$^\dagger$	& $1.21\pm 0.06$			& $2.68\pm 0.12$			& $3.76\pm 0.17$			& $1.56\pm 0.08$			& $3.18\pm 0.15$			& $2.98\pm 0.14$		\\
$a$				[AU]$^\dagger$				& $0.0379\pm 0.0009$		& $0.706\pm 0.016$			& $0.389\pm 0.009$			& $1.134\pm 0.029$			& $0.740\pm 0.018$			& $0.297\pm 0.005$		\\
$V_{r\mathrm{,\,SOPHIE}}$ [\kms]				& --						& $-70.0300 \pm 0.0009$		& --						& $-63.6852 \pm 0.0004$		& --						& --					\\
$V_{r\mathrm{,\,SOPHIE+}}$ [\kms]				& $15.9418 \pm 0.0030$		& $-69.9964 \pm 0.0006$		& $-12.5704 \pm 0.0015$		& $-63.6784 \pm 0.0003$		& $-22.5746 \pm 0.0016$		& $4.960 \pm 0.005$		\\
$N$										& 100					& 34						& 71						& 42						& 42						& 65					\\
$\sigma_\mathrm{O-C,\,SOPHIE}$ [\ms]			& --   					& 9.6						& --						& 3.7						& --						& --					\\	
$\sigma_\mathrm{O-C,\,SOPHIE+}$ [\ms]			& 10.6 					& 3.3						& 6.0						& 5.8						& 4.5						& 25.2				\\	
RV drift 			[\ms\,yr$^{-1}$]$^\ast$		&	$[-1.1;+12.5]$			&	$[-0.2;+12.0]$			&	$[-0.4;+4.5]$			&  $[-2.2;+1.5]$				&	$-35.5 \pm 1.0$	$^\ast$$^\ast$		&  $[-4.8;+32.3]$		\\
span 			[days]					& 422					& 2496					& 1261					& 2776					& 997					& 865				\\
$M_{\textrm{p}{\mathrm{\,max}}} [\rm{M}_{\odot}]$	&	14.65			&	0.43					&	1.04					&	0.27					&	0.35					&	0.96				\\ 
\hline
\multicolumn{7}{l}{$\ddagger$: time at periastron for eccentric orbits; time of possible transits in the cases of the circular or nearly circular orbits of HD\,143105b and HD\,12484b.} \\
\multicolumn{7}{l}{$\dagger$: using the stellar mass  and its uncertainty from Table~\ref{table_stellar_parameters}.} \\
\multicolumn{7}{l}{$\ast$: significant RV drift detected only for HD\,220842; 3-$\sigma$ confidence intervals for other systems.} \\
\multicolumn{7}{l}{$\ast\ast$: the RV drift of HD\,220842 is due to a substellar companion (see Sect.~\ref{sect_HD220842}).} \\
\end{tabular}
\end{table*}

Orbiting a bright star, HD\,143105b will be a precious object for follow-up in various spectral domains
despite its non-transiting nature. 
If the planet has a grazing incidence, it may be possible to detect in the ultraviolet the partial transit of an 
evaporating atmosphere, extended by the high irradiation from the F-type host star 
(see the case of 55\,Cnc\,b; Ehrenreich et al.~\cite{ehrenreich12}). 
The inclination of the HD\,143105b orbit could be constrained by measuring the polarization of 
its optical phase curve (Madhusudhan \& Burrows~\cite{madhusudhan12}). 
Such measurements are challenging in the optical 
(e.g. Charbonneau et al.~\cite{charbonneau99}; Leigh et al.~\cite{leigh03}) where hot Jupiters have usually 
low albedos and yield faint signals. A tentative detection of phase variations from the non-transiting Saturn-mass 
planet HD\,46375b was nonetheless obtained from CoRoT observations in the optical 
(Gaulme et al.~\cite{gaulme10}). Because the mass of HD\,143105b is larger and orbits a brighter star at a similar 
orbital distance, its reflected light will probably be measurable at visible wavelengths with ultra-high precision 
photometry, e.g. with the future CHEOPS satellite. 
Despite the degeneracy caused by the absence of transit (the amplitude of the modulation varies as the product 
of the planet surface and geometric albedo), the shape and offset of the phase curve could reveal the composition 
and structure of the atmosphere (see, e.g., Seager et al.~\cite{seager00}; Showman \& Guillot~\cite{showman02}; 
Heng \& Demory~\cite{heng13}). 
It might already be possible to analyse the light reflected by HD\,143105b atmosphere through cross-correlation 
of high-resolution optical spectra, 
constraining the planet orbital inclination and its true mass (Martins et al.~\cite{martins13},~\cite{martins15}). 
High-resolution spectra of HD\,143105b thermal emission from the dayside could be obtained in the future with
infrared instruments as SPIRou at CFHT or JWST. 
The identification of individual molecular lines in the infrared would then bring further constraints on the thermal 
structure, the circulation, and composition of the atmosphere (e.g. Brogi et al.~\cite{brogi12}; Snellen et al.~\cite{snellen14}).

\subsection{HIP\,109600}
\label{sect_HIP109600}

The star HIP\,109600 (BD$\,+28^{\circ}4312$) was observed before and after the SOPHIE upgrade. 
We finally secured 34
measurements. The 232-day period is obvious on the time series, and no significant drift is detected over 
the nearly seven years of observation. The residuals of the Keplerian fit shows dispersions of 9.6 and 3.3\,m/s
for SOPHIE and SOPHIE+, respectively, in agreement with the typical uncertainties of their measurements. 
SOPHIE data are redshifted by $33.6 \pm 1.1$\,m/s by comparison to SOPHIE+.

The semi-amplitude is $K=98.6 \pm 0.5$\,m/s which corresponds to 
$M_\textrm{p} \sin i =2.68\pm 0.12$\,M$_\mathrm{Jup}$.
HIP\,109600b is a giant planet orbiting a G5V star in $P=232.08 \pm 0.15$\,days 
with a low but significant eccentricity of $e=0.163 \pm 0.006$. Its orbit is thus located 
in a temperate orbit, with an insolation level around $1.5 \pm 0.3$ that of the Earth on its orbit.
The habitable zone around a star could be defined as the area where 
the insolation level would be between 0.2 and 1.8 that of the Earth (e.g. Kopparapu et al.~\cite{kopparapu13}; 
Jenkins et al.~\cite{jenkins15}). Any possible satellite orbiting HIP\,109600b could thus be habitable.
The presence of such satellites remains out of reach of present technics unfortunately.
We note also that HIP\,109600 is metal-poor~(${\rm [Fe/H]} = -0.12 \pm 0.02$), whereas most 
giant-planet hosts are metal-rich.

Several studies have shown that the Keplerian signature of a single, eccentric planet could also be 
fitted with a model including two planets on circular (or nearly circular) orbits in 2:1 resonance, 
depending on the radial velocity accuracy and their time sampling  (e.g. Anglada-Escud\'e 
et al.~\cite{anglada10}; Wittenmyer et al.~\cite{wittenmyer13}; K\"urster et al.~\cite{kurster15}).
Numerous pairs of planets near the 2:1 orbital resonance have indeed be detected in transit
with  \textit{Kepler}  (e.g. Lissauer et al.~\cite{lissauer11}; Delisle \&\ Laskar~\cite{delisle14}).
Here, adding an inner planet with a period half of the HIP\,109600b one
provides a fit with a similar residuals dispersion than the one-planet fit adopted above. 
So residuals dispersion comparison does not allow us to exclude the presence of an inner 
planet with a period of $\sim117$~days and a semi-amplitude $K = 16 \pm 2$\,m/s 
corresponding to $M_\textrm{p} \sin i =0.35\pm 0.04$\,M$_\mathrm{Jup}$. 
Adding that additional, inner planet would not significantly~change the parameters of HIP\,109600b, 
except its orbit is~circular.

To compare further
the above one- and two-planet  Keplerian models, we computed their 
Bayesian evidences using the methods of Chib \& Jeliazkov~(\cite{chib01}) and Perrakis
et al.~(\cite{perrakis14}) (see also D{\'{\i}}az et al.~\cite{diaz16b}). Both methods produce~similar 
results but the dispersion of Chib \& Jeliazkov~(\cite{chib01}) is much smaller. 
We found that the one-planet model is $21.9 \pm 0.8$ times more probable, so we 
finally adopt that~solution.

\subsection{HD\,35759}

We started observing HD\,35759 (BD$\,+64^{\circ}532$, HIP\,25883) 
in September 2011, a few weeks after the SOPHIE+ upgrade. 
71 radial velocities were finally acquired. Their Lomb-Scargle periodogram 
shows a clear peak at 82.5~days, as well as a weaker peak at the harmonic corresponding 
to half the period. There are no ambiguities however on the true period of the signal, 
measured to be $82.467 \pm 0.019$~days.	

The Keplerian fit of the radial velocities provides a semi-amplitude $K=173.9 \pm 1.3$\,m/s corresponding 
to a mass $M_\textrm{p} \sin i =3.76\pm 0.17$\,M$_\mathrm{Jup}$. The eccentricity is significantly 
detected to be $e=0.389 \pm 0.006$. 
HD\,35759b is thus a giant planet orbiting a G5V star with a period similar to that of Mercury 
but with a larger eccentricity.

The residuals of that fit show a dispersion of 6.0\,m/s, in agreement with
the typical accuracy of the measurements. Adding an inner planet in a circular orbit of half that period 
significantly degrades the fit, providing a residuals dispersion above 20\,m/s. 
Moreover, we detect no significant drift over the 3.5 years of observation neither.
So there are no hints for any additional inner nor outer companion in that system.

\subsection{HIP\,109384}
\label{sect_HIP109384}

We observed the G5V star HIP\,109384 (BD$\,+70^{\circ}1218$)
over more than seven years, before and after the SOPHIE upgrade. 
The final dataset includes 42 measurements. The periodic signal with a period near 500~days clearly appears 
in the data. The residuals of the one-planet Keplerian fit are 3.7 and 5.8\,m/s, in agreement with the 
uncertainties on the individual measurements with SOPHIE and SOPHIE+, respectively. 
The dispersion is surprisingly larger for data secured 
after the upgrade, but only seven measurements were secured before the upgrade. The SOPHIE data
are redshifted by $6.7 \pm 0.5$\,m/s by comparison to SOPHIE+, which is a small shift when compared 
to other systems observed before and after the upgrade, in the present study as well as in Moutou et al.~(\cite{moutou14}) 
or Wilson et al.~(\cite{wilson16}). This could be the signature of an additional drift
due to another companion in the system, but the current data do not allow that to be confirmed. The Keplerian 
fits made with or without that drift provide similar parameters for the planet HIP\,109384b.
Adding a circular, inner planet with a period near 250~days (half the period of HIP\,109384b) 
provides a significantly poorer fit with 
a residuals dispersion of 12\,m/s for the SOPHIE+ data, so there are no hints for such an inner, additional planet
in 2:1 resonance.

HIP\,109384b has a mass $M_\textrm{p} \sin i =1.56\pm 0.08$\,M$_\mathrm{Jup}$ ($K=56.53 \pm 0.22$\,m/s)
and a period of $499.48 \pm 0.32$~days. 
Its host star is metal-poor (${\rm [Fe/H]} = -0.26 \pm 0.03$).
The orbit is eccentric with $e=0.549 \pm 0.003$. In addition to the 
main signal of HIP\,109384b, the Lomb-Scargle periodogram shows a weak peak near 10~days which is still 
visible in the residuals of the 1-planet fit. 
No corresponding peaks are seen in the periodograms of bisectors, width, nor $\log{R'_\mathrm{HK}}$.
The residuals could be fitted at this period with an amplitude $K\simeq6$\,m/s, corresponding 
to about one Neptune mass; that small signal is not significant in our data however
so we do not claim such a detection.  
We will continue observing that star in order
to study that signal, as well as the possible drift. The presence or not of such additional inner or outer 
planet does not significantly change the parameters of the giant planet HIP\,109384b.

From periastron and apoastron of its eccentric orbit, the insolation levels of 
HIP\,109384b could be estimated between $0.4\pm0.1$ and $0.6\pm0.1$ 
of the Earth's insolation. This places HIP\,109384b 
in the outer part of its habitable zone, making particularly interesting any satellite which might orbit 
that giant~planet.

\subsection{HD\,220842}
\label{sect_HD220842}

We started observing the F8V, inactive, metal-poor star HD\,220842 
(BD$\,+56^{\circ}3009$, HIP\,115714) 
in June 2012, one year after the SOPHIE upgrade. 
The dataset includes 42 radial velocity measurements spread over 2.7~years. They display large variations 
that can be fitted with a Keplerian orbit of 218-day period.
Figure~\ref{fig_omc} could suggest that this period mainly stands on one point
(at BJD\,=\,2\,456\,847.6), and that a single planet on a period two times longer (about 440~days) 
could  fit the data as well if that point is omitted. 
However, fits without that point and with a double period show clearly a larger 
dispersion in their residuals, and we found no reasons to omit that~point.

The final fit provides a period of $218.47 \pm 0.19$~days and a semi-amplitude $K=108.1 \pm 1.2$\,m/s
corresponding to the mass $M_\textrm{p} \sin i =3.18\pm 0.15$\,M$_\mathrm{Jup}$. The residuals 
show a dispersion of 4.5\,m/s in agreement with the accuracy of the measurements. The orbit is eccentric 
with $e=0.404 \pm 0.009$. 
Including an additional,  circular planet with a period near 109~days 
(half the period of HD\,220842b) provides a significantly poorer fit with 
a residuals dispersion of 13\,m/s, so there are no hints for such inner, resonant planet.

In addition to the one-planet Keplerian model, the data clearly show an additional 
drift revealing the presence of a second companion to the star, which could be stellar or planetary. 
There are no indications of binarity of that star in the literature.
A linear trend is enough to fit the drift and we obtain $-35.5 \pm 1.0$\,\ms\,yr$^{-1}$. 
Assuming an eccentricity of $e\simeq0.5$ for that additional companion, it would correspond 
to a period of at least $\sim1100$~days and a projected mass of at least $\sim2.1$\,M$_\mathrm{Jup}$.
Solutions with longer periods and larger masses are also allowed from the data, as well as 
different values for the eccentricity. We will continue to follow that system with SOPHIE to characterize 
that second component, and in particular to determine if it is a planet or a~star.

\begin{figure}[h!] 
\begin{center}
\includegraphics[width=\columnwidth]{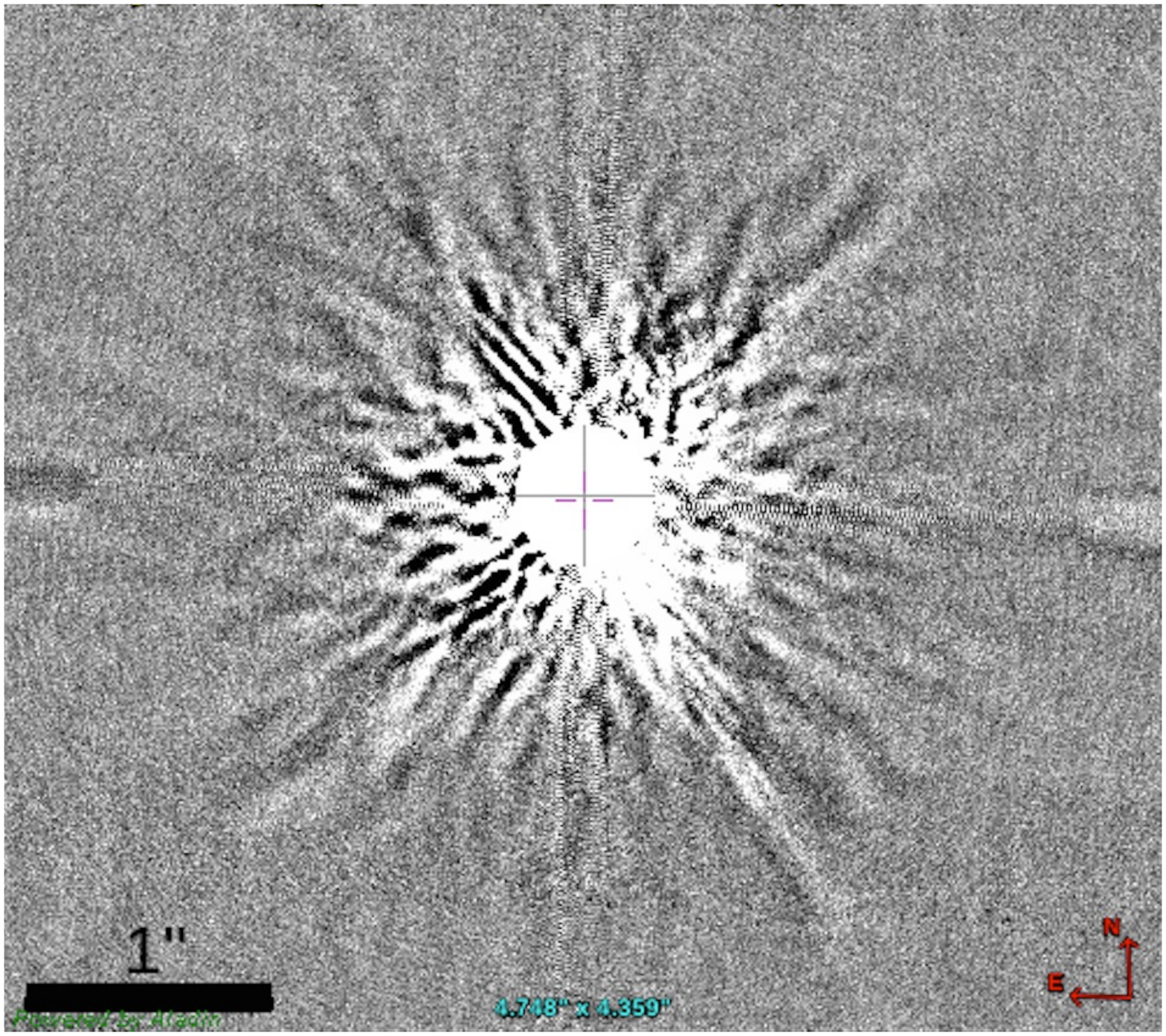}
\includegraphics[width=\columnwidth]{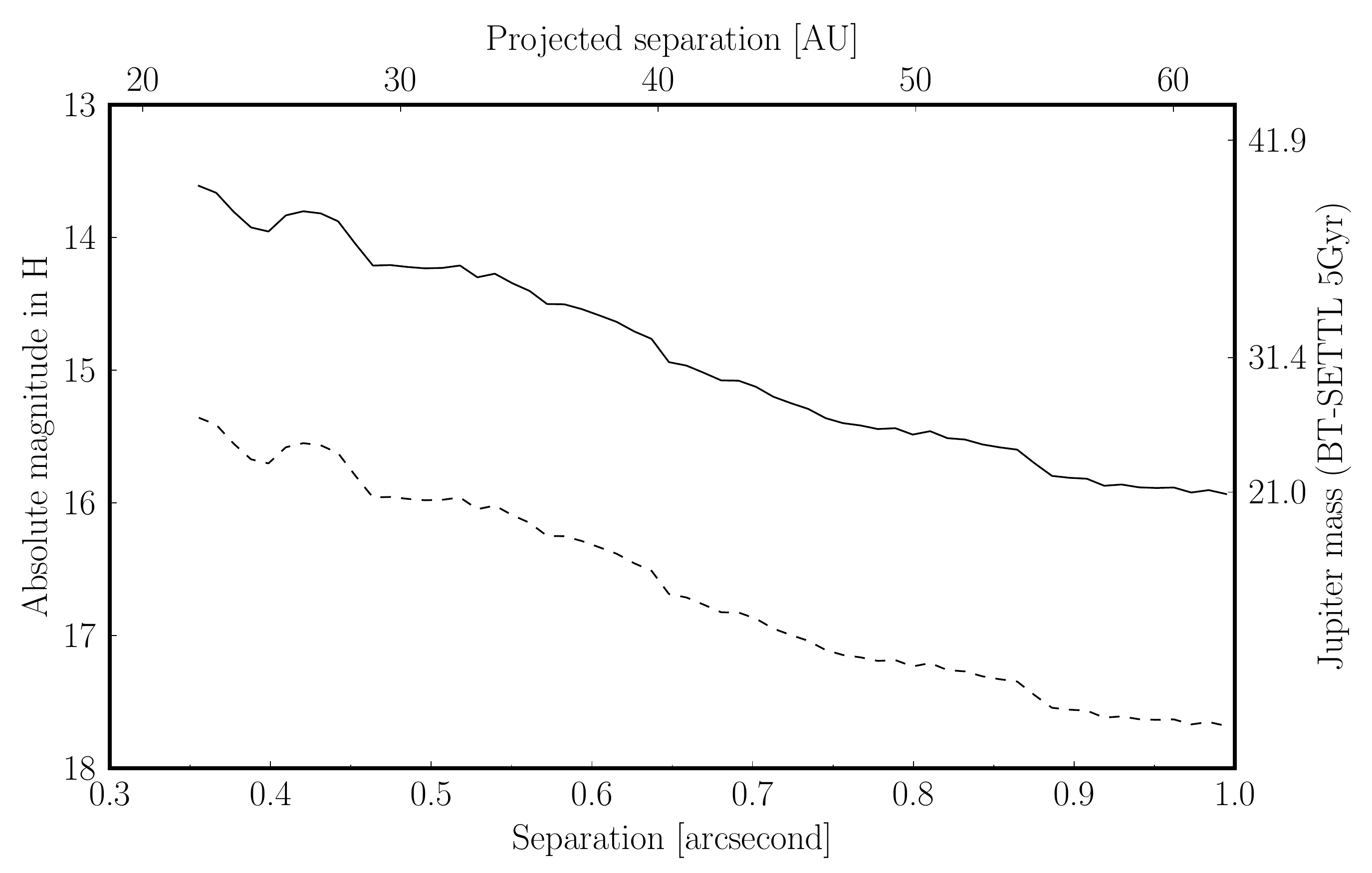}
\caption{
Subaru/SCExAO ADI observation of HD\,220842 in H-band (upper panel).
No candidate has been resolved in this snapshot. The lower panel shows
the corresponding upper limits on the companion producing the radial-velocity drift of HD\,220842.
Detections limits at 1 and 5 $\sigma$ are plotted in dashed and solid lines, respectively.  
The mass estimate is based on BT-SETTL
CIFIST2011 models (Allard et al.~\cite{allard14}). The contrast
curve has been corrected to take into account self-subtraction.
Companions above the solid line woulde have been detected at more than 5\,$\sigma$.
}
\label{fig_scexao}
\end{center}
\end{figure}

To search for the companion at the origin of the drift we observed HD\,220842 with the 
Subaru Telescope Extreme Adaptive Optics system 
(SCExAO; Jovanovic et al.~\cite{jovanovic15}).
SCExAO is a high performance coronagraph using a series of wavefront controls
at the 8-m Subaru Telescope in Mauna Kea (Hawaii, USA).
HD\,220842 was observed for 20~minutes on November 1st, 2015 in field
rotation mode, and the angular differential imaging data was reduced
using GRAPHIC (Hagelberg et al.~\cite{hagelberg16}).
No companion was detected in the resulting data but an upper limit on
the putative companion can be derived from the detection limit
(Fig.~\ref{fig_scexao}).
The detection limit is generated following a similar procedure as in
Chauvin et al.~(\cite{chauvin15}), while the mass estimates are
based on the BT-SETTL models (Allard et al.~\cite{allard14})
using a conservative age estimate of 5\,Gyr for the system.
Assuming the companion was not hidden below or in front of the star at the 
time of our observation, we can exclude at 5\,$\sigma$ that the companion 
is more massive than 20\,M$_\mathrm{Jup}$
with a period shorter than $\sim230$~years ($a \sim 60$\,AU), 
or more massive than 40\,M$_\mathrm{Jup}$
with a period shorter than $\sim75$~years ($a \sim 20$\,AU). 
We can thus conclude that the companion at the origin of the linear drift is probably sub-stellar. 
Future observations could reinforce that conclusion if the companion remains 
undetected at different epochs.
Deeper observations than the snapshot we have obtained may be able to directly
detect the companion.

\subsection{HD\,12484}
\label{sect_HD12484} 

The F8V star HD\,12484 (BD$\,+02^{\circ}320$, HIP\,9519) 
was observed with SOPHIE+ from September 2012 to January 2015.
The final dataset includes 65 radial velocities with a typical accuracy of 10\,m/s. The moderate 
level of accuracy is due to the broadening of the lines caused by the rotation of the star 
($v\sin i_\star = 8.2 \pm 1.0$\,km/s). That F8V star is active with chromospheric emissions in the 
H and K lines (Fig.~\ref{fig_caII}) corresponding to $\log{R'_\mathrm{HK}} = -4.43 \pm 0.10$.
Despite that high level of activity which implies stellar jitter, a periodic signal is clearly detected
in the data. The one-planet Keplerian fit provides a period of $58.83 \pm 0.08$~days and a semi-amplitude 
$K = 155 \pm 5$\,m/s corresponding to $M_\textrm{p} \sin i =2.98\pm 0.14$\,M$_\mathrm{Jup}$.
No significant eccentricity is detected and we obtained the formal value $e = 0.07 \pm 0.03$. 
The period being outside the domain range of hot Jupiters, there are 
no reasons  to fix the orbit as circular, and we let the eccentricity free in the fit.

\begin{figure}[h] 
\begin{center}
\includegraphics[scale=0.56]{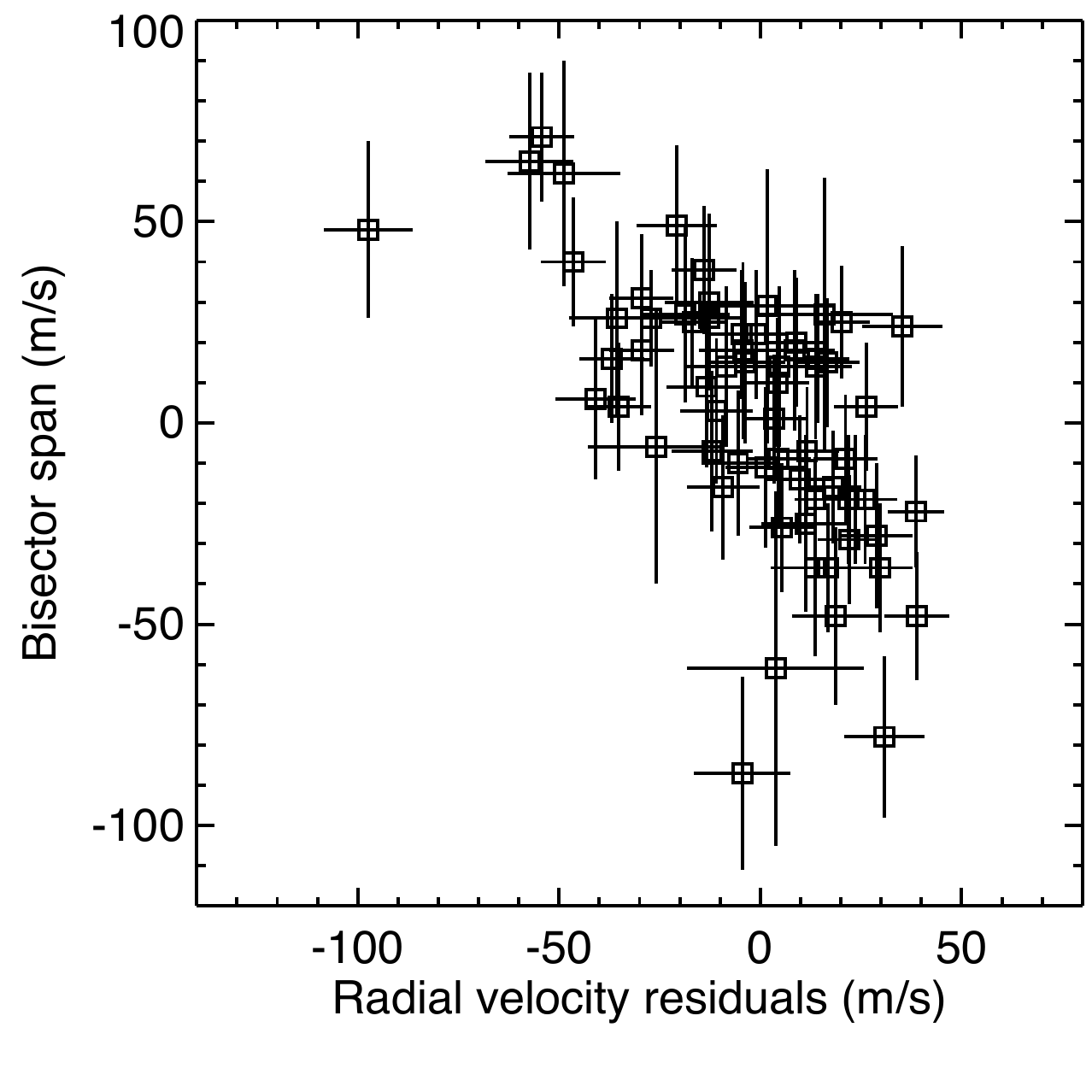}
\vspace{-0.5cm}
\caption{Bisector span as a function of the the radial 
velocity residuals after the Keplerian fit for HD\,12484. 
The clear anti-correlation indicates that the residuals
are mainly due to the stellar activity.
}
\label{fig_bis_HD12484}
\end{center}
\end{figure}

The dispersion of the residuals of the fit is 25.2\,m/s which is large in comparison to the accuracy 
of the data. They agree however with the stellar jitter of $\sim$20\,m/s expected for an F-type star 
at that level of activity (Santos et al.~\cite{santos00b}). Figure~\ref{fig_bis_HD12484} shows the clear 
anti-correlation between the bisector spans and the radial velocity residuals after the Keplerian fit, 
which is the typical signature of  variations induced by stellar activity
(Queloz et al.~\cite{queloz01}; Boisse et al.~\cite{boisse09}). We thus conclude that the extra noise in 
the radial velocities is due to the stellar activity of HD\,12484. No periodic signals were detected in the 
residuals, even by considering only short time spans of data. The rotation period deduced from the 
activity level of the star is $P_\mathrm{rot} \simeq 7 \pm 4$\,days. It is different enough from the orbital 
period of the planet to assume that the activity noise does not significantly perturb the parameters 
of the planet.
Indeed, the parameters are deduced from the Keplerian model which considers the stellar jitter 
as an uncorrelated noise, which is not necessary the case~here.

The inclination of the orbit being unknown, the transit probability of HD\,12484b is $\sim1.5$\,\%.
Among the new single planets presented here, this is the second highest transit 
probability after the hot Jupiter HD\,143105b (Sect.~\ref{sect_HD143105}). 
The four other single planets presented above have transit probabilities included between 0.5 and 1.0\,\%.
HD\,143105b is the only one for which we have attempted a transit detection up to now.

Adding an inner planet with half the orbital period of HD\,12484b provides a fit with a similarly high 
residuals dispersion of 25\,m/s.
This dispersion comparison does not exclude 
the presence around HD\,12484 of an inner planet with a period 
of $\sim 29$~days and semi-amplitude $K = 15 \pm 5$\,m/s corresponding to 
$M_\textrm{p} \sin i =0.22\pm 0.08$\,M$_\mathrm{Jup}$. 
However, Bayesian evidences comparison for both models (see Sect.~\ref{sect_HIP109600})
indicates that the one-planet model is $7.8 \pm 0.5$ times more probable
than the two-planet model.
So there is no evidence for the presence of such speculative inner, resonant planet, 
and we adopt the one-planet model for HD\,12484.

Active stars known to host planets are sparse. This is likely to be a selection bias.
Indeed, the stellar jitter caused by activity makes them unfavorable objects to detect 
planets so they are frequently excluded from radial velocity surveys searching 
extrasolar planets. 
Only a few planetary systems are known around stars with $\log{R'_\mathrm{HK}} > -4.5$.
They include hot Jupiters as WASP-52b, 59b, or 84b (H\'ebrard et al.~\cite{hebrard13}; 
Anderson et al.~\cite{anderson14}), giant planets on longer periods as HD\,192263b, GJ\,3021b, 
and HD\,81040b (Santos et al.~\cite{santos00a}; 
Naef et al.~\cite{naef01}; Sozzetti et al.~\cite{sozzetti06}), 
or the multi-planet systems around HD\,128311 and HD\,9446 (Vogt et al~\cite{vogt05}; 
H\'ebrard et  al.~\cite{hebrard10}). The new giant planet HD\,12484b increases that population.

\section{Multi-planet systems detection and characterization}
\label{sec_multi_planet_systems}

The two last systems presented in this paper show the signature of multi-planet systems:
two planets around HIP\,65407 and four around HD\,141399.
They add new cases among the few systems hosting several giant planets, 
whereas most of the low-mass planets are in multiple systems. 
Over the 535 systems from exoplanets.org
known to include planet(s) with masses 
above $0.3$\,M$_\mathrm{Jup}$,  only 50 are multiple.
After detection-bias corrections, 
Mayor et al.~(\cite{mayor11}) report multi-planetary rates of 26\,\% 
for giant planets and over 70\,\%  for planets below $0.1$\,M$_\mathrm{Jup}$.

We fitted the datasets of both 
multiple systems and determined the uncertainties on their parameters 
using the same Keplerian procedure as above for singe planets (Sect.~\ref{sect_method}), 
with five extra free parameters per additional planet. These Keplerian analyses do not take into account for any  
gravitational interactions between planets, which could be significant in some cases.
So we subsequently complement the purely Keplerian analyses by additional dynamical 
studies and compare the results of both approaches. 
Our different analyses of both multi-planet systems are discussed~below.

\begin{figure}[t!] 
\begin{center}
\includegraphics[width=\columnwidth]{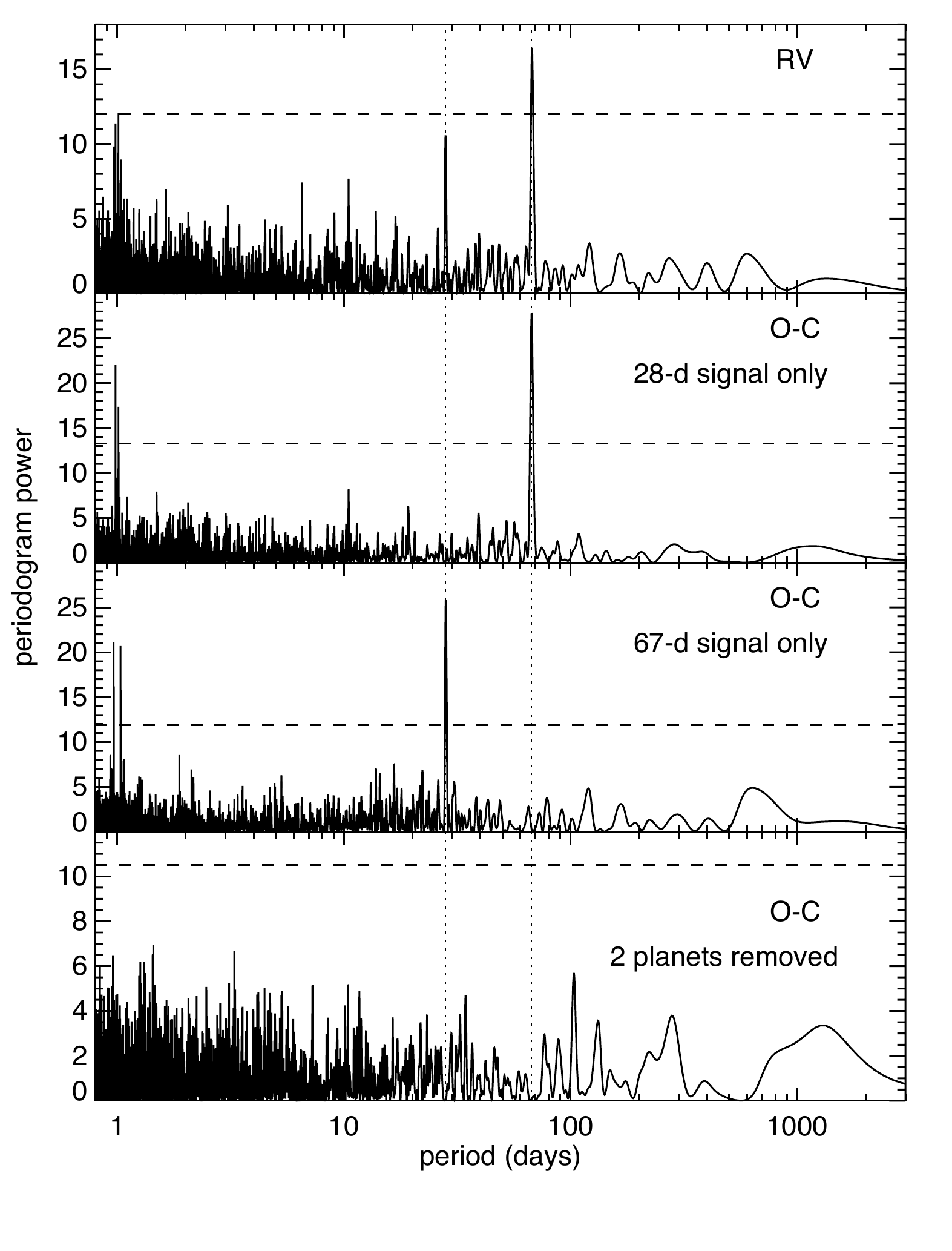}
\vspace{-0.9cm}
\caption{Lomb-Scargle periodograms of the SOPHIE radial velocities of HIP\,65407. 
The upper panel shows the periodogram computed on the radial velocities without any fit 
removed. The second and third panels show the periodograms computed on the residuals 
of the fits only including HIP\,65407b or HIP\,65407c, respectively. 
The bottom panel shows the periodogram after the subtraction of the two-planet fit. 
The two vertical dotted lines show the periods of the two planets.
The horizontal dashed lines correspond to the false-alarm probability of~$1\times10^{-3}$.}
\label{fig_periodogram_HIP65407}
\end{center}
\end{figure}

\subsection{HIP\,65407}

\subsubsection{Planets detection}

We obtained 65 radial velocity measurements of HIP\,65407 (BD$\,+49^{\circ}2212$) 
over more than four years, both with SOPHIE and SOPHIE+. 
The data soon revealed  the signature of two periodic variations in the planet-mass range, 
which were confirmed by subsequent SOPHIE observations. 

Figure~\ref{fig_periodogram_HIP65407} shows Lomb-Scargle periodograms 
of the measurements and residuals of HIP\,65407.
In the upper panel of Fig.~\ref{fig_periodogram_HIP65407} that presents the 
periodogram of the  radial velocities of HIP\,65407,
periodic signals at 28~days and 67~days are clearly detected with 
peaks at the periods corresponding to the two planets. 
The peaks around one day correspond to the aliases of both detected signals, as the sampling 
is as usual biased with the nights and days succession. 

In the second panel of Fig.~\ref{fig_periodogram_HIP65407} is plotted the periodogram 
of the residuals after subtraction of 
a fit including only the 28-day-period planet. The peak at 
67~days is visible in those residuals. In the same manner, the third panel of 
Fig.~\ref{fig_periodogram_HIP65407} shows the periodogram of the residuals after a fit 
including only the 67-day-period planet. The peak at 67~days is no longer 
visible and this time the peak at 28~days is visible.  The 
bottom panel of Fig.~\ref{fig_periodogram_HIP65407} shows the periodogram of the residuals 
after subtraction of the Keplerian fit including both HIP\,65407b and HIP\,65407c. There are no remaining 
strong peaks on this periodogram; even the 1-day alias disappeared, showing that most 
of the periodic signals have been removed from the data. 
False-alarm probabilities of~$1\times10^{-3}$ are also reported in Fig.~\ref{fig_periodogram_HIP65407};
they are based on bootstrap experiments, and comparisons of perio\-dograms peaks among 
all periodograms obtained for each radial-velocity~permutation.

This periodogram analysis shows that the two planets are clearly detected in our data with unambiguous 
periods, and that there are no significant additional periodic signals. 
In particular, there are no signs of additional 
planets at longer periods on the time span of our data (more than four years), which is also seen by the 
fact we do not detect any significant linear drift in the radial velocities in addition to the signature of 
the two planets.

\begin{table*}[ht]
  \centering 
  \caption{Fitted Keplerian orbits and planetary parameters for the two multi-planetary system, with 1-$\sigma$ error bars.}
  \label{table_parameters_multi}
\begin{tabular}{l|cc|cccc}
\hline
\hline
Parameters  								&	HIP\,65407b				&	HIP\,65407c			&	HD\,141399b		 &	HD\,141399c			 &	HD\,141399d		&	HD\,141399e		 \\
\hline
$P$ 				[days]					& $28.125 \pm 0.019$			&  $67.30 \pm 0.08$			& $94.44 \pm 0.05	$	&  $201.99 \pm 0.08	$		& $1069.8 \pm 6.7$		& $5000^{+560}_{-2000}$$^\ast$\\
$e$										& $0.14 \pm 0.07$				&  $0.12 \pm 0.04$			& $0.04 \pm 0.02	$	&  $0.048 \pm 0.009	$		& $0.074 \pm 0.025$		& $0.26 \pm 0.22 $$^\ast$		\\
$\omega$ 		[$^{\circ}$]					& $50 \pm 150$				&  $-19 \pm 25$			& $-90 \pm 80		$	&  $-140 \pm 40	$		& $-140 \pm 30	$		& $-10 \pm 20$			\\
$T_0$			[BJD] $\ddagger$			& $56990.8 \pm 2.5$				&  $57047 \pm 6$	 		& $56998 \pm 15	$	&  $56838 \pm 10	$		& $56923 \pm 65$		& $58900 \pm 800$		\\
$K$				[\ms]						& $30.5 \pm 1.6$				&  $41.5 \pm 1.9$			& $19.23 \pm 0.47 	$	&  $44.20 \pm 0.50	$		& $22.63 \pm 0.59$		& $8.8 \pm 0.9$		\\
$M_\textrm{p} \sin i$	[M$_\mathrm{Jup}$]$^\dagger$	& $0.428 \pm 0.032$				&  $0.784\pm 0.054$			& $0.451 \pm 0.030	$	&  $1.33 \pm 0.08	$		& $1.18 \pm 0.08$		& $0.66 \pm 0.10$		\\
$a$				[AU]$^\dagger$				& $0.177 \pm 0.005$				&  $0.316 \pm0.008$			& $0.415 \pm 0.011$		&  $0.689 \pm 0.020$		& $2.09 \pm 0.06$		& $5.0 \pm 1.5$		\\
$V_{r\mathrm{,\,SOPHIE}}$ [\kms]				& 		\multicolumn{2}{c|}{$-8.748 \pm 0.005$} 				& 					\multicolumn{4}{c}{$-21.566 \pm 0.002$} 				\\
$V_{r\mathrm{,\,SOPHIE+}}$ [\kms]	[\kms]		& 		\multicolumn{2}{c|}{$-8.720 \pm 0.003$} 				& 					\multicolumn{4}{c}{$-21.548 \pm 0.001$} 				\\
$N$										& 		\multicolumn{2}{c|}{65}	 						& 					\multicolumn{4}{c}{139}	 						\\
$\sigma_\mathrm{O-C,\,SOPHIE}$ [\ms]			&		\multicolumn{2}{c|}{11.9}	 						&					\multicolumn{4}{c}{12.8}	 						\\	
$\sigma_\mathrm{O-C,\,SOPHIE+}$ [\ms]			&		\multicolumn{2}{c|}{7.5}	 						&					\multicolumn{4}{c}{4.8}	 						\\	
RV drift 			[\ms\,yr$^{-1}$]				&		\multicolumn{2}{c|}{$[-0.6;+7.8]$} 					&					\multicolumn{4}{c}{$[-1.3;+2.4]$} 					\\
span 			[days]					&		\multicolumn{2}{c|}{1517}                     				&					\multicolumn{4}{c}{2566}							\\
$M_{\textrm{p}{\mathrm{\,max}}} [\rm{M}_{\odot}]$	&		8.11				&  2.25					&	0.22				&	0.12					&	0.04				&	--		\\ 
\hline
\multicolumn{7}{l}{$\ddagger$: time of possible transits.} \\
\multicolumn{5}{l}{$\dagger$: using the stellar mass and its uncertainty from Table~\ref{table_stellar_parameters}.} \\
\multicolumn{7}{l}{$\ast$: stability analyses of HD\,141399e provide $P_e=3370 \pm 90 $~days and $e_e<0.1$ (see Sect.~\ref{saHD}).} \\
\end{tabular}
\end{table*}

\begin{figure}[b!] 
\begin{center}
\includegraphics[width=\columnwidth]{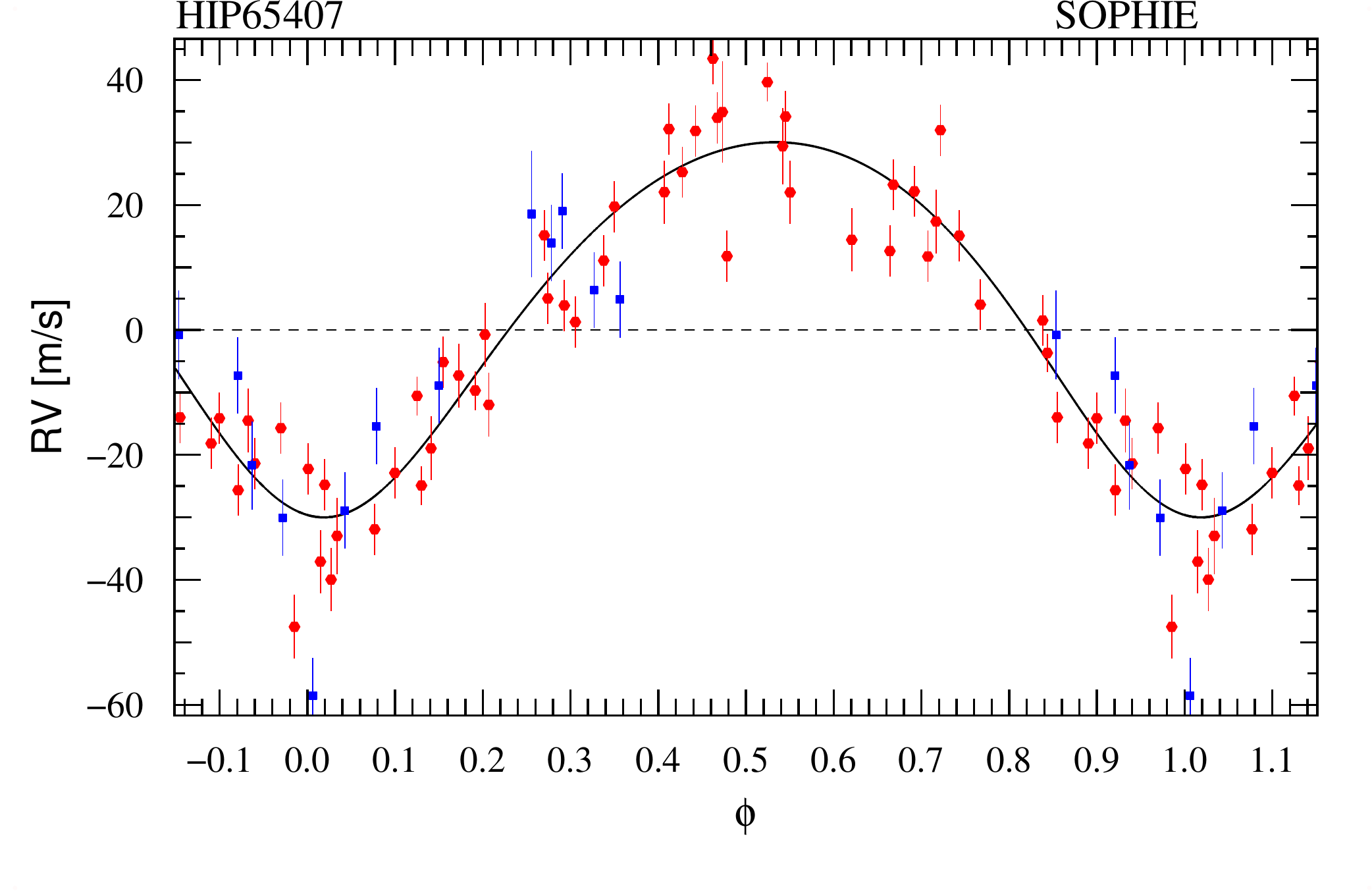}
\includegraphics[width=\columnwidth]{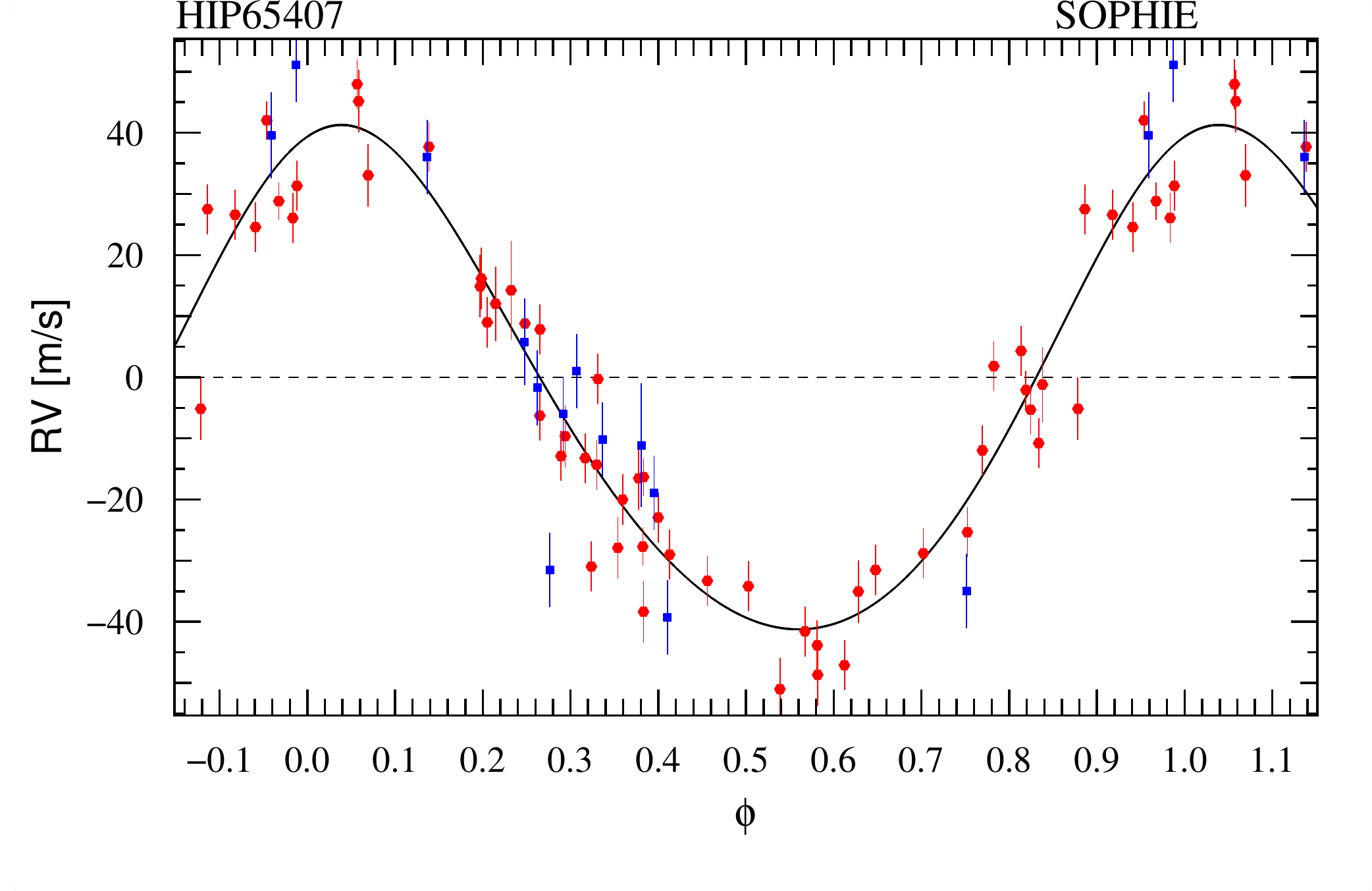}
\caption{Phase-folded radial velocity curves for HIP\,65407b ($P=28.1$\,d, top) 
and HIP\,65407c ($P=67.3$\,d, bottom) after removing the effect of the other planet. 
The SOPHIE data obtained before/after the June-2011 upgrade are plotted in blue/red 
with 1-$\sigma$~error bars, and the Keplerian fits are the solid lines. 
Orbital parameters corresponding to the 
fits are reported in Table~\ref{table_parameters_multi}.}
\label{fig_orb_phas2}
\end{center}
\end{figure}

\subsubsection{Keplerian fit}
\label{Keplerian_fit_HIP65407}

The radial velocities were first fitted with a Keplerian model, i.e. neglecting any possible gravitational 
interactions between both planets.
The two-planet Keplerian best fit  is plotted in Fig.~\ref{fig_omc} as a function of 
time (lower, left panel) and in Fig.~\ref{fig_orb_phas2} as function of the orbital phase. 
The  different phases of both orbits are well covered by our observations. The derived 
parameters are reported in Table~\ref{table_parameters_multi}. 
The inner planet, HIP\,65407b, produces radial velocity variations with a semi-amplitude 
$K=30.5\pm1.6$\,m/s corresponding to a planet with a minimum mass 
$M_\textrm{p} \sin i   = 0.428 \pm 0.032$~M$_\mathrm{Jup}$. 
Its orbit has a period of 
$28.125 \pm 0.019$~days and no significant detection of any eccentricity ($e=0.14\pm 0.07$).
The outer planet, HIP\,65407c, yields a semi-amplitude $K=41.5\pm1.9$\,m/s corresponding 
to a planet with a projected mass $M_\textrm{p} \sin i = 0.784 \pm 0.054$~M$_\mathrm{Jup}$.
The orbital period is $67.30\pm 0.08$~days and there is a hint of a possible small eccentricity detection
($e=0.12\pm 0.04$).
No astrometric motions corresponding to these planets were detected in the Hipparcos~data 
(Sect.~\ref{sect_single_planet_systems}), in agreement with the non-detection of any secondary 
peak in the CCFs. The derived upper limits $M_{\textrm{p}{\mathrm{\,max}}} $ 
on the planetary masses are above one Solar mass and thus not~constraining. 

SOPHIE data are redshifted by $20.0 \pm 5.8$\,m/s by comparison to SOPHIE+.
There are mainly SOPHIE+ data for that 
system and their residuals show a  better dispersion (7.5\,m/s) than that of SOPHIE residuals
(11.9\,m/s). That dispersion is slightly  larger than the one expected for SOPHIE+. This could be partly due 
to stellar jitter as the host star is active. Its $\log{R'_\mathrm{HK}} = -4.60 \pm 0.10$ could imply
stellar jitter of the order of 10\,m/s (Sect.~\ref{sect_stellar_properties}). 
In addition, the period of the inner planet ($28.125 \pm 0.019$~days) is just slightly longer than the
stellar rotation period, roughly estimated in Sect.~\ref{sect_stellar_properties} to be 
$19\pm7$\,days.
Confusions between stellar activity signals and Doppler wobble due to the planet could be suspected here.
We note however that the bisectors show no significant variations, periodicity, or correlation as a function of the 
radial velocities, or as a function of the radial velocity residuals after 
fits including both planets or just one of them. 
Such correlation could have been the signature of changes in the stellar line profiles due to the activity of the star, 
as in the case of HD\,12484 (Sect.~\ref{sect_HD12484}).
We do not see either any significant periodicities in the periodograms of Fig.~\ref{fig_periodogram_HIP65407}
in addition to those of the two planets. This could have been the signature of stellar rotation.
So despite the fact that the star is active, we detect no significant signs of stellar jitter in our radial velocity 
data, except possibly dispersion residuals that are slightly larger than expected. 
Moreover, the parameters of both planets are stable over the more than four years of observations; 
they also remain unchanged when only short-time sequences are analyzed instaed of the whole dataset.
Together with the absence of periodicities in the periodograms of other activity indicators as the width of the 
CCF or the $\log{R'_\mathrm{HK}}$ index, it is thus unlikely that 
the activity of the star does significantly affect the orbital parameters that we derived from their Keplerian fit.
These parameters appear thus~reliable.

The transit probabilities for HIP\,65407b and HIP\,65407c are $\sim2.6$\,\%\ and $\sim1.3$\,\%, 
respectively. We have not attempted any transit detection for them up to now.

As the system shows a planetary system composed of two planets in compact orbits, 
we assessed below the gravitational interactions between both planets in order to evaluate 
their possible effects and differences with the Keplerian fit which assumes both planets do not interact. 
The stability of the system is not straightforward since the minimum masses of the planets are of~the 
same order as the mass of Jupiter. 
As a consequence, mutual gravitational interactions between planets 
may give rise to some~instabilities.

\subsubsection{Secular coupling}

The ratio between the orbital periods determined by the Keplerian fit 
(Table~\ref{table_parameters_multi}) is 
$ P_c / P_b = 2.4 $, which is close to a 12:5 mean motion resonance.
We performed a frequency analysis of the 
Keplerian orbital solution listed in Table~\ref{table_parameters_multi} computed over 10\,kyr.
The orbits of the planets are integrated with the 
symplectic integrator SABAC4 of Laskar \& Robutel~(\cite{laskar01}), 
using a step size of 0.01\,yr.
We concluded that, in spite of the proximity of the 12:5 mean motion resonance, the two planets 
in the HIP\,65407 system are not trapped inside, at least when we adopt the minimum values for the masses.

\begin{figure}[b!]
    \includegraphics*[width=\columnwidth]{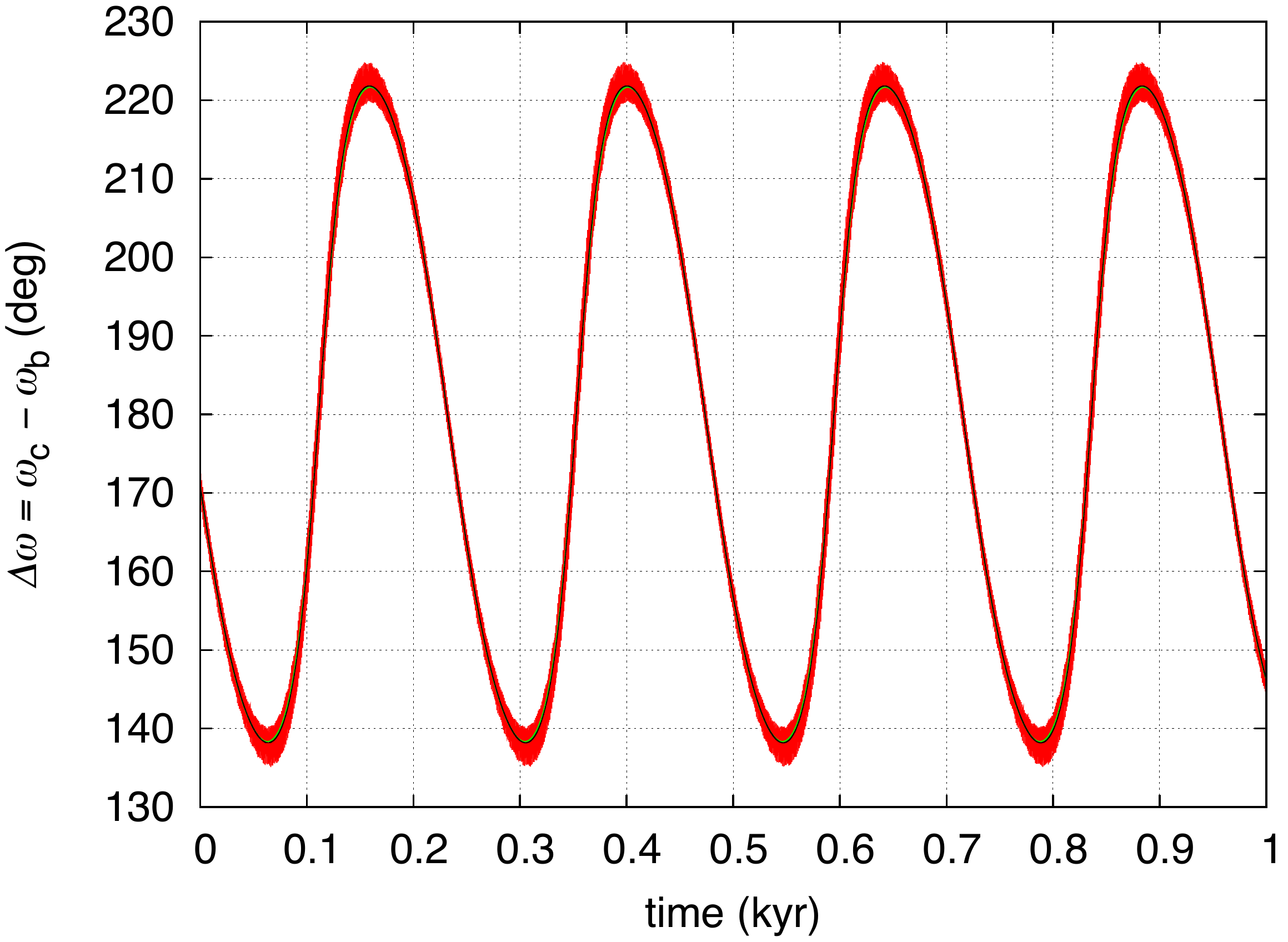} 
  \caption{Evolution in the HIP\,65407 two-planet system of the angle  
  $\Delta \omega = \omega_c-\omega_b$ 
  (red line), which oscillates around $180^\circ$ with a maximal amplitude of $45^\circ$.  
  The green and black lines (nearly over-plotted) also gives the 
  $\Delta \omega $ 
  evolution, but obtained with the linear secular models 
  (Eqs.~\ref{eq.lape} and \ref{eq.lape2},~respectively).
   \label{Fdyn1}   }
\end{figure}

The fundamental frequencies of the systems are then
the mean motions $n_b$ and $n_c$, and the two secular frequencies of the 
periastrons $g_1$ and $g_2$ (Table~\ref{Tdyn1}).
Because of the proximity of the two orbits, there is  a strong  coupling 
within the secular system (see Laskar~\cite{laskar90}).
Both planets HIP\,65407b and HIP\,65407c precess with  the same precession frequency 
$g_1$, which has a period of 206~years. 
The two periastrons are thus locked and 
$\Delta \omega = \omega_c-\omega_b$
oscillates around $180^\circ$ (anti-aligned ellipses) with a maximal amplitude of about $45^\circ$ 
(see Fig.\,\ref{Fdyn1}).

\begin{table}[t!]
 \caption{Fundamental frequencies for the Keplerian orbital solution in
 Table~\ref{table_parameters_multi}. $n_b$ and $n_c$ are the mean motions, and $g_1$ and $g_2$ are the
 secular frequencies of the periastrons.
 \label{Tdyn1}} 
 \begin{center}
 \begin{tabular}{crrr}
 \hline\hline
      & Frequency   & Period & Angle ($\phi_\lp$) \\
      & ($^\circ/yr$) & (yr) & (deg) \\
 \hline
 $n_b$ & 4675 &             0.0770 &       21.89 \\
 $n_c$ &   1954 &           0.1843 &   -125.4 \\
 $g_1$ &          1.747 &       206.1 & 161.6 \\
 $g_2$ &          0.2568 & 1402 & -32.66 \\ \hline
\end{tabular}
\end{center}
\end{table}

\begin{figure}[b!]
    \includegraphics*[width=\columnwidth]{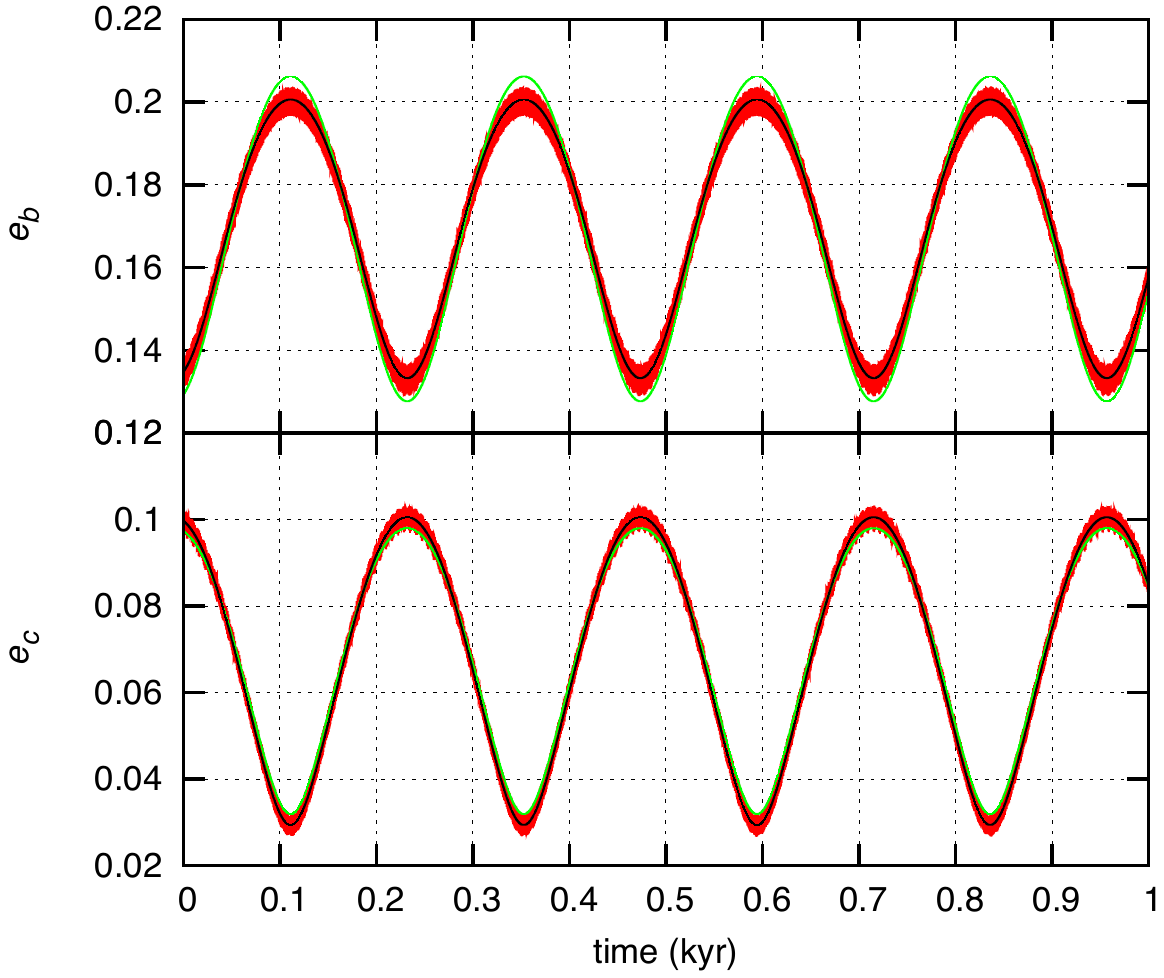} \\    
  \caption{Evolution  with time of the eccentricities $e_b$ and $e_c$ 
  of both planets HIP\,65407b and HIP\,65407c, starting with the orbital solution 
  from~Table~\ref{table_parameters_multi}. The red lines give the complete solutions for the 
  eccentricities~obtained numerically, while the green and black curves are the associated 
  values obtained with 
  the linear secular models (Eqs.~\ref{eq.lape} and~\ref{eq.lape2},~respectively). 
   \label{Fdyn2}}   
\end{figure}

To present the solution in a clearer way, it is useful to make a linear change of variables 
into eccentricity proper modes (Laskar~\cite{laskar90}).
In the present case, due to the proximity of the 12:5 mean motion resonance and due 
to the possible eccentric orbits, 
the linear transformation is numerically 
obtained by a frequency analysis of the solutions.
Using the classical complex notation 
$z_\kp = e_\kp \mathrm{e}^{i \omega_\kp}$,
for $\kp = b, c$, we have for the linear Laplace-Lagrange solution:
\begin{equation}
\left(\begin{array}{c} z_b\\ z_c \end{array}\right)= 
\left(\begin{array}{rrr} 
 0.1669 & \   0.03924 \\
-0.06496 &  \  0.03307 \\
\end{array}\right)
\left(\begin{array}{c} u_1\\ u_2 \end{array}\right) \ .
\label{eq.lape}
\end{equation}
The proper modes $u_\lp$ (with $\lp = 1, 2$) are obtained from the $z_\kp$ by inverting the above linear relation. 
To good approximation, we have $u_\lp \approx  \mathrm{e}^{i (g_\lp t +\phi_\lp)}$, 
where $g_\lp$ and $\phi_\lp$ are given in Table~\ref{Tdyn1}.
From Eqs.~\ref{eq.lape}, the meaning of the observed 
libration between the periastrons $\omega_b$ and $\omega_c$ could be understood. 
Indeed, for both planets  HIP\,65407b and HIP\,65407c, the dominant term is $u_1$  
with frequency $g_1$, and they thus both precess with an average value of $g_1$ 
(green line in Fig.\,\ref{Fdyn1}). 

Equations~\ref{eq.lape} usually also provide a good approximation for the long-term evolution of the eccentricities. 
In Fig.~\ref{Fdyn2} we plot the eccentricity evolution
with initial conditions from Table~\ref{table_parameters_multi}. 
Simultaneously, we plot with green lines the evolution of the same elements given by the above secular, 
linear approximation. 
We see nevertheless that there is a small deviation from the linear approximation.
Going back to the decomposition of $z_\kp$ in the proper modes, we notice that there 
is an additional forcing frequency at $2 g_1 - g_2$ with significant power.
We can correct the linear Laplace-Lagrange solution (Eqs.\,\ref{eq.lape}) to include the contribution from this term as:
\begin{equation}
\left(\begin{array}{c} z_b\\ z_c \end{array}\right)= 
\left(\begin{array}{rrr} 
 0.1670 & \   0.03924 &  -0.005583 \\
-0.06496 & \   0.03307 & 0.002529 
\end{array}\right)
\left(\begin{array}{c} u_1\\ u_2 \\ u_1^2u_2^{-1} \end{array}\right) \ .
\label{eq.lape2}
\end{equation}
In Figs.~\ref{Fdyn1} and~\ref{Fdyn2} we plot with black lines the evolution 
using the corrected secular approximation. 
We see there is now a perfect agreement with the numerical solution, 
so the term in $2 g_1 - g_2$ is essential for the correct description of the secular evolution of the system.
The eccentricity variations are limited and described well by the secular approximation (Eqs.~\ref{eq.lape2}). 
The eccentricity of planets HIP\,65407b and HIP\,65407c are within the ranges 
$ 0.129 < e_b < 0.204 $ and 
$ 0.027 < e_c < 0.103 $, respectively. These variations are  
driven mostly by  the  secular frequency $g_1-g_2$, of period approximately 242~years (Table~\ref{Tdyn1}). 

There is an increasing difference with time between the present model taking into 
account for mutual gravitational interactions between planets one a first hand,
and on a second hand the Keplerian model which does not (Sect.~\ref{Keplerian_fit_HIP65407}).
Extrapolated in 2050, both models should show differences of the order of $\pm10$\,m/s. 
A detection of that difference would thus be feasible in next~decades.

\subsubsection{Stability analysis}
\label{saBD}

In order to analyze the stability of the Keplerian solution (Table~\ref{table_parameters_multi}),
we performed a global frequency analysis (Laskar~\cite{laskar93}) in the vicinity of this solution, 
in the same way as achieved for other planetary systems 
(see, e.g., Correia et al.~\cite{correia05}, \cite{correia09}, \cite{correia10}).

\begin{figure}[t!]
  \centering
    \includegraphics*[width=\columnwidth]{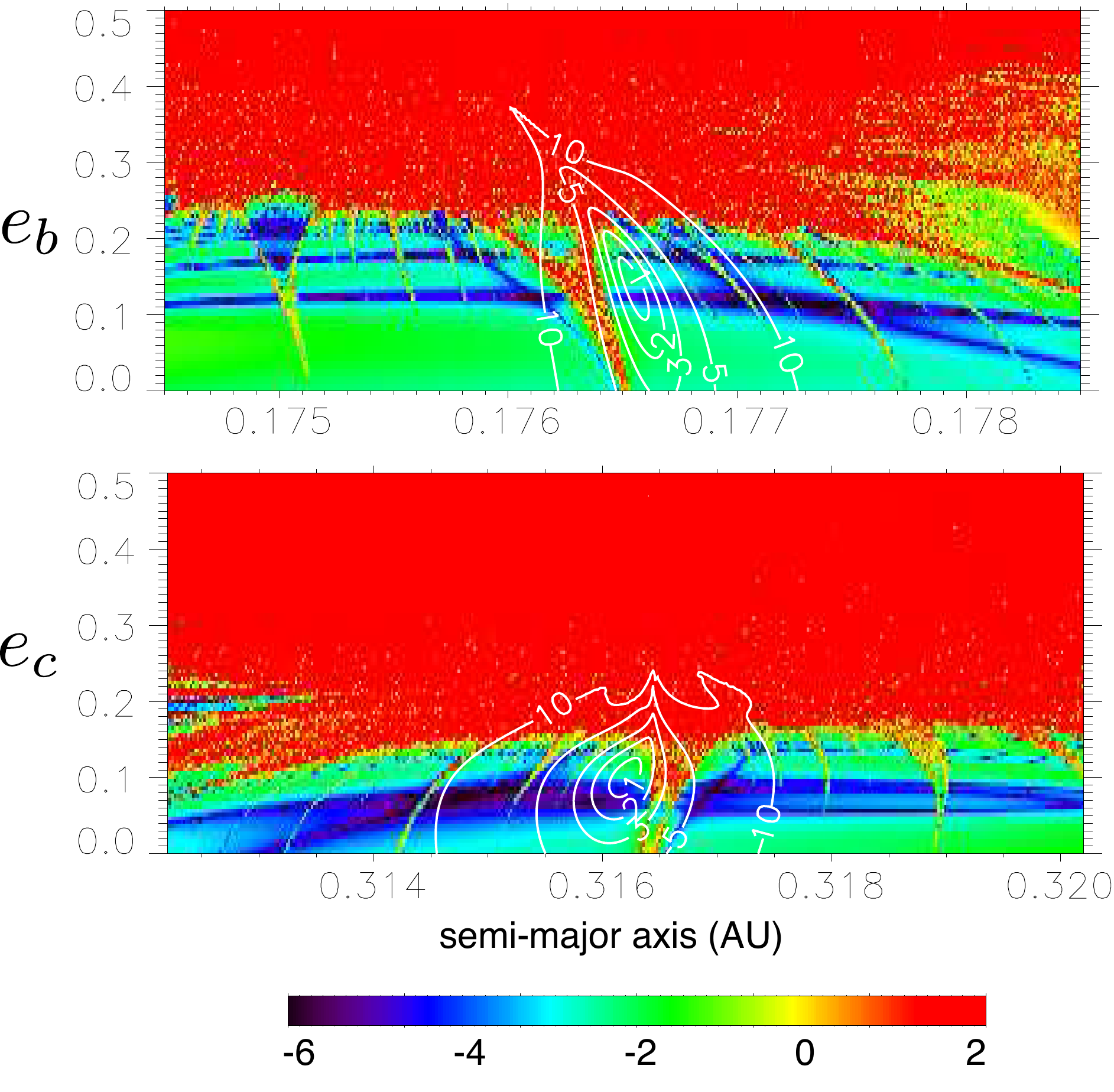} 
  \caption{
  Stability analysis of the Keplerian fit of the HIP\,65407 planetary system. For fixed initial conditions  
  (Table~\ref{table_parameters_multi}), the phase space of the system is explored by varying the 
  semi-major axis $a_\kp$ and eccentricity $e_\kp$ of each planet, respectively HIP\,65407b 
  and~HIP\,65407c. The red zone corresponds to highly unstable orbits while  the dark-blue region 
  can be assumed to be stable on timescales of a billion years
(the color scale shows the $ D $ values; see Sect.~\ref{saBD}). 
  The contour curves indicate the 1, 2, 3, 5, 10-$\sigma$ confidence~intervals from~$\Delta\chi^2$.
  \label{Fdyn3}}   
\end{figure}

For each planet the system is integrated on a regular 2D mesh of initial conditions, 
with varying semi-major axis and eccentricity, while the other parameters are retained at their 
Keplerian values (Table~\ref{table_parameters_multi}). The solution is integrated over 1000~years 
for each initial condition. 
There are typically 400 steps for the semi-major axis and 100 for the eccentricity. 
All the stability analyses below have similar numbers of steps in their meshes.
The stability criterion is derived with the frequency analysis of the mean longitude 
over the two consecutive 500-year intervals of time (Laskar~\cite{laskar90},~\cite{laskar93}).
As in Correia et al.~(\cite{correia05}, \cite{correia09},~\cite{correia10}), the chaotic diffusion 
is measured by the variation of the mean motion in each time interval, $n_i$ and $n_i'$, given by the 
index $D = \log(|n_i-n_i'|/n_i)$, where $i$ corresponds to the orbit whose initial
conditions are not fixed. 
For regular, quasi-periodic motions, there are no variations in the mean  motion along the trajectory,
whereas a variation of the mean  motion is the signature of chaotic behavior.
The result is reported using a color index in Fig.~\ref{Fdyn3}, where the red area 
represents the strongly chaotic trajectories ($ D > -2$ in the present case), 
and the dark-blue area the extremely stable ones ($ D < -4$).

In Fig.~\ref{Fdyn3} we show the wide vicinity of the best fitted solution, the minima of the $\chi^2$ 
level curves corresponding to the Keplerian parameters (Table~\ref{table_parameters_multi}). 
We observe the presence of the large 12:5 mean motion resonance, which is particularly unstable.
We clearly see that the present system is outside this resonance, in a more stable area at the 
bottom right side (Fig.\,\ref{Fdyn3}, {\it top}) or at the bottom left side (Fig.\,\ref{Fdyn3}, {\it bottom}).

From the previous stability analysis, it is clear that the HIP\,65407 planetary system listed in 
Table~\ref{table_parameters_multi} is stable over Gyr timescale. Nevertheless, we also tested 
directly this by performing a numerical integration of the orbits over 1\,Gyr. We used the symplectic 
integrator SABAC4 of Laskar \&\ Robutel~(\cite{laskar01}) with a step size of 0.01\,yr, including 
general relativity corrections. The orbits evolve in a regular way, and remain stable throughout the simulation.

\subsubsection{Additional constraints from dynamic}
\label{bdac}

\begin{figure}[h]
\includegraphics[width=\columnwidth,angle=0]{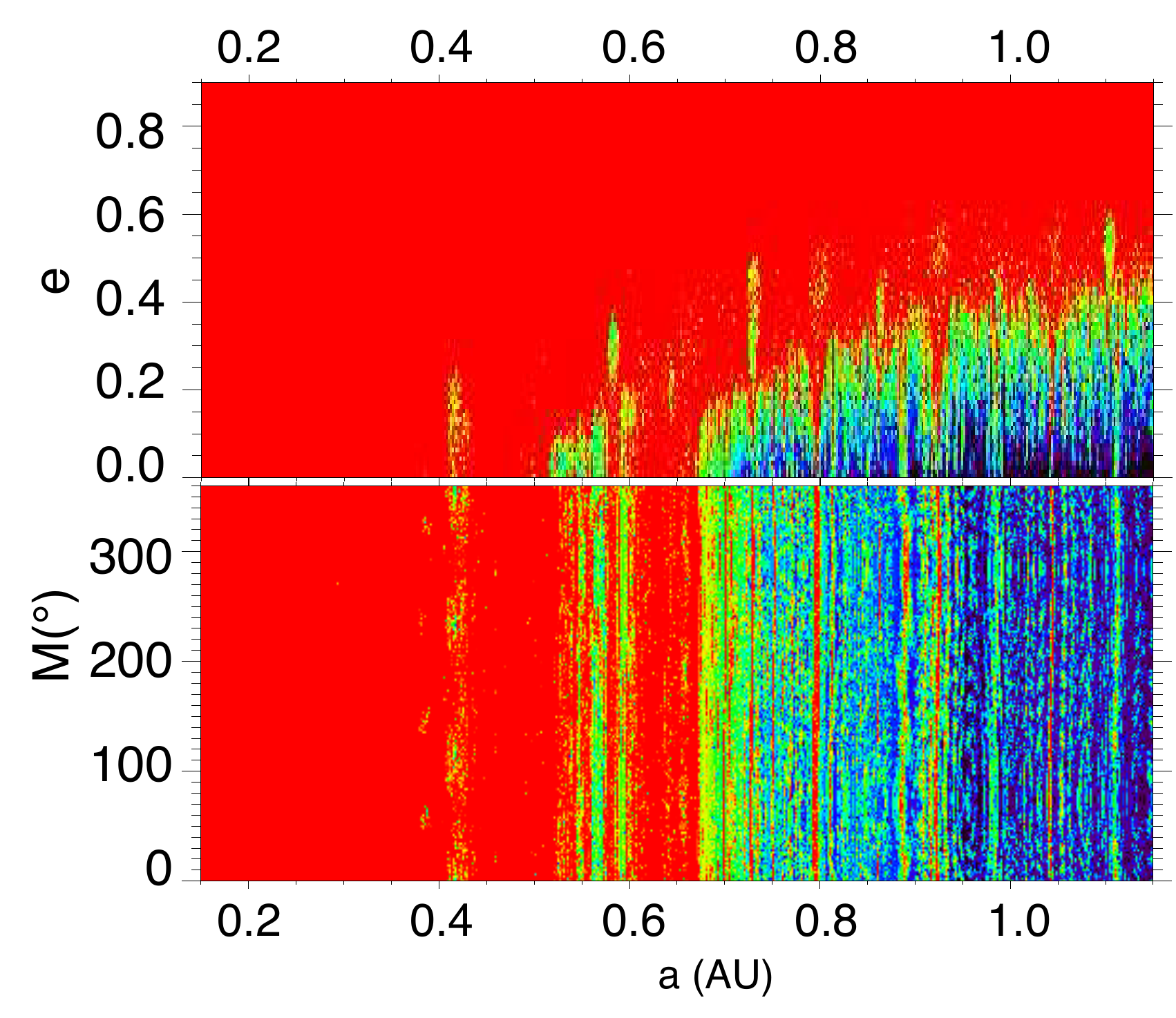} 
  \caption{Possible location of an additional third planet in the HIP\,65407 system.
  The stability of an Earth-mass planet is analyzed, for various semi-major axis versus 
  eccentricity ({\it top}), or mean anomaly ({\it bottom}). 
  All the angles of the putative planet are set to $0^\circ$ (except the mean anomaly in the bottom panel), 
  and in the bottom panel its eccentricity is 0.
 As in Fig.~\ref{Fdyn3}, the color scale corresponds to $ D $ values between $-6$ and $2$.
  The only stable zones  where additional planets can be found are the dark-blue regions.
    \label{Fdyn8}}   
\end{figure}

We assume that the dynamics of the two planets is not 
disturbed much by the presence of an additional low-mass, close-by planet.
We can thus test the possibility of an additional third planet in the system
by varying its semi-major axis, eccentricity, and longitude of the
periastron over a wide range, and performing a stability analysis as in Fig.~\ref{Fdyn3}. 
The test was completed for a fixed value  $K=0.1$\,m/s, corresponding to
a possible Earth-mass planet HIP\,65407d at approximately 1\,AU, whose radial-velocity amplitude 
is at the edge of detection. 
From the analysis of the stable areas in Fig.~\ref{Fdyn8}, one can see that 
additional planets are only possible beyond 0.7\,AU, which corresponds to 
orbital periods longer than 0.6\,yr (the orbital period of Venus).

We can also try to find constraints on the maximal masses of the current
two-planet system if we assume co-planarity of the orbits.
Indeed, up to now we have been assuming that the inclination of the system to
the line-of-sight is $90^\circ$, which gives minimum values for the planetary
masses (Table~\ref{table_parameters_multi}).
By decreasing the inclination of the orbital plane of the system, we increase
the values of the mass of all planets and repeated a stability analysis of the orbits,
like in Fig.~\ref{Fdyn3}.
As we decrease the inclination, the stable dark-blue areas become narrower,
to a point that the minimum $\chi^2$ of the best fit solution lies outside the
stable zones.
At that point, we conclude that the system cannot be stable anymore.
It is not straightforward to find a transition inclination between the two
regimes, but we can infer from our plots that stability of the whole system is
still possible for an inclination as low as $10^\circ$, but becomes impossible for an
inclination around $7^\circ$. 
Therefore, we conclude that the maximum masses of the planets may be
most probably computed for an inclination around $10^\circ$, corresponding to a 
scaling factor of about $6$ to $8$ for the possible masses. This 
corresponds to maximum masses 3.4\,M$_\mathrm{Jup}$ and 6.2\,M$_\mathrm{Jup}$
for HIP\,109600b and HIP\,109600c, respectively.

\subsection{HD\,141399}

\subsubsection{Keplerian fits}

We started observing the low-activity star HD\,141399 
(BD$\,+47^{\circ}2267$, HIP\,77301) with SOPHIE in March 2008. 
Two inner planets HD\,141399b and HD\,141399c 
appeared near resonance 2:1 at periods around 94 and 202~days. We continued the follow-up of that interesting 
system with SOPHIE+ starting in July 2011. The improved accuracy of SOPHIE+ allow the third planet HD\,141399d 
to be detected at period near 1070~days. We were following up HD\,141399 to characterize the additional drift due 
to a possible fourth, external planet when Voigt et al.~(\cite{vogt14}) announced their independent 
detection of a four-planet 
system around that star based on HIRES and APF data.

\begin{figure}[] 
\begin{center}
\includegraphics[width=\columnwidth]{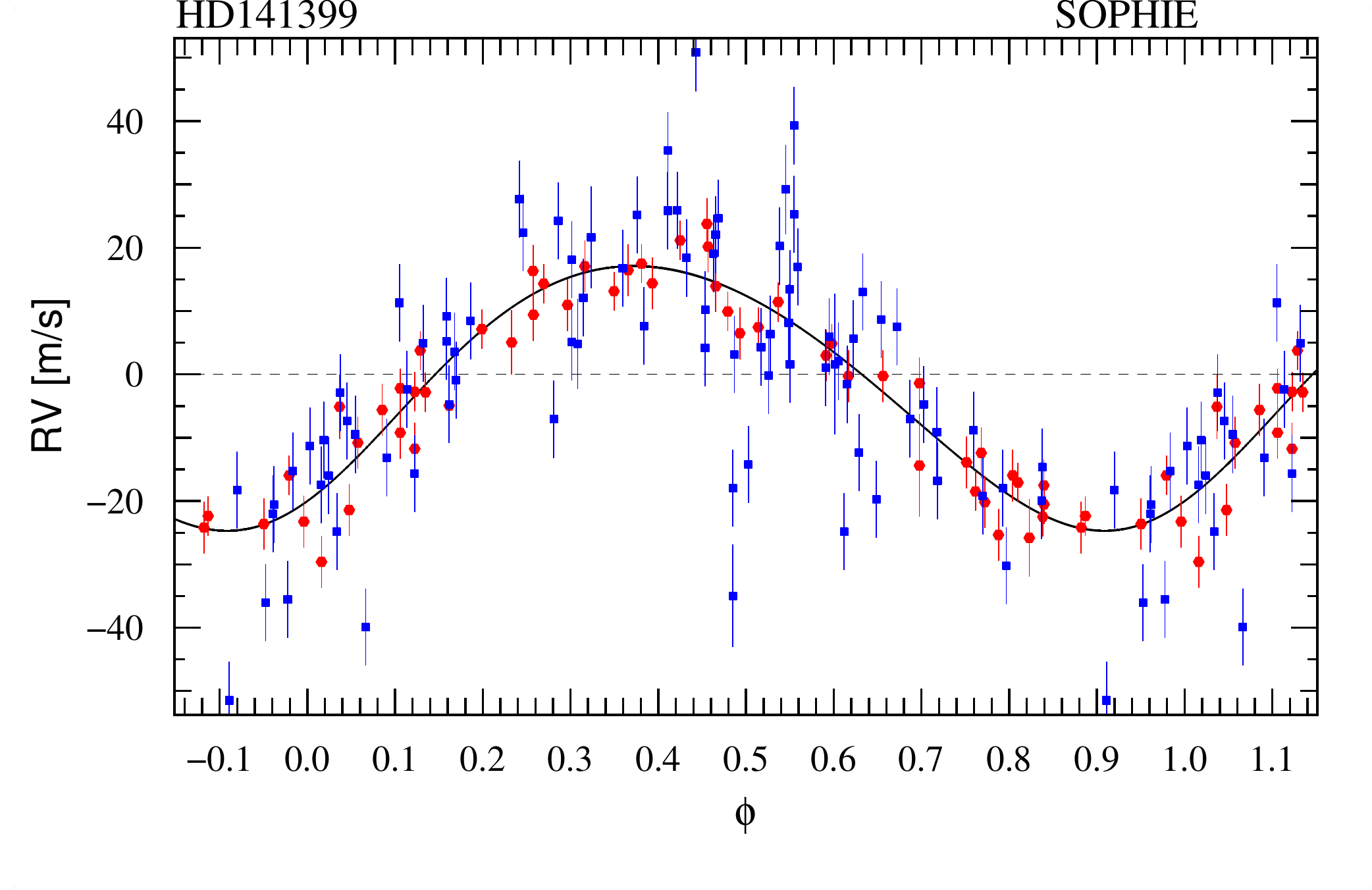}
\includegraphics[width=\columnwidth]{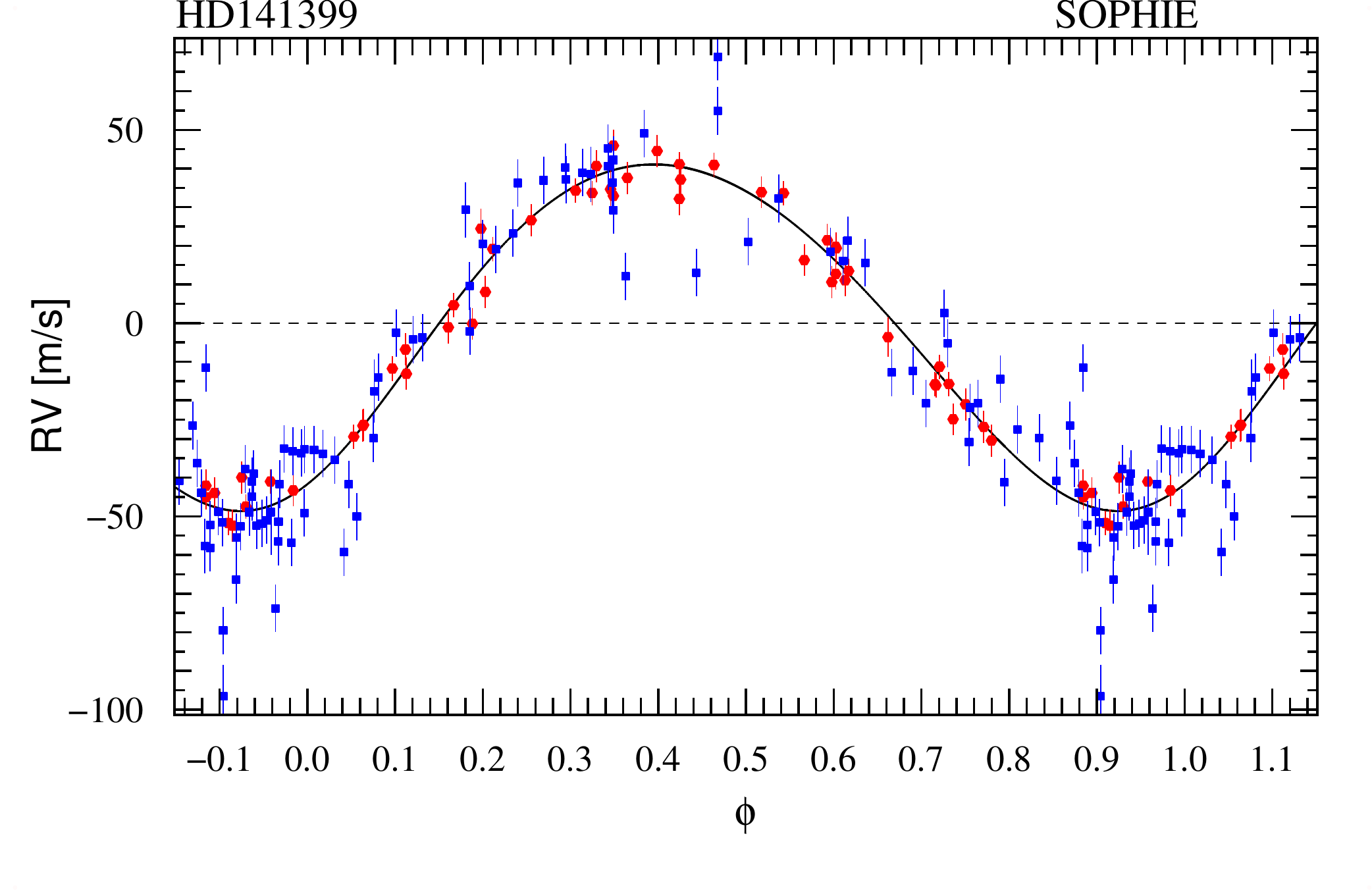}
\includegraphics[width=\columnwidth]{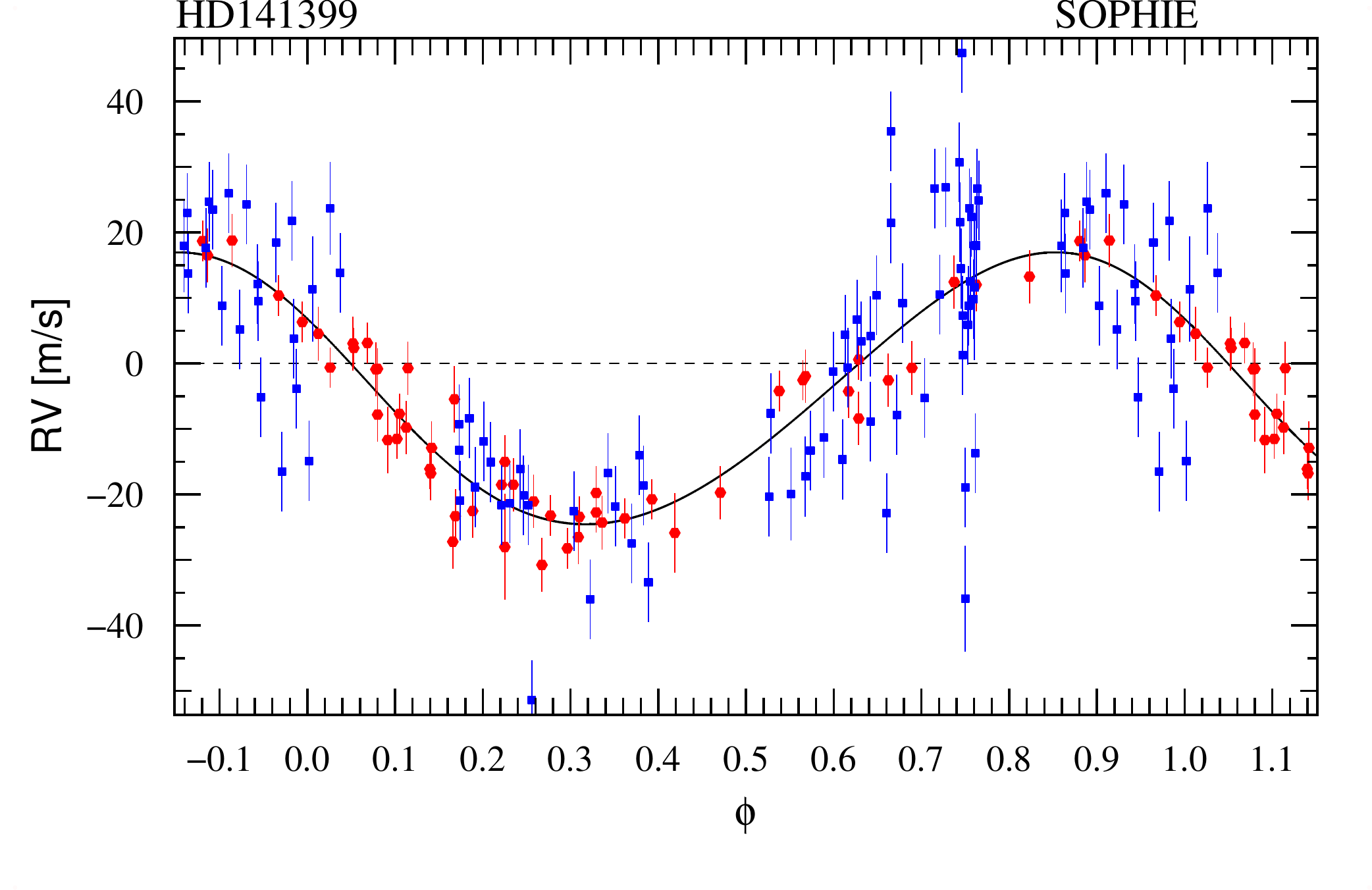}
\includegraphics[width=\columnwidth]{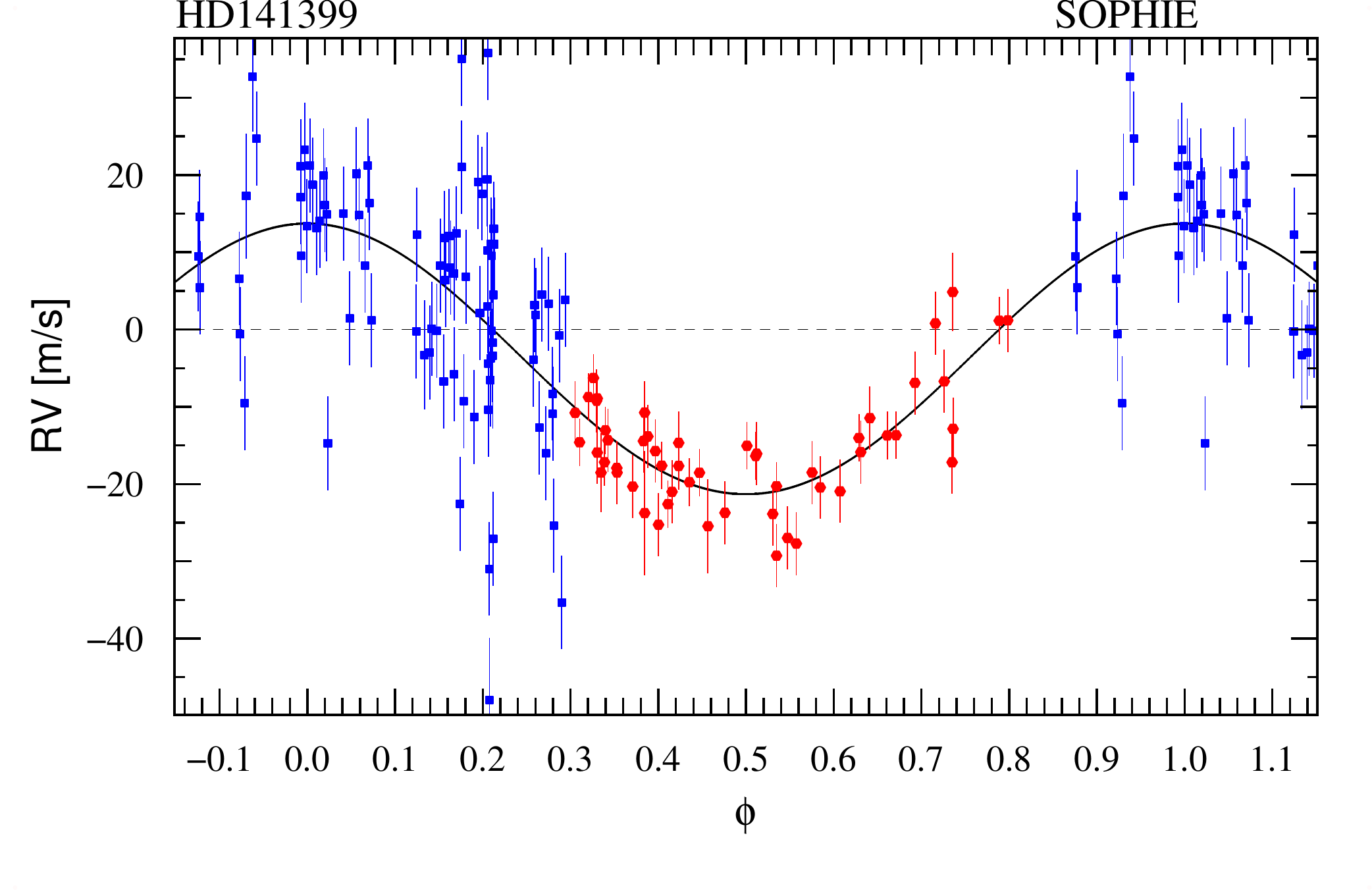}
\caption{Phase-folded radial velocity curves for each of the four planets around HD\,141399
(planets b, c, d, and e from top to bottom) after removing the effect of the three other planets.
The SOPHIE data obtained before/after the June-2011 upgrade are plotted in blue/red 
with 1-$\sigma$~error bars, and the Keplerian fits are the solid lines. 
Orbital parameters corresponding to the 
fits are reported in Table~\ref{table_parameters_multi}.}
\label{fig_orb_phas3}
\end{center}
\end{figure}

Voigt et al.~(\cite{vogt14}) provided stellar mass and activity index in agreement with what we obtained. We however 
derived a slightly hotter $T_{\rm eff}$ and a more accurate metallicity.
The SOPHIE radial velocities and their four-planet Keplerian model are plotted in Fig.~\ref{fig_omc} as a function 
of time. The four phase-folded curves are presented in Fig.~\ref{fig_orb_phas3}. The 139 SOPHIE and SOPHIE+ 
measurements cover seven~years of time span and have a relative shift of $25.9 \pm 5.8$\,m/s. 
Their dispersions around the four-planet Keplerian fit are 12.8 and 4.8~m/s 
respectively. By comparison, the 77 HIRES measurements from Voigt et al.~(\cite{vogt14}) cover ten~years 
with a better dispersion around the Keplerian fit of 2.3~m/s, 
whereas their 14 APF measurements cover half a year with a similar accuracy (2.6~m/s).
The four-planet Keplerian fit of our data present a solution similar to the one presented by Voigt et al.~(\cite{vogt14}).
Our periodograms are similar to those they present in their Fig.~4. The peaks in the SOPHIE data are slightly 
stronger and narrower because of our higher number of measurements allowing an improved phase sampling.

The parameters we present in Table~\ref{table_parameters_multi} are those obtained from the four-planet Keplerian 
fit of the whole available dataset, i.e. including our new data as well as those of Voigt et al.~(\cite{vogt14}). The 
parameters are similar to those reported by Voigt et al.~(\cite{vogt14}), with some improvements in the uncertainties 
thanks to the additional data. So we confirm the four-planet system around HD\,141399 as presented by 
Voigt et al.~(\cite{vogt14}). 
We note that with its period of 202 days, the insolation level received by HD\,141399b is
$1.3 \pm 0.3$ that received by the Earth; the giant planet HD\,141399b is thus located in the habitable 
zone of its host star.
The transit probabilities for HD\,141399b and HD\,141399c are around 1\,\%\ but did not 
attempt any transit detection up to now. The transit probability of the two outer planets is 
one order of magnitude smaller.

No astrometric motions corresponding to the four planets were detected in the Hipparcos data 
(Sect.~\ref{sect_single_planet_systems}), in agreement with the non-detection of any secondary 
peak in the CCFs.  The derived upper limits $M_{\textrm{p}{\mathrm{\,max}}}$ on the masses 
of the inner planets b and c are $0.22$ and $0.12\;\rm{M}_{\odot}$, respectively. No upper limit 
could be derived for the outer planet e, the time span of the Hipparcos data covering only a small 
portion of the orbit. In the case of HD\,141399d, the Hipparcos data cover slightly more than an orbital 
period (about three years) and the derived upper limit is $0.04\;\rm{M}_{\odot}$; we can thus conclude 
this companion is clearly substellar, with a true mass below 40\,M$_\mathrm{Jup}$.

The main issue of discussion is the characterization of the outer planet, HD\,141399e, which is a Jupiter near-twin.
Its eccentricity and orbital period remain poorly constrained. 
Even if a complete revolution could have been covered thanks to our new data, 
similar radial-velocity amplitudes can be produced by simultaneously increasing the eccentricity 
and the orbital period  (as $ K^{-1} \propto P^{1/3} \sqrt{1-e^2} $).
Voigt et al.~(\cite{vogt14}) announced a period 
of $3717 \pm 555$~days whereas our new data alone provide $3000^{+1300}_{-120}$~days. Fitting the whole 
data set provides us the conservative estimation of $5000^{+560}_{-2000}$~days. We obtained an eccentricity~of 
$0.26 \pm 0.22$ by the simultaneous fit of the whole dataset, so we can not significantly conclude from the Keplerian fits 
whether~the orbit is circular or not. Voigt et al.~(\cite{vogt14}) assumed a circular orbit for HD\,141399e due to the lack of 
constraint, acknowledging~their are no reasons for this long-period orbit to be perfectly~circular.

Following these Keplerian approaches, we made 
studies of the gravitational interactions between planets, in particular to try to put constraints 
on the period and eccentricity of the outer planet HD\,141399e.

\begin{figure}[t!]
  \centering
    \includegraphics*[width=\columnwidth,angle=0]{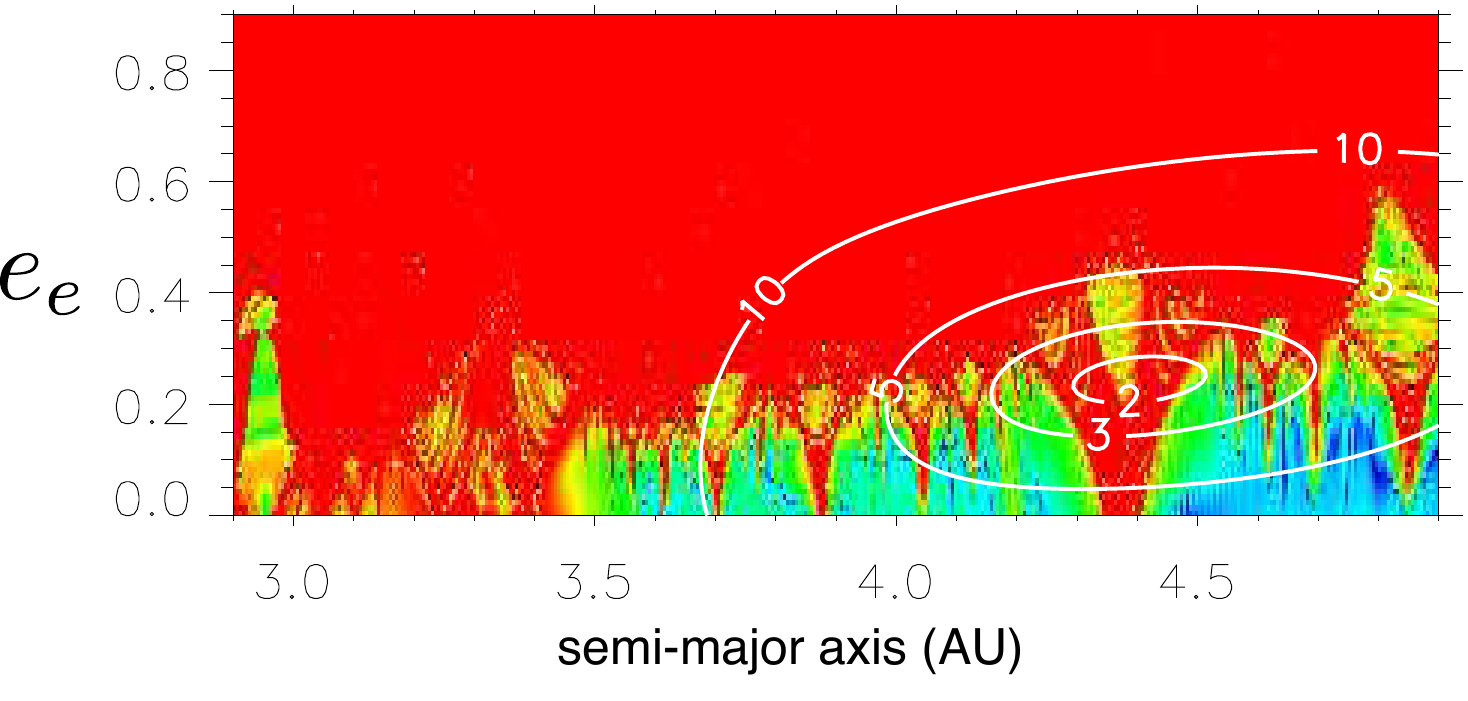} 
  \caption{
  Stability analysis of the outermost planet HD\,141399e for the 
  Keplerian fit (Table~\ref{table_parameters_multi}). For fixed initial conditions, the phase space of the 
  system is explored by varying the semi-major axis $a_e$ 
    and the eccentricity $e_e$. As in Fig.~\ref{Fdyn3}, the color scale shows $ D $ values between $-6$ and $2$.  
  The red zone corresponds to highly unstable 
    orbits while  the dark-blue region can be assumed to be stable on a billion-years timescale.
  The contour curves indicate the 2, 3, 5, 10-$\sigma$ confidence intervals from~$\Delta\chi^2$.
 The best fit lies well inside a red chaotic region.
  \label{Fdyn4}}   
\end{figure}

\subsubsection{Stability analysis}
\label{saHD}

The solution corresponding to the Keplerian fit (Table~\ref{table_parameters_multi}) 
allows high values for the eccentricity  of the outer planet HD\,141399e ($e_e = 0.26 \pm 0.22 $).
We can test the reliability of this solution by performing a global stability study 
in its vicinity as explained above for the HIP\,65407 system (Sect.~\ref{saBD}).  
Figure~\ref{Fdyn4} shows the wide vicinity of the best fitted solution.
For fixed initial conditions, the phase space of the 
system is explored here again by varying the semi-major axis $a_e$ and eccentricity $e_e$.
  For each initial condition, the system is integrated over 1000~years
  and the stability criterion is derived  with the frequency analysis of the mean longitude. 
We observe that the minima of the $\chi^2$ level curves corresponding to the Keplerian parameters 
(Table~\ref{table_parameters_multi}) lies deep inside an extremely chaotic region, so we can conclude 
that the Keplerian solution is highly unstable.

\begin{figure}[t!]
  \centering
    \includegraphics*[width=\columnwidth]{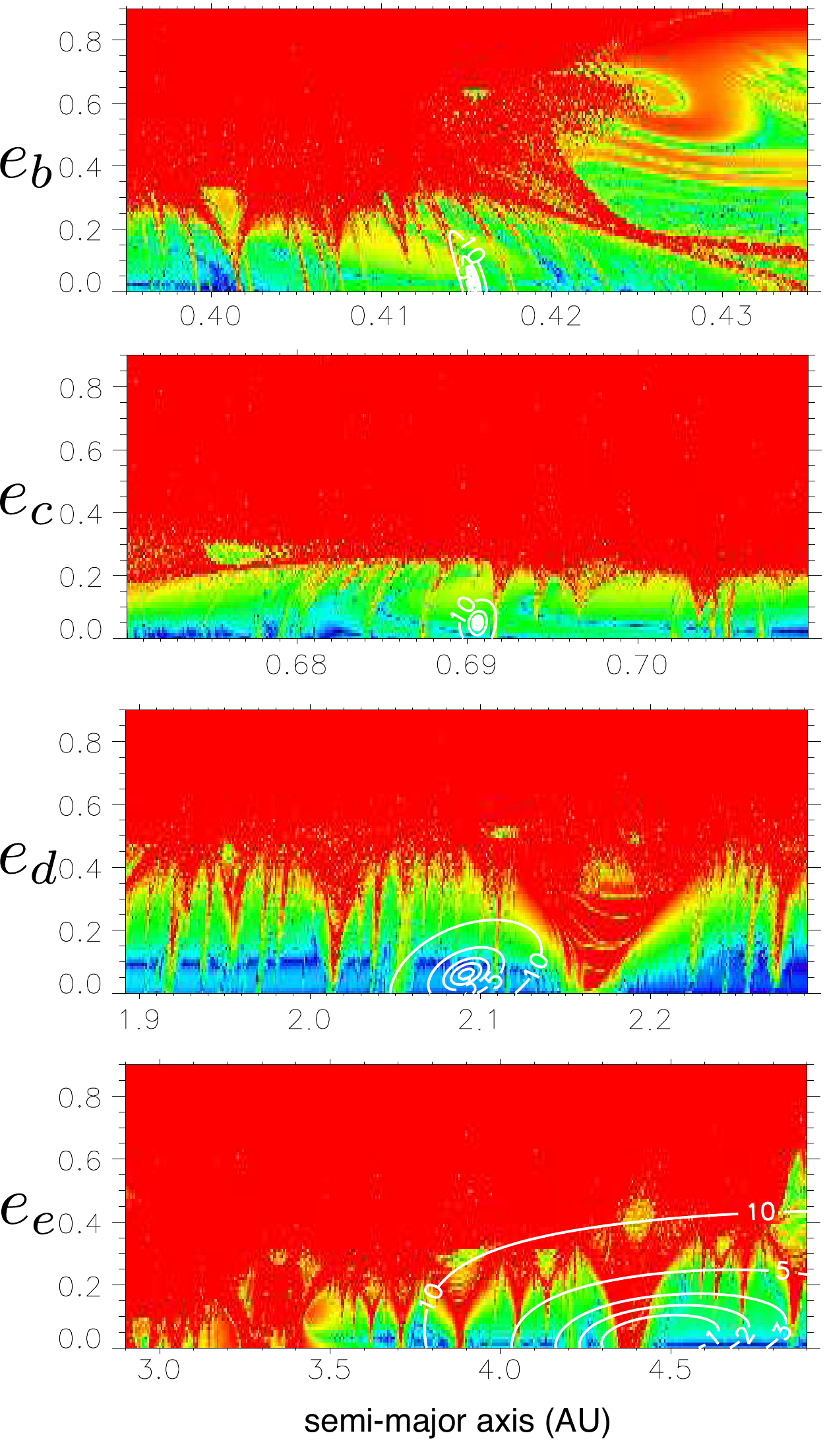} 
  \caption{
 Stability analysis of the   dynamically stable orbital solution for the HD\,141399 planetary system
  (planets HD\,141399b, c, d, and e from top to bottom).
  As in Fig.~\ref{Fdyn3}, the color scale shows $ D $ values between $-6$ and $2$.
  The red zone corresponds to highly unstable 
  orbits while  the dark-blue region can be assumed to be stable on a
  billion-years timescale.
  The contour curves indicate the 1, 2, 3, 5, 10-$\sigma$ confidence intervals from~$\Delta\chi^2$.
  The minima always lie in a stable region for all four planets in the system.
  \label{Fdyn4b}}   
\end{figure}

In Fig.~\ref{Fdyn4} we also observe that any solution with eccentricity above 0.1 is unstable.
Therefore, stable solutions for the HD\,141399 system require a small eccentricity value for
 the outermost planet.
For simplicity, we proceed as in Vogt et al.~(\cite{vogt14}) and fix the eccentricity at zero ($e_e = 0$).
We refit the observational data holding this parameter constant, which provides a new orbital 
configuration for the system.
The orbits of the three inner planets are almost unchanged with respect to the Keplerian solution, 
but the orbital period of the outermost planet is now better constrained. 
In order to check the robustness of this new solution, we then redo a stability study
 in the vicinity of the outermost planet. Again, the minima of the $\chi^2$ level curves lie inside a chaotic region, 
namely, the resonant island corresponding to a 3:1 mean motion resonance between planets HD\,141399d 
and HD\,141399e.

Contrarily to other planetary systems (see, e.g., Laskar \& Correia~\cite{laskar09}), 
the 3:1 mean motion resonances is unstable in this case.
However, just outside this resonance the system appears to be stable.
It is then likely that the period ratio between the two outermost planets 
is slightly shorter or longer than the 3:1 ratio, longer periods being even more stable.

We can test this new scenario by additionally fixing the orbital period of 
the outer planet slightly above the resonant ratio (we chose $P_e = 3412$~days here).
We fitted the radial velocities holding this parameter constant 
(together with $e_e = 0$), which provides a new orbital configuration for the system. 
The parameters are all similar to the Keplerian ones presented in Table~\ref{table_parameters_multi}, 
except the period and eccentricity of HD\,141399e.
Both solutions provide similar fits in terms of dispersion of their residuals, but the 
new solution is stable whereas the Keplerian one is not.
We perform a global stability study in the vicinity of this new solution, which is shown 
in Fig.~\ref{Fdyn4b}. We verify that the minima of the $\chi^2$ level curves lie 
inside stable regions for all four planets in the system.
At $e_e=0$, the intersection between $\chi^2$ level curves and stabilities zones provides 
the two intervals corresponding to $a_e = 4.33 \pm 0.02$ and $a_e = 4.50 \pm 0.08$
for  HD\,141399e, below and above the 3:1 resonance respectively.
We finally adopt the  solution above the resonance, which is the most stable; 
it corresponds to the orbital period $P_e=3370 \pm 90 $~days.

\begin{figure}[h]
\includegraphics[width=\columnwidth,angle=0]{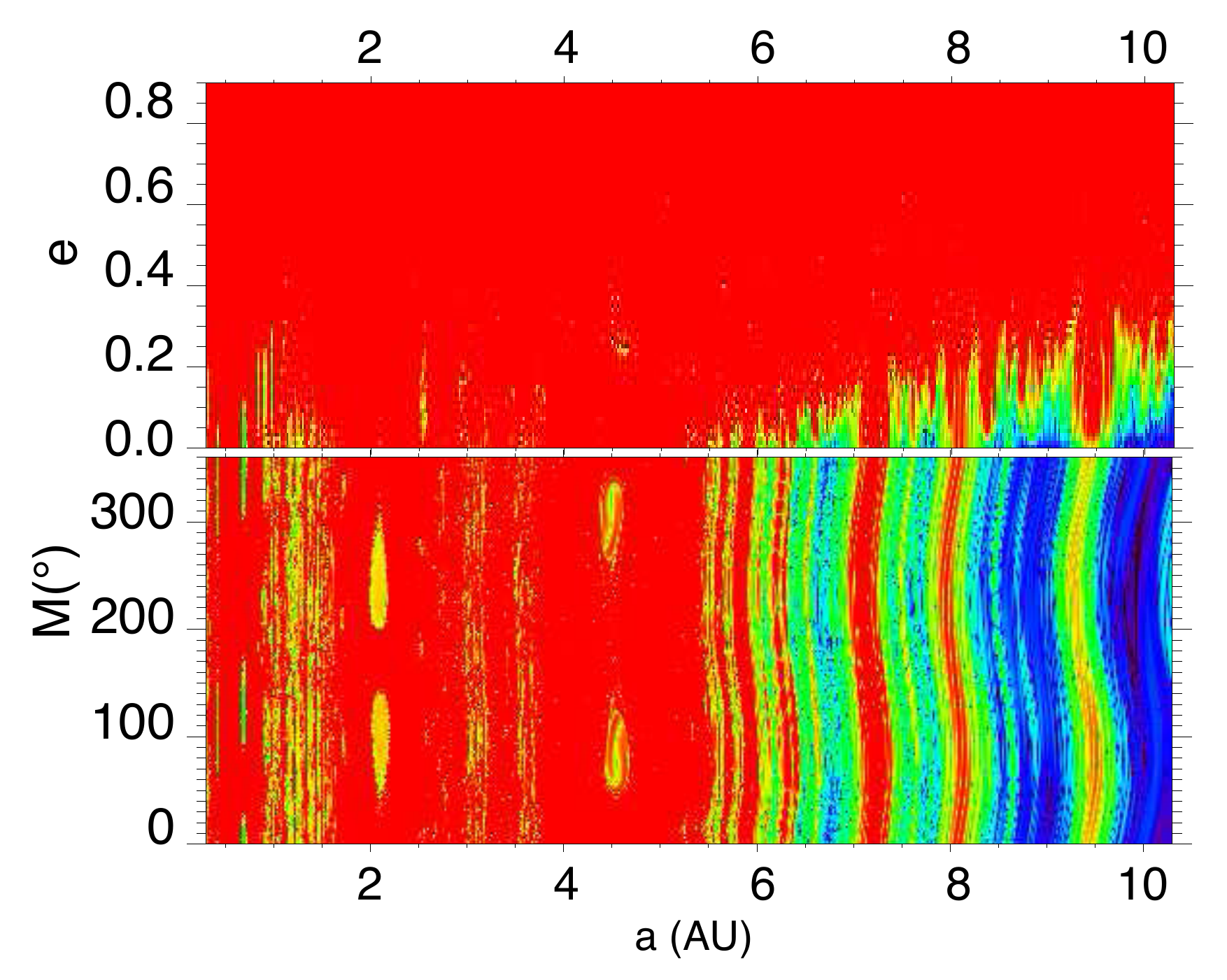} 
  \caption{Possible location of an additional fifth planet in the HD\,141399 system.
  The stability of an Earth-mass planet is analyzed, for various semi-major axis versus 
  eccentricity ({\it top}), or mean anomaly ({\it bottom}). All the angles of the putative planet are set 
  to $0^\circ$ (except the mean anomaly in the bottom panel), and in the bottom panel its eccentricity is 0.
  As in Fig.~\ref{Fdyn3}, the color scale corresponds to $ D $ values between $-6$ and $2$.
  The stable zones  where additional planets can be found are the dark-blue regions.
    \label{Fdyn9}}   
\end{figure}

\subsubsection{Additional constraints from dynamic}

Similarly to the HIP\,65407 system (Sect.~\ref{bdac}), we can assume that 
the dynamics of the already known planets is not disturbed much by the presence of an additional 
fifth, low-mass still undetected planet.
This test is again completed 
for a fixed value  $K=0.1$\,m/s, corresponding to
a possible Earth-mass object at approximately 1\,AU, whose radial-velocity amplitude is at the edge of detection. 
From the analysis of the stable areas in Fig.~\ref{Fdyn9}, one can see that additional planets are  possible beyond 6\,AU, 
which corresponds to orbital periods longer than 14~years.
We also spot some small stability islands at the same semi-major axes of planets HD\,141399d and HD\,141399e;
they correspond to the Lagrange co-orbital resonance of both planets, i.e. Trojan~planets.

We can also try to find constraints on the maximal masses of the current system if 
we assume co-planarity of the orbits, 
as we did above in the case of BD+492211 (Sect.~\ref{bdac}).
We conclude that the maximum masses of the planets may be
most probably computed for an inclination around $10^\circ$, corresponding to 
4, 11, 9, and 5\,M$_\mathrm{Jup}$ for planet HD\,141399b, c, d, and e, respectively.

\section{Conclusions}
\label{sect_conclusion}

\begin{figure}[h] 
\begin{center}
\includegraphics[width=\columnwidth]{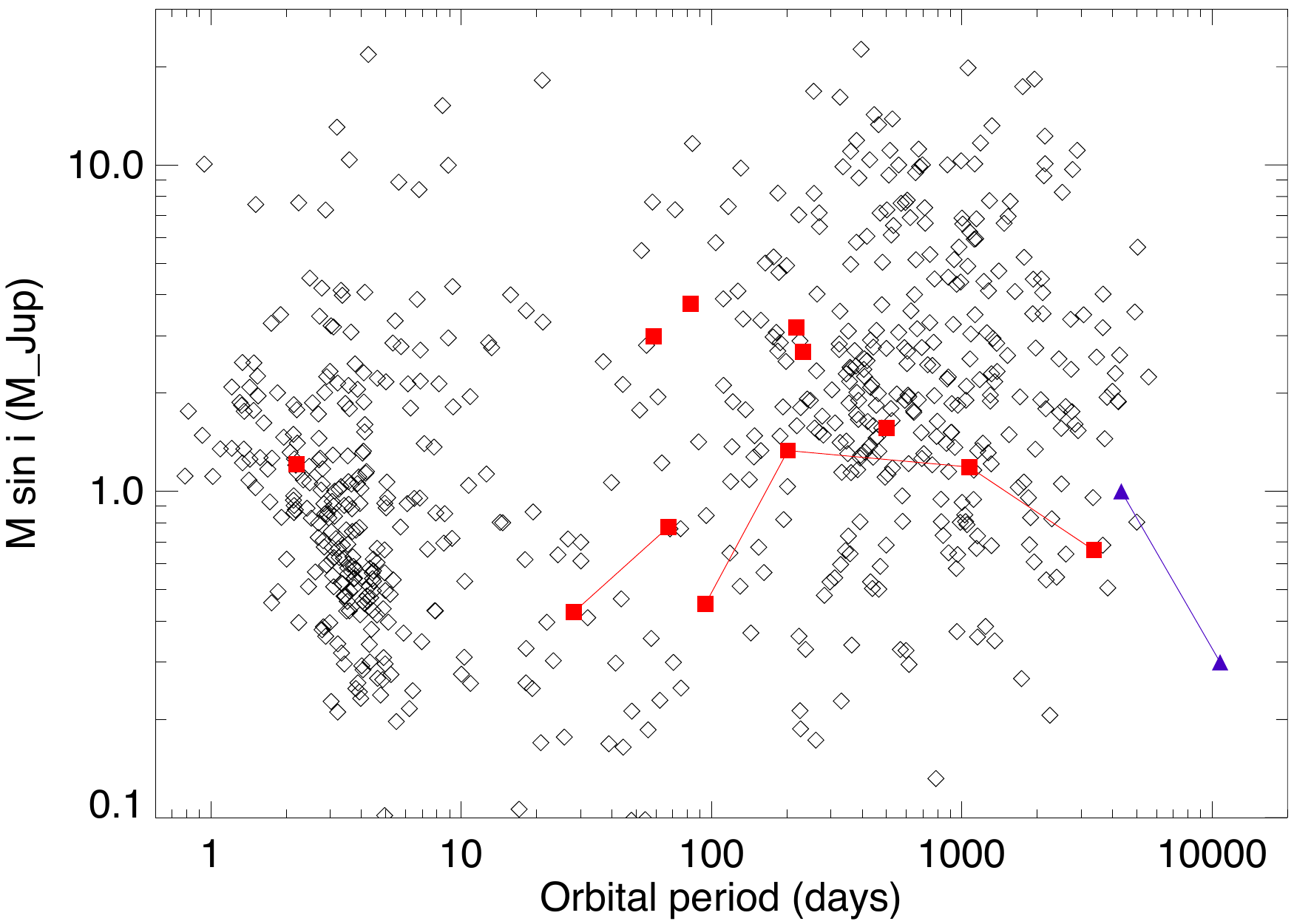}
\caption{Sky-project masses and orbital periods of known giant planets (diamonds, from exoplanets.org).
The 12 planets from the present study are plotted in red squares. The planets from the multiple systems 
HIP\,65407 and HD\,141399 are linked by lines. Jupiter and Saturn are also plotted and linked (blue triangles).
With the adopted range in $y$-axes the plot only shows giant planets; planets with sky-projected masses below 
$0.1\,$M$_{\mathrm{Jup}} = 32\,$M$_{\mathrm{Earth}}$ are not displayed here.
}
\label{fig_concle}
\end{center}
\end{figure}

We have presented new radial velocity measurements of eight stars secured with the spectrograph 
SOPHIE at the 193-cm telescope of the Haute-Provence Observatory
allowing the detection and characterization of new giant extrasolar planets. 
Seven of the stars allowed us to discover and characterize six new single planets as well as 
a system including two planets. The eighth star was already known to host a multi-planetary system
and we published here our independent measurements and discovery. 
The 12 giant planets studied here are compared in Fig.~\ref{fig_concle} 
with the other known giant exoplanets. We measured their masses with an 
accuracy of the order of $\pm5$\,\%.
They allow the diversity of extrasolar planets populations to be 
explored, putting constraints on models of planetary systems formation and evolution. 

We summarize below some of their notable properties:

\begin{itemize}

\item
HD\,143105 is one of the brightest stars known to host a hot Jupiter, 
which could allow numerous follow-up studies to be conducted despite this is not a transiting system;

\item
HIP\,109600b, HD\,35759b, HIP\,109384b, HD\,12484b are new giant, single planets with orbital
periods of several~months, with no hints of additional companions in the~systems;

\item
the  data of the new single-planet systems allow us to exclude the possibility that  they 
actually are composed of two circular planets in 2:1 resonance;
\item
HD\,220842 hosts a massive, giant planet, as well as another substellar companion which need long-term 
follow-up to be~characterized; 

\item
the hot Jupiter HD\,143105b and the warm Jupiter HD\,12484b are among the few planets 
detected around moderately fast rotating stars,  HD\,12484 being in addition a particularly active~star;

\item
the two multi-planetary systems around HIP\,65407 and~HD\,141399 have host stars with particularly high metallicities;

\item
HIP\,109600, HIP\,109384, and HD\,220842 are metal-poor stars, which is unusual for hosts of giant planets;

\item
the system around HIP\,65407 includes two planets 
just outside the instable 12:5 resonance, both precessing with a period of 206~years;

\item
in order to be stable, the nearly Jupiter-twin HD\,141399e should have an eccentricity
below 0.1 and an orbital period $P = 3370 \pm 90$~days;

\item
the giant planets HIP\,109600b, HIP\,109384b, and HD\,141399c are located in the habitable zone of their host star, 
making particularly interesting any potential satellite orbiting these planets, even if such moons remain now 
out of reach of standard detection technics.

\end{itemize}

\begin{acknowledgements}
We thank all the staff of Haute-Provence Observatory for their 
support at the 1.93-m telescope and on SOPHIE.
This work is also based in part on data collected at Subaru Telescope, which is operated
by the National Astronomical Observatory of Japan, and on 
observations taken with the Remote Observatory 
Theoretical Astrophysics Tuebingen (ROTAT) located at OHP (France). 
ROTAT is operated by the Foundation Interactive Astronomy and 
Astrophysics (Stiftung Interaktive Astronomie und Astrophysik; 
www.stiftung-astronomie.de).
This work was granted access to the HPC resources of MesoPSL financed
by the Region \^Ile-de-France and the project Equip@Meso (reference
ANR-10-EQPX-29-01) of the programme \textit{Investissements d'Avenir} supervised
by the Agence Nationale pour la Recherche.
This work was also supported by the ``Programme National de Plan\'etologie'' (PNP) of CNRS/INSU, 
the Swiss National Science Foundation, 
the French National Research Agency (ANR-12-BS05-0012),
and the Funda\c{c}\~ao para a Ci\^encia e a Tecnologia (FCT) through the research grants 
UID/FIS/04434/2013 and PTDC/FIS-AST/1526/2014. 
N.\,C.\,S.  acknowledges support from FCT 
through Investigador FCT contracts of reference IF/00169/2012, and POPH/FSE (EC) by FEDER 
funding through the program ``Programa Operacional de Factores de Competitividade - COMPETE''.  
A.\,S. is supported by the European Union under a Marie Curie Intra-European Fellowship for Career 
Development with reference FP7-PEOPLE-2013-IEF, number 627202.
J.\,H. is supported by the Swiss National Science Foundation (SNSF).
A.\,C.\,M.\,C. and J.\,L. acknowledge support from the ``conseil scientifique'' of the Observatory of 
Paris and CIDMA strategic project UID/MAT/04106/2013.
P.\,A.\,W acknowledges the support of the French Agence~Nationale de la Recherche (ANR), 
under program ANR-12-BS05-0012 "Exo-Atmos". 
\end{acknowledgements}

\end{document}